%% file: nested-long.tex
\documentclass{LMCS}

\def\doi{7 (3:21) 2011}
\lmcsheading%
{\doi}
{1--42}
{}
{}
{Mar.~\phantom08, 2011}
{Sep.~28, 2011}
{}   

\usepackage{enumerate}
\usepackage{amsmath,amssymb,xspace}
\usepackage{caption}
\usepackage{stmaryrd}
\usepackage{mathpartir}
\usepackage{hyperref}

\newcommand{\myskip}[1]{}

\newtheorem{theorem}{Theorem}[section]
\newtheorem{lemma}[theorem]{Lemma}

\newtheorem{definition}[theorem]{Definition}
 
\theoremstyle{remark}
\newtheorem{remark}[theorem]{Remark}
\newtheorem{example}[theorem]{Example}

\newcommand{\ALMOSTPURE}{\text{pseudo pure}\xspace}
\newcommand{\Heap}{\ensuremath{\mathit{Heap}}\xspace}

\newcommand{\Val}{\ensuremath{\mathit{Val}}\xspace}

\newcommand{\Com}{\ensuremath{\mathit{Com}}\xspace}

\newcommand{\ERR}{\ensuremath{\mbox{\it error}}\xspace}

\newcommand{\SEXP}{{\mathit{Exp}}\xspace}

\newcommand{\SCOM}{{\mathit{Com}}\xspace}
\newcommand{\QUOTE}[1]{\textnormal{`\ensuremath{#1}'}}
\newcommand{\UNQUOTE}[1]{\SYN{eval}\,{#1}}
\newcommand{\SYN}[1]{\ensuremath{\texttt{#1}}}

\renewcommand{\l}{\ensuremath{\ell}\xspace}

\newcommand{\fvar}{\ensuremath{\mathsf{fv}}}
\newcommand{\fv}[1]{\ensuremath{\mathit{fv}(#1)}}

\newcommand{\bnfeq}{\ensuremath{::=}}
\newcommand{\hmid}{\ensuremath{\;|\;}}
\newcommand{\defiff}{\ensuremath{\stackrel{\mbox{\tiny \it def}}{\Leftrightarrow}}}
\newcommand{\defeq}{\ensuremath{\stackrel{\mbox{\tiny \it def}}{=}}}
\newcommand{\ignore}[1]{}
\newcommand{\Rec}[1]{\ensuremath{\mathit{Rec}(#1)}}
\newcommand{\recselect}[2]{\ensuremath{{#1}({#2})}}  

\newcommand{\record}[1]{\ensuremath{\left.\!\{\!|{#1}|\!\}\right.\!}}
\newcommand{\dom}[1]{\ensuremath{\mathsf{dom}({#1})}}

\newcommand{\id}{\ensuremath{\mathit{id}}}

\newcommand{\unlift}[1]{\ensuremath{#1_\downarrow}}
\newcommand{\lift}[1]{\ensuremath{#1_\bot}}
\newcommand{\COMB}{\ensuremath{\cdot}}
\newcommand{\DISJ}{\ensuremath{\mathop{\#}}}

\newcommand{\TERR}{\ensuremath{T_\textit{err}}\xspace}

\newcommand{\Cppo}{\ensuremath{\textbf{Cppo}}\xspace}

\newcommand{\Env}{\ensuremath{\textit{Env}}\xspace}
\newcommand{\Var}{\ensuremath{\mathit{Var}}\xspace}

\newcommand{\Nats}{\ensuremath{\mathit{Nats}^{\scriptsize +}}\xspace}
\newcommand{\Ints}{\ensuremath{\mathit{Integers}}\xspace}

\newcommand{\IMPLIES}{\ensuremath{\Rightarrow}}
\newcommand{\IFF}{\ensuremath{\Leftrightarrow}}

\newcommand{\sto}{\ensuremath{\multimap}}
\newcommand{\key}[1]{\ensuremath{\mathrm{#1}}\xspace}

\newcommand{\triple}[3]{{\ensuremath{\!\left.\{ #1 \}\, #2\, \{  #3 \}\!\right.}}}

\newcommand{\True}{\ensuremath{\mathit{true}}}
\newcommand{\False}{\ensuremath{\mathit{false}}}
\newcommand{\pointsto}{\ensuremath{\mapsto}}

\newcommand{\ASR}{{\mathit{Assn}}\xspace}
\newcommand{\EMP}{\ensuremath{\textit{emp}}}
\newcommand{\wemp}{\textnormal{\textit{emp}}}

\newcommand{\rnk}[1]{\textit{rnk}(#1)}

\newcommand{\den}[1]{%
  \left\llbracket #1
  \right\rrbracket}                            

\newcommand{\EXPden}[1]{\ensuremath{\den{#1}}}

\newcommand{\mimp}{\,{-\!\!*}\,}

\newcommand{\ntria}[4]{{{#1}\,{\mapsto}\,{{\{{#2}\}}{\, #3\, }{\{{#4}\}}}}\xspace}

\newcommand{\commentout}[1]{}

\newcommand{\Pred}{\ensuremath{\mathrm{Pred}}\xspace}

\newcommand{\Ad}{\mathrm{Ad}}  
\newcommand{\DCl}{\Ad}  
\newcommand{\UAdm}{\mathit{UAdm}\xspace}
\newcommand{\CBUlt}{\mathit{CBUlt}\xspace}
\newcommand{\RANK}[2]{\ensuremath{{#1}_{[#2]}}}
\newcommand{\RR}{\ensuremath{\mathbb R}}
\newcommand{\W}{\ensuremath{W}}
\newcommand{\FOLD}{\ensuremath{\iota}}
\newcommand{\UNFOLD}{\ensuremath{\iota^{-1}}}
\newcommand{\nequiv}[1]{\ensuremath{\mathrel{\stackrel{#1}{=}}}}
\newcommand{\PHI}[1]{\ensuremath{\mathop{\overline{#1}}}}

\newcommand{\HOLE}{\ensuremath{\cdot}}

\newcommand{\adash}{\ensuremath{\vdash}} 

\renewcommand{\clubsuit}{\boxempty}

\newcommand{\X}{\ensuremath{\Xi}}

\title[Nested Hoare Triples and Frame Rules for Higher-order Store]{Nested Hoare Triples and Frame Rules for Higher-order Store\rsuper*} 
  
\author[J.~Schwinghammer]{Jan Schwinghammer\rsuper a}
\address{{\lsuper a}Programming Systems Lab, Saarland University, 66123 Saarbr\"ucken, Germany}
\email{jan@ps.uni-saarland.de}

\author[L.~Birkedal]{Lars Birkedal\rsuper b} 
\address{{\lsuper b}IT University of Copenhagen, Rued Langgaards Vej 7, 2300 K{\o}benhavn S., Denmark}
\email{birkedal@itu.dk}

\author[B.~Reus]{Bernhard Reus\rsuper c}
\address{{\lsuper c}School of Informatics, University of Sussex, Brighton BN1 9QH, U.K.}
\email{bernhard@sussex.ac.uk}

\author[H.~Yang]{Hongseok Yan\rsuper dg} 
\address{{\lsuper d}Department of Computer Science, University of Oxford, Oxford OX1 3QD, U.K.}
\email{Hongseok.Yang@cs.ox.ac.uk}

 \keywords{Higher-order store, Hoare logic, separation logic, semantics.}
\subjclass{F.3.1, F.3.2}
\titlecomment{{\lsuper*}A preliminary version of this work was presented at the 18th EACSL Annual Conference on Computer Science Logic (CSL'09), 7--11 September 2009, Coimbra, Portugal \cite{schwinghammerBRY09}.}

\begin{document}

\begin{abstract}
\noindent
Separation logic is a Hoare-style logic for reasoning about programs with heap-allocated mutable data structures.  
As a step toward extending separation logic to high-level languages with ML-style general (higher-order) storage, we investigate the compatibility of nested Hoare triples with several variations of higher-order frame rules. 

The interaction of nested triples and frame rules can be subtle, and the inclusion of certain  frame rules is in fact unsound. 
A particular combination of rules can be shown consistent by means of a Kripke model where worlds live in a recursively defined ultrametric space. 
The resulting logic allows us to elegantly prove programs involving stored code. In particular, using recursively defined assertions, it leads to natural specifications and proofs of invariants  required for dealing with recursion through the store.
\end{abstract}

\maketitle

\section{Introduction}
\label{sec:intro}

Many programming languages permit not only the storage of first-order data, but also forms of higher-order store. Examples are code pointers in C, and ML-like general references. 
It is therefore important to have modular reasoning principles for these language features. 
Separation logic is  an effective formalism for modular reasoning about pointer programs, in low-level C-like programming languages and, more recently, also in higher-level languages 
\cite{Krishnaswami:ML-SL07,Nanevski:Morrisett:Shinnar:Govereau:Birkedal:08,Parkinson:Biermann:08,Reynolds:02a}.
However, its assertions are usually limited to talk about first-order data. 

In previous work, we have begun the study of separation logic for languages with higher-order store \cite{Birkedal:Reus:Schwinghammer:Yang:08,Reus:Schwinghammer:06}. A challenge in this research is the combination of proof rules from separation logic for  modular reasoning, and proof rules for code stored on the heap. 
 Ideally, a program logic for higher-order store provides sufficiently expressive proof rules that, e.g., can deal with recursion through the store, and at the same time interact well with (higher-order) frame rules, which enable modular program verification. 

Our earlier work \cite{Birkedal:Reus:Schwinghammer:Yang:08,Reus:Schwinghammer:06} shows that separation logic is consistent with higher-order store. However, the formulation in this earlier work has a shortcoming: code is treated like any other data in that  assertions can only mention concrete commands. In order to obtain modular, open and reusable reasoning principles,  it is clearly desirable to abstract from particular code and instead (partially) specify its behaviour. For example, when verifying mutually recursive procedures on the heap, one would like to consider each  procedure in  isolation, relying on properties but not the implementations of the others. The recursion rule given by Birkedal et al.\ \cite{Birkedal:Reus:Schwinghammer:Yang:08} and Reus and Schwinghammer  \cite{Reus:Schwinghammer:06} does not achieve this. 
A second, and less obvious consequence of  lacking  behavioural specifications for code in assertions is that one cannot take full advantage of the frame rules of separation logic. For instance, 
the programming language in \cite{Birkedal:Reus:Schwinghammer:Yang:08} 
can simulate higher-order procedures by passing arguments through the heap, but the available (higher-order) frame rules are not useful here because an appropriate specification for this encoding is missing. 

In this article, we address these shortcomings by investigating a program logic in which stored code can be specified using Hoare triples, i.e., an assertion language with \emph{nested triples}. 
This is an obvious idea, but the combination of nested triples and frame rules turns out to be tricky: the most natural combination is in fact unsound. 

The main technical contributions of this article are therefore: 
\begin{enumerate}[(1)]
\item the observation that certain ``deep'' frame rules can be unsound, 
\item the suggestion of a ``good'' combination of nested Hoare triples and frame rules,  and 
\item the verification of those  rules  by means of   an elegant Kripke model, based on a denotational semantics of the programming language, where the worlds are themselves world-dependent sets of heaps. 
\end{enumerate}
The worlds form a complete metric space and  (the denotation of) the operation $\otimes$, needed to generically express higher-order  frame rules, is contractive; as a consequence, our logic permits recursively defined assertions. 

\paragraph{Outline} After introducing the syntax of programming language and assertions in Section~\ref{sec:syntax} we discuss some unsound combinations of rules in Section~\ref{sec:ProgramLogic}. This section also contains the suggested set of rules for our logic.   The soundness of the logic is then shown  in Section~\ref{sec:Semantics}.
Section~\ref{sec:discussion-of-proof-rules} discusses further proof rules for nested triples. Finally the conclusion  addresses  related work and   the differences between the model presented here and a step-indexed model.

\section{Syntax of Programs and Assertions}
\label{sec:syntax}
 
This section presents the syntax of the programming language and that of assertions.

\subsection{Programming language}
We consider a simple imperative programming language extended with operations for 
stored code and heap manipulation. The syntax of the language is shown in 
Figure~\ref{fig:LanguageSyntax-alt}. The expressions in the language are
integer expressions, variables, and the quote expression
$\QUOTE{C}$ for representing an unevaluated command $C$. The integer
or code value denoted by expression $e_1$ can be stored in a heap cell
$e_0$ using $[e_0]{:=}e_1$, and this stored value can later be looked up
and bound to the (immutable) variable $y$ by $\SYN{let}~y\SYN{=}[e_0]~\SYN{in}~D$.
In case the value stored in cell $e_0$ is code $\QUOTE{C}$, we can
run (or ``evaluate'') this code by executing $\UNQUOTE [e_0]$. Our language also
provides constructs for allocating and disposing heap cells such as $e_0$
above. 

We point out that, as in ML, all variables $x,y,z$ in
our language are \emph{immutable}, so that once they are bound to a value, their values do not change. 
This property of the language lets us avoid side conditions on variables when studying frame rules.
Finally, we do not include while loops in our language; these could be added easily, and they can also be expressed by stored code (using Landin's knot).\footnote{To obtain the original while rule of Hoare logic one needs to be able to hide the additional pointer storing the body of the while loop. This can be achieved  using anti-frame rules as discussed e.g.\ in \cite{schwinghammerYBPR10}.}
 
\begin{example}[Iterate procedure] 
An iterator that calls its parameter function as well as itself through the store can be programmed as follows. 
\[
\begin{array}{lcl}
C_{\mathit{it},f,c} & \equiv & \SYN{let}\ n\, \SYN{=}\, [c]\ \SYN{in} \\
&& \SYN{if}\ n\,\SYN{=0}\ \SYN{then}\ \SYN{skip}\  \SYN{else}\  (\,  \UNQUOTE{[f]} \SYN{;}\ [c]\, \SYN{:=}\, n \SYN{-1} \SYN{;}\ \UNQUOTE{[\emph{it}]} \, ) 
\end{array}\] 
Here we assume that cells \emph{it}, $f$ and $c$ are some fixed global constants, and that the iterator code is stored in the cell \emph{it}.
Command $C_{\mathit{it},f,c}$ then calls the code in $f$ as many times as the value of counter cell $c$ prescribes.
\end{example}

\begin{figure*}[t]
\hrule
$$
\begin{array}{@{}r@{\;}c@{\;}l@{\;\;}l@{}}
e\in\SEXP 
& \bnfeq &
  \SYN{0}\hmid 
  \SYN{-1}\hmid
  \SYN{1}\hmid
  \dots\hmid  
  e_1{+}e_2 \hmid 
  \dots \hmid
  x 
& \mbox{integer expressions, variable} 
\\
& \hmid & 
  \QUOTE{C} 
& \mbox{quote (command\,as\,expression)} 
\\[1ex]
C\in\SCOM 
& \bnfeq &
   [e_1]\SYN{:=}e_2 \hmid
  \SYN{let}~y\SYN{=}[e]~\SYN{in}~C \hmid
  \UNQUOTE [e] 
  & \mbox{assignment, lookup, unquote}
\\
& \hmid &
  \SYN{let}~x{=}\SYN{new}~(e_1,\ldots,e_n)~\SYN{in}~C\hmid\SYN{free}~e
  & \mbox{allocation, disposal}
\\
& \hmid &
  \SYN{skip}\! \hmid \!
  C_1\SYN{;}C_2 
  & \mbox{no op, sequencing } 
\\
& \hmid &
  \SYN{if}\,(e_1{=}e_2)\,\SYN{then}\,C_1\,\SYN{else}\,C_2   
  & \mbox{conditional} 
\\[1ex]
P,Q 
\,{\in}\, \ASR 
& \bnfeq & 
   \False \hmid 
   \True \hmid 
   P\,{\vee}\, Q \hmid 
   P\,{\wedge}\,Q \hmid 
   P\,{\Rightarrow}\, Q
   & \mbox{intuitionistic-logic connectives} 
\\
& \hmid &
   \forall x. P \hmid
   \exists x. P \hmid 
   e_1 {=} e_2 \hmid 
   e_1 {\leq} e_2 
  & \mbox{quantifiers, atomic formulas} 
\\
& \hmid &  
   e_1\,{\mapsto}\, e_2 \hmid 
   \EMP \hmid 
   P*Q 
& \mbox{separating connectives} 
\\
& \hmid &  
  \triple{P}{e}{Q} \hmid 
   P\otimes Q 
& \mbox{Hoare triple, invariant extension}
\\
&\hmid & 
  X(\vec e)\hmid (\mu X(\vec x).P)(\vec e)\hmid 
   \dots
   &\mbox{relation variable, recursion}
\end{array}
$$
\hrule
\caption{\label{fig:LanguageSyntax-alt}Syntax of expressions, commands and assertions}
\end{figure*}

\subsection{Assertions and distribution axioms}

Our assertion language is standard first-order intuitionistic logic,
extended with
separating connectives $\EMP$ and $*$, the points-to predicate 
${\pointsto}$~\cite{Reynolds:02a},
and recursively defined assertions $(\mu X(\vec x).P)(\vec e)$. 
The syntax of assertions appears in Figure~\ref{fig:LanguageSyntax-alt}.
Each assertion describes a property of states, which consist of an immutable stack and 
a mutable heap. Formula $\EMP$ means that the heap component of the state is
empty, and $P*Q$ means that the heap component can be split
into two, one satisfying $P$  and the other satisfying $Q$, both evaluated with respect to the same
stack. The spatial implication operator (``magic wand'') is omitted here for reasons explained
later in Remark~\ref{remark:wand}. The points-to predicate $e_0\pointsto e_1$
states that the heap component consists of only one cell $e_0$ whose 
content is  $e_1$ or, in case $e_1$ is a command, an approximation  $e'$ of $e_1$ which is defined (terminates) 
for less heaps than $e_1$. This is in line with the fact that we consider \emph{partial correctness} only.

One interesting aspect of our assertion language is that it includes Hoare triples 
$\triple{P}{e}{Q}$ and invariant extensions $P\otimes Q$; previous work
\cite{BirkedalL:semslt-lmcs,Birkedal:Reus:Schwinghammer:Yang:08} does not treat them as assertions 
but as so-called \emph{specifications}, which form a different syntactic category. 
A consequence of having these new constructs as assertions
is that they allow us to study proof rules for exploiting locality of stored code
systematically, as we will describe shortly.

Intuitively, $\triple{P}{e}{Q}$ means that $e$ denotes code 
satisfying $\triple{P}{\_}{Q}$, and $P\otimes Q$ denotes a modification of $P$ where 
all the pre- and post-conditions of triples inside $P$ are $*$-extended with $Q$. In other words,
all code specified by pre- and postconditions inside $P$ must preserve  invariant $Q$. For instance,
the assertion
$(\exists k.\, (1\,{\pointsto}\, k) \wedge \triple{\EMP}{k}{\EMP}) \otimes (2{\pointsto}0)$
is equivalent to
$(\exists k.\, (1\,{\pointsto}\, k) \wedge \triple{2{\pointsto}0}{k}{2{\pointsto}0})$.
This assertion says that cell $1$ is the only cell in the heap
and it stores code $k$ that satisfies the triple $\triple{2{\pointsto}0}{\_}{2{\pointsto}0}$.
This intuition about the $\otimes$ operator is made precise in the set of
axioms in Figure~\ref{fig:DistRules}, which let us distribute $\otimes$ through
  the constructs of the assertion language.

\begin{figure*}[t]
\hrule
$$
\begin{array}{@{}r@{}c@{}l@{\quad}l@{}}
P \circ R &\;\defeq\;& (P\otimes R) * R
\\[1ex]
  \triple{P}{e}{Q} {\otimes} R
&
  \;\Leftrightarrow\;
&
  \triple{P \,{\circ}\, R}{e}{Q \,{\circ}\, R}
\\
  (P\,{\otimes}\,R')\,{\otimes}\, R
&
  \;\Leftrightarrow\;
&
  P\,{\otimes}\,(R'\,{\circ}\, R)
\\
  (\kappa x. P)\,{\otimes}\, R
&
  \;\Leftrightarrow\;
&
  \kappa x. (P \,{\otimes}\, R)
& (\kappa \,{\in}\, \{\forall,\exists\}, x\notin\fv{R})
\\
  (P \,{\oplus}\, Q)\,{\otimes}\, R
&
  \;\Leftrightarrow\;
&
  (P \,{\otimes}\, R) \,{\oplus}\, (Q \,{\otimes}\, R)
&
  (\oplus \,{\in}\, \{\Rightarrow, \wedge, \vee, *\})
\\
  P \otimes R
&
  \;\Leftrightarrow\;
&
  P
&
(\mbox{$P$ is one of $\True$, $\False$, $\EMP$, $e\mathop{=}e'$, $e\mathop{\pointsto} e'$
})
\end{array}
$$
\hrule
\caption{Axioms for distributing $-\otimes R$}
\label{fig:DistRules}
\end{figure*}

Note that since triples are assertions, they can appear in pre- and post-conditions of
triples. This \emph{nested} use of triples is useful in reasoning, because it allows one to specify
stored code behaviourally, in terms of properties that it satisfies. 
Typically, 
a program logic consists of both an assertion logic and a specification logic (e.g.\ \cite{Reynolds:82}). 
With the introduction of nested triples, assertions and specifications necessarily become mutually recursive; 
for simplicity, we have chosen to identify our
specification and assertion logics and just work with a single logic of assertions. 


A second interesting aspect of our assertion language is that assertions include ($n$-ary) relation variables $X(\vec e)$, and that assertions can be defined recursively: the assertion $(\mu X(\vec x).P)(\vec e)$ binds $X$ and $\vec x=x_1\ldots x_n$ in $P$ and satisfies the axiom 
\begin{align}
\label{eqn:unfolding-axiom}
(\mu X(\vec x).P)(\vec e) \Leftrightarrow P[X:=\mu X(\vec x).P,\, \vec x:=\vec e]\ .
\end{align}
In the case where $X$ has arity 0 we will simply write $X$ in place of $X()$. 

\begin{example}[Specification of the iterator via recursion through the store]\label{example:spec}
The previously given command
\[\begin{array}{lcl}
C_{\mathit{it},f,c} & \equiv & \SYN{let}\ n\, \SYN{=}\, [c]\ \SYN{in} \\
&& \SYN{if}\ n\,\SYN{=0}\ \SYN{then}\ \SYN{skip}\  \SYN{else}\  (\,  \UNQUOTE{[f]} \SYN{;}\ [c]\, \SYN{:=}\, n \SYN{-1} \SYN{;}\ \UNQUOTE{[\emph{it}]} \, ) 
\end{array}\]
can be specified as follows, if we assume that the called procedure in $f$ does preserve some invariant $I$ that does not access the counter and iterator cells $c$ and $\mathit{it}$, respectively. For instance, $I$ could be $\EMP$ (in case $f$ has no side effects) or \ $\exists m.\, x\mapsto m \, *\, y\mapsto n!/m!$ when the factorial  of $n$ is computed in $y$. If $x$ contains the content of the counter then, like with a while loop, upon termination $y$ contains the expected result. In the following, to keep the triples simple, we assume that $I=\EMP$. 
$$
\triple{c\, {\pointsto}\_ * f\,{\pointsto}\triple{\EMP}{\_}{\EMP} * R_{\mathit{it}}}
{\QUOTE{C_{\mathit{it},f,c}}}
{c\,{\pointsto}0  * f\,{\pointsto}\triple{\EMP}{\_}{\EMP} * R_{\mathit{it}}}\ .$$
Here, we use the abbreviation 
$e\,{\pointsto}\,\triple{P}{\_}{Q}$ for 
$(\exists k.\, (e\,{\pointsto}\, k) \wedge \triple{P}{k}{Q})$, and
 $R_{\mathit{it}}$ is a recursive specification for the iterator itself:
$$R_{\mathit{it}} \equiv \mu X.\, \mathit{it} \, {\pointsto}\,
 \triple{c\,{\pointsto}\_ * f\,{\pointsto}\triple{\EMP}{\_}{\EMP} * X}
 {\_}
{c\,{\pointsto}0 * f\,{\pointsto}\triple{\EMP}{\_}{\EMP} * X}\ .
$$
Consequently, heap $\mathit{it}\,{\pointsto}\, \QUOTE{C_{\mathit{it},f,c}}$ is in $R_{\mathit{it}}$ and thus one can prove (see Example~\ref{example:eval}) that
\[\begin{array}{l}
\{\, c\,{\pointsto}\_ * f\,{\pointsto}\triple{\EMP}{\_}{\EMP} * \mathit{it}\,{\pointsto}\_ \, \}
\\
\, [\mathit{it}]  \SYN{:=}\,  \QUOTE{C_{\, \mathit{it},f,c}}  \SYN{;}\, \SYN{\UNQUOTE{[$\mathit{it}$]}}\\
\{\, c\, {\pointsto}0 * f\,{\pointsto} \triple{\EMP}{\_}{\EMP} * R_{\mathit{it}} \}\ .
\end{array}
\]
The   specification for the iterator in $\mathit{it}$ is recursive since the iterator calls itself through the store and any recursive call through the store
requires the same specification as the original call.
Assuming that  procedure $f$ has no side effect,
we guarantee that the iterator will  have no other side effect than setting the counter to $0$. 
The iterator specification   also works with more sophisticated behaviour of $f$: in Example~\ref{example:DFR}   below we will   discuss  how to deal with situations where  $f$ has side effects on some heap space (but preserves an invariant $I$). It will turn out that we can generalise from invariant $\EMP$ to $I$ without even having to reprove the original  side-effect free specification given here, using the so-called deep frame rule.
\end{example}

Analogously to the definition of equi-recursive types in typed lambda calculi,  for the assertion $(\mu X(\vec x).P)(\vec e)$ to be well-formed we require that $P$ is \emph{(formally) contractive in X} \cite{Pierce:02}. This means that $X$ can occur in $P$ only in subterms of the form $\triple{P'}{e}{Q'}$ or $P''\otimes R'$ where $P''$ is formally contractive in $X$. (We omit the straightforward inductive definition of formal contractiveness.) 
Semantically, this requirement ensures that  $\mu X(\vec x).P$ is well-defined as a unique fixed point. 
Note that in particular all assertions of the form $P\otimes X$ and $\triple{P'*X}{e}{Q'*X}$ 
are formally contractive in $X$, provided $X$ does not appear in $P$. 
Thus,  $\mu X. P\otimes X$, and $\mu X. \mathit{it} {\pointsto} \triple{P*X}{e}{Q*X}$ are well-formed (in particular, $R_{\mathit{it}}$ above). 
Let $R$  abbreviate  the latter assertion. Then, with the help of Axiom~\ref{eqn:unfolding-axiom} and the distribution axioms of Figure~\ref{fig:DistRules}
one can show that $R$ is equivalent to $\mathit{it} {\pointsto} \triple{P* R}{e}{Q*R}$ which in turn is  equivalent to
 $\mathit{it} {\pointsto} \triple{P*  \mathit{it} {\pointsto}   \triple{P*R}{e}{Q*R}}{e}{Q*   \mathit{it} {\pointsto}   \triple{P*R}{e}{Q*R}}$ and one can keep unfolding $R$ as many times as one wishes. A successful invocation of the code in $\mathit{it}$ thus requires a heap satisfying $P$ as well as containing  $\mathit{it}$ again pointing to  code that satisfies the very same specification. It is this potentially infinite unfolding that frees one from having to prove triples by  various forms of induction on the number of recursive calls as  in \cite{Honda:Yoshida:Berger:05,Birkedal:Reus:Schwinghammer:Yang:08}.

More generally, in order to deal with mutually recursive stored procedures we may need to compute fixpoints of  mutually recursively defined  assertions. 
For brevity we omit formal syntax for mutual recursion.  
We will say more about
the use of recursively defined predicates 
and their existence in Sections~\ref{sec:ProgramLogic} 
and~\ref{sec:Semantics}. 
In particular, the semantics in Section~\ref{sec:Semantics} can be used to interpret mutually recursive families of assertions. 

Finally, note that we have not included an axiom for distributing $\otimes$ through a recursive type in Figure~\ref{fig:DistRules}. In particular, the axiom 
$(\mu X .P)\otimes R \Leftrightarrow \mu X .(P\otimes R)$ 
does not hold in the presence of nested triples. 
Instead, one has to use the   axiom 
$\mu X .P \Leftrightarrow P[X:=\mu X.P]$ 
and unfold the recursive type to exhibit a ``proper'' connective through which $\otimes R$ can be distributed.   
  
We shall make use of two abbreviations. The first is $Q\circ R$, which stands for $(Q\otimes R)  * R$
 and which has already been used in Figure~\ref{fig:DistRules}.
This abbreviation describes the  combination of two invariants $Q$ and $R$ into a single invariant in the axiom $(P\,{\otimes}\,Q)\,{\otimes}\, R\Leftrightarrow P\,{\otimes}\,(Q\,{\circ}\, R)$. It is also 
used to add an invariant $R$ to a Hoare triple $\triple{P}{e}{Q}$,
so as to obtain $\triple{P\circ R}{e}{Q\circ R}$. We use the asymmetric 
$\circ$ instead of the symmetric $*$ here to extend not only $Q$ ($P$ and $Q$ resp.) by $R$ but also ensure, via $\otimes$, that all Hoare triples nested inside $Q$ ($P$ and $Q$, resp.) preserve $R$ as an invariant. The $\circ$ operator has been introduced in \cite{Pottier:08}, where it is credited to Paul-Andr\'{e} Melli\`{e}s and Nicolas Tabareau. 
The second abbreviation is for the points-to operator  of separation logic:
$e_1\,{\pointsto}\, P[e_2] 
\;\defeq\; 
e_1\,{\pointsto}\, e_2 \wedge P[e_2]$
and 
$e_1\,{\pointsto}\,P[\_] 
\;\defeq\; 
\exists x.\,e_1\,{\pointsto}\,P[x]$. 
Here $x$ is a fresh (logic) variable 
and $P[\HOLE]$ is an assertion with an expression hole, such as
$\triple{Q}{\HOLE}{R}$, 
$\HOLE = e$ or $\HOLE\leq e$.\footnote{These abbreviations do not necessarily lead to a unique reading, e.g.\  $x{\pointsto} 1\leq 2$ could mean $x{\pointsto} 1  \wedge  1\leq 2$  or $x{\pointsto} 2  \wedge  1\leq 2$, but we will only use them when the $P$ in question is uniquely defined.}

\section{Proof Rules for Higher-order Store}
\label{sec:ProgramLogic}

In our formal setting, reasoning about programs
is done by deriving  judgements of the form 
$\X;\Gamma \;\adash\; P$, 
where $P$ is an assertion expressing properties of programs, 
$\X$ is a list of (distinct) relation variables $X_1,\ldots,X_n$ containing all the free relation variables in $P$,  
and $\Gamma$ is a list of (distinct) variables $x_1,\ldots,x_n$ containing all the free
variables in $P$. For instance, to prove that command $C$ 
stores at cell $1$ the code that initializes cell $10$ to $0$, we need
to derive
$\X;\Gamma \,\adash\,
\triple{1\,{\pointsto}\,\_}{\QUOTE{C}}{\ntria{1}{10\,{\pointsto}\,\_}{\_}{10\,{\pointsto}\,0}}$. 
(One
concrete example of such a command $C$ is $[1]{:=}\QUOTE{[10]{:=}0}$.) 
Below, we will sometimes omit the contexts $\X$ and $\Gamma$ when they are empty.

In this section, we describe inference rules and axioms
for assertions that let one efficiently reason about 
programs. We focus on those related to higher-order store.

\subsection{Standard proof rules}
The proof rules include the standard proof rules for intuitionistic\footnote{A classical interpretation of the assertion language is inconsistent, see Section~\ref{subsec:ClassicalAssertionLogic}.} logic and the logic of bunched implications~\cite{OHearn:Pym:99} (not repeated here).
Moreover, the proof rules include variations of standard 
separation logic proof rules, see Figures~\ref{fig:SLRules} and \ref{fig:Nonsyntax:SLRules}. {The ({\sc Update}), 
({\sc Free}) and ({\sc Skip}) rules in the figure are not the usual 
small axioms in separation logic, since they contain an assertion $P$
that describes the unchanged part. Since we have the standard frame rule
for $*$, we could have used small axioms instead here. We chose not to do this, because the current non-small axioms make it easier
to follow our discussions on frame rules and higher-order store in
the next subsection.}
We added a specific version of ({\sc Update}), called ({\sc UpdateInv}), which  will turn out not to be derivable from  ({\sc Update}) because triples cannot be used in the ({\sc Invariance}) rule. (This will be explained in Section~\ref{sec:discussion-of-proof-rules}).  The side condition of ({\sc Invariance}) 
``$\psi$ is pure''  ensures that $\psi$ is an assertion denoting a predicate that is actually \emph{independent} of the heap. Examples for pure predicates are arithmetic formulae like $x=1$.

The figure neither includes the rule for executing stored code with $\UNQUOTE{[e]}$
nor  the frame rule for adding invariants to triples. 
The reason for this omission is that these two rules raise nontrivial 
issues in the presence of higher-order store and nested triples,
as we shall discuss below.
We also omit the conjunction axiom for triples:
\begin{gather*}
\inferrule[Conj]{ }  { \X;\Gamma \;\adash\; \triple{P_2}{e}{Q_2} \wedge  \triple{P_1}{e}{Q_1} \Rightarrow
  \triple{P_1{\wedge} P_2}{e}{Q_1{\wedge} Q_2} 
}
\end{gather*}
as it is \emph{not} sound (neither as a rule) in the presence of higher-order or deep frame rules, for the reasons given in \cite{OHearn:Yang:Reynolds:04}. If we wanted to use it we would need to restrict to precise assertions, as they do.

\begin{figure*}[!t]
\hrule
\begin{gather*}
\inferrule[Deref]{
  \X;\Gamma,x\,{\adash}\,\triple{P\,{*}\,e\,{\mapsto}\,x}{\QUOTE{C}}{Q}
}{
  \X;\Gamma\,{\adash}\, \triple{\exists x.P\,{*}\,e\,{\mapsto}\,x}{\QUOTE{\SYN{let}\,{x{=}[e]}\,\SYN{in}\,{C}}}{Q}
}(x \not\in \fvar(e,Q))
\\[2ex] 
\inferrule[Update]{
}{
  \X;\Gamma\,{\adash}\,\triple{e\,{\mapsto}\,\_ \,{*}\, P}{\QUOTE{[e] \,{:=}\, e_0}}{e\,{\mapsto}\,e_0 \,{*}\, P}
} \\[2ex] 
\inferrule[UpdateInv]{
}{
  \X;\Gamma\,{\adash}\,\triple{e\,{\mapsto}\,\_ \,{*}\, (e_1{\mapsto}e_0 \wedge  \triple{A}{e_0}{B})}{\QUOTE{[e] \,{:=}\, e_0}}{(e\,{\mapsto}\,e_0\wedge  \triple{A}{e_0}{B}) \,{*}\, (e_1{\mapsto}e_0 \wedge \triple{A}{e_0}{B}) }
}
\\[2ex]
\inferrule[New]{
  \X;\Gamma,x \,{\adash}\, \triple{P*x\,{\mapsto}\,e}{\QUOTE{C}}{Q}
}{
  \X;\Gamma \,{\adash}\, \triple{P}{\QUOTE{\SYN{let}\,{x {=} \SYN{new}\,e}\,\SYN{in}\,{C}}}{Q}
}(x \not\in \fvar(P,e,Q))
\qquad
\inferrule[Free]{
}{
  \X;\Gamma\adash\triple{e\,{\mapsto}\,\_ * P}{\QUOTE{\SYN{free}(e)}}{P}
}
\\[2ex]
\inferrule[If]{
  \X;\Gamma\adash \triple{P \,{\wedge}\, e_0{=}e_1}{\QUOTE{C}}{Q} \quad \X;\Gamma\adash \triple{P \,{\wedge}\, e_0{\not=}e_1}{\QUOTE{D}}{Q}
}{
  \X;\Gamma\adash \triple{P}{\QUOTE{\SYN{if}\;(e_0{=}e_1)\;\SYN{then}\;C\;\SYN{else}\;D}}{Q}
}
\\[2ex]
\inferrule[Skip]{
}{
  \X;\Gamma\adash\triple{P}{\QUOTE{\SYN{skip}}}{P}
}
\qquad
\inferrule[Seq]{
  \X;\Gamma\adash \triple{P}{\QUOTE{C}}{R} \quad \Gamma\adash \triple{R}{\QUOTE{D}}{Q}
}{
  \X;\Gamma\adash \triple{P}{\QUOTE{C;D}}{Q}
}
\end{gather*}
\hrule
\caption{Proof rules from separation logic}
\label{fig:SLRules}
\end{figure*}

\begin{figure*}[!t]
\hrule
\begin{gather*}
\inferrule[Conseq]{
  \X;\Gamma \,\adash\, P'{\Rightarrow}\,P 
  \quad \X;\Gamma \,\adash\, Q\,{\Rightarrow}\,Q'
}{
  \X;\Gamma \,\adash\, {\triple{P}{e}{Q}} \Rightarrow {\triple{P'}{e}{Q'}}
}
\quad
\inferrule[Disj]{
}{
  \X;\Gamma \;\adash\;
  \triple{P}{e}{Q}\wedge{\triple{P'}{e}{Q'}} \Rightarrow \triple{P\vee P'}{e}{Q\vee Q'}
}
\\[2ex]
\inferrule[ExistAux]{
}{
  \X;\Gamma \;\adash\; (\forall x.\triple{P}{e}{Q})\Rightarrow\triple{\exists x.P}{e}{\exists x.Q}
}(x \not\in \fvar(e))
\\[2ex]
\inferrule[Invariance]{
}{
 \X;\Gamma \;\adash\; \triple{P}{e}{Q}
 \Rightarrow
\triple{P\wedge \psi}{e}{Q \wedge \psi}
  }(\mbox{$\psi$ is pure})
\end{gather*}
\hrule
\caption{Non-syntax driven proof rules}
\label{fig:Nonsyntax:SLRules}
\end{figure*}

\subsection{Proof rule for recursive assertions}\label{subsubsec:runique}
Besides the axiom \eqref{eqn:unfolding-axiom} which lets us unfold recursive assertions, we include a proof rule that expresses the uniqueness of recursive assertions, 
\begin{align*}
\inferrule[RUnique]{
\X;\Gamma\adash R\Leftrightarrow P[X:=R]\\ 
\X;\Gamma\adash  S\Leftrightarrow P[X:=S]\\ 
  }
{ \X;\Gamma\adash R \Leftrightarrow S }
\end{align*}
for any $P$ formally contractive in $X$. 
Using this rule, the equivalence of (possibly recursively defined) assertions $R$ and $S$ can be proved by finding a suitable assertion $P$ that has both $R$ and $S$ as fixed points.

\subsection{Frame rule for higher-order store}

The frame rule is the most important rule in separation logic, and it formalizes the intuition of local reasoning, where proofs focus on the footprints of the programs we verify. 
For instance, in Example~\ref{example:spec}, we have said we can prove
\begin{align} \label{example_dfr}
\triple{c\, {\pointsto}\_ * f\,{\pointsto}\triple{\EMP}{\_}{\EMP} * R_{\mathit{it}}}
{\QUOTE{C_{\mathit{it},f,c}}}
{c\,{\pointsto}0 * f\,{\pointsto}\triple{\EMP}{\_}{\EMP} * R_{\mathit{it}}}
\end{align}
But if we wanted now to prove a similar result for an $f$ that had some side effect like
$$f \pointsto \QUOTE{\SYN{let}\ r \SYN{=} [x]  \ \SYN{in let}\ v \SYN{=} [y]\  \SYN{in}\ [y] \SYN{:=}\, r \SYN{*} v; [x] \SYN{:=}\, r{-}1}$$
then setting $I \defeq \exists m.\, x{\pointsto} m * y{\pointsto} n!/m!$ we can prove $\triple{I}{f}{I}$ but now we need to  show
\begin{align}\label{IIexDFR_eq}
\triple{c\, {\pointsto}\_ * f\,{\pointsto}\triple{I}{\_}{I} * (R_{\mathit{it}}\otimes I) * I}
{\QUOTE{C_{\mathit{it},f,c}}}
{c\,{\pointsto}0 * f\,{\pointsto}\triple{I}{\_}{I} * (R_{\mathit{it}} \otimes I)* I}
\end{align}
The so-called ``deep frame rule'' will allow us to  do just that, to prove triple \eqref{IIexDFR_eq} from triple \eqref{example_dfr} in one reasoning step, such that we can re-use our original proof. This rule will be discussed below and details of its concrete usage can be seen in  Example~\ref{example:DFR}.
Note also that the first-order (or shallow) frame rule does not achieve this, it would only give us
\begin{align}\label{shallow}
\triple{c\, {\pointsto}\_ * f\,{\pointsto}\triple{\EMP}{\_}{\EMP} * R_{\mathit{it}} * I}
{\QUOTE{C_{\mathit{it},f,c}}}
{c\,{\pointsto}0 * f\,{\pointsto}\triple{\EMP}{\_}{\EMP} * R_{\mathit{it}} * I}
\end{align}
 which is   not useful here.

Establishing  ``deep'' frame rules in our setting is challenging, because nested triples allow for several choices regarding the shape of the rule.  Moreover, the recursive nature of the higher-order store complicates matters and it is difficult to see which choices actually make sense (i.e., do not lead to inconsistency).

To see this problem more clearly, consider
the rules below:
\begin{align*}
\inferrule{
  \X;\Gamma \vdash \triple{P}{e}{Q}
}{
  \X;\Gamma \vdash \triple{P \,\clubsuit\, R}{e}{Q \,\clubsuit\, R}
}
\ \ 
\mbox{and}
\ \ 
\inferrule{
}{
  \X;\Gamma \vdash \triple{P}{e}{Q} \Rightarrow \triple{P \,\clubsuit\, R}{e}{Q \,\clubsuit\, R}
}
\  \mbox{ for $\clubsuit \in \{*,\circ\}$}. 
\end{align*}
Note that we have four choices, depending on whether
we use $\clubsuit=*$ or $\clubsuit=\circ$ and on whether
we have an inference rule or an axiom. If we choose the separating conjunction $*$ for $\clubsuit$,
we obtain \emph{shallow} frame rules that  add $R$ to 
the outermost triple $\triple{P}{e}{Q}$ only; 
they do not add $R$ in nested triples
appearing in pre-condition $P$ and post-condition $Q$. On the other hand,
if we choose $\circ$ for $\clubsuit$, since $(A \circ R) = (A \otimes R * R)$,
we obtain \emph{deep} frame rules that add
the invariant $R$ not just 
to the outermost triple but also to all the nested triples in $P$ and $Q$. 

The distinction between inference rule and axiom has  
some bearing on where the frame rule can be applied. With
the axiom version, we can apply the frame rule not just to valid triples,
but also to nested triples appearing in pre- or post-conditions
which is not possible with the inference rule

Ideally, we would like to have the axiom versions of the frame rules
for both the $*$ and $\circ$ connectives. Unfortunately, this is not possible for $\circ$: 
adding the axiom version for $\circ$  makes our logic unsound.
The source of the problem is that with the axiom 
version for $\circ$, one can add invariants selectively to some, but not necessarily all,
nested triples. This flexibility can be abused to derive 
incorrect conclusions. 

Concretely, with the axiom version for $\circ$ (\textsc{DeepFrameAxiom}) we can make the following derivation:

\[ \hspace{-1.95cm}
\inferrule*[Right=ModusPon.]{
    \inferrule*[Right={$\otimes$-Dist}$^r$]{ 
      \X;\Gamma \adash \triple{P \,{\circ}\, S}{e}{Q \,{\circ}\, S} 
    }{
      \X;\Gamma \adash \triple{P}{e}{Q} \,{\otimes}\, S
    }
    \quad \ \;\;
    \\
    \inferrule*[Right=$\otimes$-Mono]{
      \inferrule*[Right=DeepFrameAx.]{
      }{
        \X;\Gamma \adash \triple{P}{e}{Q} \Rightarrow \triple{P\,{\circ}\, R}{e}{Q\,{\circ}\, R}
      }
    }{
      \X;\Gamma \adash 
      \triple{P}{e}{Q}\,{\otimes}\, S \Rightarrow \triple{P\,{\circ}\, R}{e}{Q\,{\circ}\, R}\,{\otimes}\, S
    }
}{
  \inferrule*[Right={$\otimes$-Dist}$^r$]{ 
    \X;\Gamma \adash \triple{P\,{\circ}\, R}{e}{Q\,{\circ}\, R} \,{\otimes}\, S
  }{
    \X;\Gamma \adash \triple{(P\,{\circ}\, R) \,{\circ}\, S}{e}{(Q\,{\circ}\, R) \,{\circ}\, S}
  }
}
\vspace{0.2cm}
\]
Here we use the monotonicity of $-\otimes R$ in the form of rule (\textsc{$\otimes$-Mono}), cf.~Figure~\ref{fig:proofrulesummary} in the Appendix.
The steps annotated  \textsc{$\otimes$-Dist}$^r$ use the first equivalence    $\triple{P}{e}{Q}\; {\otimes}\; R
  \;\Leftrightarrow\;
  \triple{P \,{\circ}\, R}{e}{Q \,{\circ}\, R}$
of the distribution axioms for $\otimes$ in Fig.~\ref{fig:DistRules} 
(in $\Leftarrow$ and $\Rightarrow$ direction, respectively).
We annotate the application of an axiom between triples with $^r$ to indicate that we apply  it actually as a rule via the application of (\textsc{ModusPonens}). So, for instance,   (\textsc{Conseq})$^r$, used frequently below,   denotes   a sub-derivation of the following form:

\[ \hspace{-1cm}
\inferrule*[Right=ModusPonens]
{ \triple{A}{e}{B}  \quad
 \inferrule*[Right=Conseq]
    {A'\Rightarrow A \quad B\Rightarrow B'}
    {\triple{A}{e}{B} \Rightarrow \triple{A'}{e}{B'}}
}
{ \triple{A'}{e}{B'}
}
\vspace{0.1cm}
\]
where we  will usually omit the implications  $A'\Rightarrow A$ and $ B\Rightarrow B'$ when they are obvious from the context.

The fact we could derive $\triple{(P\,{\circ}\, R) \,{\circ}\, S}{e}{(Q\,{\circ}\, R) \,{\circ}\, S}$ means that
when adding $R$ to nested triples, we can skip the triples in the $S$ part of
the pre- and post-conditions of $\triple{P \,{\circ}\, S}{e}{Q \,{\circ}\, S}$. 
This flexibility leads to the unsoundness:
\begin{prop}
\label{prop:DeepFrameUnsound}
Adding the axiom version (\textsc{DeepFrameAxiom}) of the frame rule for $\circ$ renders our logic
unsound.
\end{prop}
\proof
Let $R$ be the recursive assertion $\mu X. (\ntria{3}{1{\pointsto}\_}{\_}{1{\pointsto}\_}) \otimes X$, and note  that this means   
$R \Leftrightarrow (\ntria{3}{1{\pointsto}\_}{\_}{1{\pointsto}\_}) \otimes R$ holds.
Then, we can derive the triple:
 
 \begin{equation}\hspace{-0.1cm} \vspace{0.2cm}
   \inferrule*[Left=Conseq$^r$]{
    \inferrule{
          k\,\adash 
          \triple
            {\ntria{2}{1{\pointsto}\_}{\_}{1\;{\pointsto}\; \_} \circ R}
            {k}
            {2\; {\pointsto}\; \_  \circ R}
    }{
      k\,\adash       \triple
        {\bigl(\ntria{2}{1{\pointsto}\_}{\_}{1{\pointsto}\_} \circ 1\,{\pointsto}\,\_\bigr) \circ R}
        {k}
        {\bigl(2\,{\pointsto}\,\_ \circ 1 \,{\pointsto}\,\_ \bigr) \circ R}
    }
  }{
    k\,\adash 
    \triple
      {2{\pointsto}\QUOTE{\SYN{free}(-1)} * 1{\pointsto}\_ * R}
      {k}
      {2{\pointsto}\_ \,{*}\, 1 \,{\pointsto}\,\_ \,{*}\, \ntria{3}{1{\pointsto}\_\; {*}\, R}{\_}{1{\pointsto}\_\;{*}\, R}}
  }\hspace{-0.88cm} \label{derivation_dagger}
  \vspace{0.1cm}
  \end{equation}
Here the first step uses the derivation above for adding invariants 
selectively, and the last step uses the consequence rule with
the following two implications:

\[ 
\begin{array}{@{}r@{}c@{}l@{}}
{\ntria{2}{1{\pointsto}\_}{\_}{1{\pointsto}\_} \,{\circ}\, 1{\pointsto}\_ \,{\circ}\, R}
& \;{\Longleftrightarrow}\; &
{ \ntria
    {2}
    {1{\pointsto}\_ \,{*}\, 1{\pointsto}\_\,{*}\,R}
    {\_}
    {1{\pointsto}\_ \,{*}\, 1{\pointsto}\_\,{*}\,R} 
\,{*}\, 1{\pointsto}\_ 
\,{*}\, R}
\\
& \;{\Longleftrightarrow}\;&
{ \ntria{2}{\False}{\_}{\False} * 1{\pointsto}\_ * R}
\\
& \;{\Longleftarrow}\; &
{2\,{\pointsto}\QUOTE{\SYN{free}(-1)} * 1{\pointsto}\_ * R}
\end{array} \vspace{0.1cm}
\]
where the second equivalence follows from  the fact that $1{\pointsto}\_ * 1{\pointsto}\_ \Leftrightarrow \False$ (use axioms ($\star$-\textsc{Overlap}), ($\star$-\textsc{Zero}), and ($\star$\textsc{-Mono}) of   Separation Logic from  Figure~\ref{fig:proofrulesummary})       with (\textsc{Conseq})\footnote{Note that it is important here that (\textsc{Conseq})  derives an implication between triples.}, and
\[ 
\begin{array}{@{}r@{}c@{}l@{}}
{2\,{\pointsto}\_ \circ 1 {\pointsto}\_ \circ R}
& \,\;{\Longleftrightarrow}\;\, &
{2\,{\pointsto}\_ \,{*}\, 1 {\pointsto}\_ \,{*}\, R}\\
& \,\;{\Longleftrightarrow}\;\, &
{2\,{\pointsto}\_ \,{*}\, 1 {\pointsto}\_ \,{*}\,
((\ntria{3}{1{\pointsto}\_}{\_}{1{\pointsto}\_})\,{\otimes}\, R)}
\\
& \,\;{\Longleftrightarrow}\;\, &
{2\,{\pointsto}\_ \,{*}\, 1 {\pointsto}\_ \,{*}\,
 \ntria{3}{1{\pointsto}\_\,{*}\,R}{\_}{1{\pointsto}\_\,{*}\,R}}.
\end{array}
 \vspace{0.1cm}
 \] 
in which the distribution axioms of  Figure~\ref{fig:DistRules} are used, again in concert with (\textsc{Conseq}) and Separation Logic rules like ($\star$\textsc{-Mono}).
 
Consider $C \;\equiv\;\SYN{let}~x \SYN{=} [2]~\SYN{in}~[3]\SYN{:=} x$, i.e., the program that copies the contents from cell $2$ to cell $3$.
When $P[y] \;\equiv\; \triple{1{\pointsto}\_}{y}{1{\pointsto}\_}\,{\otimes}\,R$ such that $R\Leftrightarrow 3{\pointsto}P[\_]$ holds,

\[ 
\inferrule*[Right=Conseq$^r$]{
\inferrule*[Right=Conseq$^r$]{
\inferrule*[Right=Conseq$^r$]{
  \inferrule*[Right=Deref]{
  \inferrule*[Right=Conseq$^r$]{
   \inferrule*[Right=UpdateInv]{ }{
   x\, {\adash}   \triple
     { 3{\pointsto}\_ *( 2{\pointsto}x \wedge P[x]) }
    {\QUOTE{[3] \SYN{:=}x}}
     {(3{\pointsto}x \wedge P[x]) * (2{\pointsto}x   \wedge P[x])}
  }}{  x\, {\adash}   \triple
     { 3{\pointsto}\_ *( 2{\pointsto}x \wedge P[x]) }
    {\QUOTE{[3] \SYN{:=}x}}
     {3{\pointsto}  P[\_] * 2{\pointsto} P[\_]}
  }}{
    {\adash}\;
    \triple
     {\exists x.\, 3{\pointsto}\_ *  (2{\pointsto}x   \wedge P[x])}
     {\QUOTE{\SYN{let}~x \SYN{=}[2]~\SYN{in}~[3] \SYN{:=}x}}
     {3{\pointsto}  P[\_] * 2{\pointsto} P[\_]}
  }
}{
  {\adash}\;
  \triple
     {3{\pointsto}P[\_] *  2{\pointsto} P[\_]}
     {\QUOTE{C}}
     {3{\pointsto}  P[\_] * 2{\pointsto} P[\_]}
}}
{ {\adash}\;
  \triple
     {R *  2{\pointsto} P[\_]}
     {\QUOTE{C}}
     {R *  2{\pointsto} P[\_]}
}}
{{ {\adash}\;
  \triple
     {\ntria{2}{1{\pointsto}\_}{\_}{1{\pointsto}\_} \circ R}
     {\QUOTE{C}}
     {2\,{\pointsto}\,\_  \circ R}
}
}
\vspace{0.2cm}
\] 
Now we instantiate $k$ in $\eqref{derivation_dagger}$ with $\QUOTE{C}$,
discharge the premise of the resulting derivation
with the above derivation for $C$, and obtain
\[ \ \vspace{0.2cm}
  \inferrule{ \vdots
  }{
    \adash\,
    \triple
      {2\,{\pointsto}\,\QUOTE{\SYN{free}(-1)} * 1\,{\pointsto}\,\_ * R}
      {\QUOTE{C}}
      {2\,{\pointsto}\,\_ \,{*}\, 1 \,{\pointsto}\,\_ \,{*}\, \ntria{3}{1{\pointsto}\_\,{*}\,R}{\_}{1{\pointsto}\_\,{*}\,R}}
  }
 \vspace{0.1cm}
 \] 
But the post-condition of the conclusion here is equivalent to
$2{\pointsto}\_\,{*}\,1{\pointsto}\_\,{*}\,R$ by the definition of $R$ and 
the distribution axioms for $\otimes$. Thus, as our rule for $\SYN{eval}$
will show later, we should be able to conclude that
\[
  {\adash}\,
  \triple
    {2\,{\pointsto}\,\QUOTE{\SYN{free}(-1)} \,{*}\, 1{\pointsto}\_ \,{*}\, R}
    {\QUOTE{C;
    \UNQUOTE{[3]}}}
    {2\,{\pointsto}\_\,{*}\,1 {\pointsto}\_ \,{*}\, \ntria{3}{1{\pointsto}\_\,{*}\,R}{\_}{1{\pointsto}\_\,{*}\,R}}
\]
However, since $-1$ is not even an address, the program $(C;\
\UNQUOTE{[3]})$ which executes the code \SYN{free}(-1)   now stored in cell 3 always faults, contradicting the requirement of separation logic that proved programs
run without faulting.
\qed

\begin{remark}[Counterexample for the Deep Frame Axiom]
Notice that in the derivation above  it is essential that $R$ is a recursively defined assertion, otherwise we would not obtain that the locations $2$ and $3$ point to code satisfying the same  assertion $P$.

While the above counterexample has been the first such counterexample historically, there is also another   form of counterexample discovered later which  uses the same ideas as the above but works ``through the store.'' More precisely,  in this alternative counterexample the copying code $\QUOTE{C}$ resides on the heap where the frame axiom  can  be applied directly on a nested triple, and not through the derivation 
\[
\inferrule
{\triple{P\circ S}{e}{Q\circ S}}
{\triple{(P\circ R)\circ S}{e}{(Q\circ R) \circ S}
}
\]
This rather follows the style of  \cite{Pottier09}\footnote{However, the antiframe rule is used there.} and \cite{CharltonReusLola10}\footnote{This uses a version where the copied code accesses a cell that is then disposed of before the code itself is executed later.}. For this counterexample, let $R$ be as above and let
 $$P_1[y] \equiv  \triple{1{\pointsto}\_}{y}{1{\pointsto}\_}  \ .$$
  First, observe that the following triple can be derived with  a rule for eval (this rule  \textsc{(Eval)} will be explained in detail in Section~\ref{subsec:eval-rule}):
 
\begin{equation}\label{eval_dagger}
\begin{array}{ll}
\{ \ntria{2}{\False}{\_}{\False}\, {*}\, \ntria{c}{\ntria{2}{\False}{\_}{\False}}{\_}{\ntria{2}{\False}{\_}{\False}} \}&\\
   {\QUOTE {\UNQUOTE{[c]} }}   \\
    \{ 2\,{\pointsto}\_\, {*}\,  c\,{\pointsto} \, \_ \} &
\end{array} \vspace{0.2cm}
\end{equation}
 But the  (\textsc{DeepFrameAxiom}) (the axiom version for $\circ$)  can be used to derive
 $$ \ntria{c}{ {2\, {\pointsto} P_1[\_]}}{\_}{2\,{\pointsto} P_1[\_] }  \Longrightarrow 
 \ntria{c}{ {2\, {\pointsto} P_1[\_]} \circ  (1\, {\pointsto}\_ ) }{\_}{2\,{\pointsto} P_1[\_]   \circ (1\, {\pointsto}\_ )} 
 $$
 which then  by applying distribution axioms unfolding the definition of $P_1$ yields:
  $$ \ntria{c}{ {2\, {\pointsto} P_1[\_]}}{\_}{2\,{\pointsto} P_1[\_] }  \Longrightarrow 
 \ntria{c}{\, \ntria{2}{\False}{\_}{\False}\, }{\_}{\, \ntria{2}{\False}{\_}{\False}\,} 
 $$
 Applying this to triple \eqref{eval_dagger} with the help of an appropriate (\textsc{Conseq}$^r$) step we can therefore derive
$$
  {\adash}\,
  \triple
    { \ntria{2}{\False}{\_}{\False}\, {*}\, \ntria{c}{ {2\, {\pointsto} P_1[\_]}}{\_}{2\,{\pointsto} P_1[\_] }}
    {\QUOTE{\UNQUOTE{[c]}}}
    {  2\,{\pointsto}\_\, {*}\,  c\,{\pointsto} \, \_}
$$
and thus by the shallow frame rule again
$$
  {\adash}\,
  \triple
    {1\,{\pointsto} \_  \, {*} \, \ntria{2}{\False}{\_}{\False}\, {*}\, \ntria{c}{ {2\, {\pointsto} P_1[\_]}}{\_}{2\,{\pointsto} P_1[\_] }}
    { \QUOTE{\UNQUOTE{[c]} }}
    {1\,{\pointsto} \_  \, {*} \, 2\,{\pointsto}\_\, {*}\,  c\,{\pointsto} \, \_}
$$
This  triple   should not hold for all heaps since actually now the code in $2$ has been laundered to work with its caller code in $c$ although  the code in $c$, to function properly, might depend on the code in $2$ meeting the specification $P_1$.  Using the above derivation, we can now construct a program that is provably safe but crashes, showing that (\textsc{DeepFrameAxiom}) cannot be correct (as the other used rules and axioms clearly are). First, with the rule version for $\circ$ (\textsc{DeepFrameRule}) to add $R$ one   gets

\[
\begin{array}{l}
    \{\,  1\,{\pointsto} \_  \, {*} \, \ntria{2}{\False}{\_}{\False}\, {*}\, \ntria{c}{ {2\, {\pointsto} P_1[\_]}\circ R}{\_}{2\,{\pointsto} P_1[\_]\circ R } \, {*}\, R\, \}\\
    \QUOTE{\UNQUOTE{[c]}} \\
    \{\, 1\,{\pointsto} \_  \, {*} \, 2\,{\pointsto}\_\, {*}\,  c\,{\pointsto} \_  \, {*} \, R\, \}
\end{array} \vspace{0.2cm}
\]
so that by definition of  $\circ$, $P_1$, and $R$ we obtain
\[ \vspace{0.2cm}\begin{array}{l}
    \{\,  1\,{\pointsto} \_  \, {*} \, \ntria{2}{\False}{\_}{\False}\, {*}\, \ntria{c}{ {2\, {\pointsto} P[\_]}\, {*}\, R}{\_}{2\,{\pointsto} P[\_] \, {*}\, R } \, {*}\, R\, \}\\
   \QUOTE{ \UNQUOTE{[c]}}\\
    \{\, 1\,{\pointsto} \_  \, {*} \, 2\,{\pointsto}\_\, {*}\,  c\,{\pointsto} \_  \, {*} \, R\, \}
\end{array} \vspace{0.1cm}
\]
where $P[y]$ is the assertion 
$\triple{1{\pointsto}\_}{y}{1{\pointsto}\_}\,{\otimes}\,R$
(also used in the proof of Proposition~\ref{prop:DeepFrameUnsound}).
 From that  one can easily derive with the rules  \textsc{(Seq)}, \textsc{(Eval)} and \textsc{(Conseq)}  that
 \[ \ \vspace{0.2cm}\begin{array}{l}
    \{1\, {\pointsto} \_  \, {*} \,\ntria{2}{\False}{\_}{\False}\, {*}\, \ntria{c}{ {2\, {\pointsto} P[\_]} \, {*} \, R}{\_}{2\,{\pointsto} P[\_]\, {*}\,  R }\, {*}\, R \}\\
   \QUOTE{ \UNQUOTE{[c]}\SYN{;}    \UNQUOTE{[3]}} \\
    \{ 1\,{\pointsto} \_  \, {*} \, 2\,{\pointsto}\_\, {*}\,  c\,{\pointsto} \_  \, {*} \, R\, \} \quad .
\end{array} \vspace{0.1cm}
\] 
Yet, if $c\pointsto \QUOTE{\SYN{let}~x{=}[2]~\SYN{in}~[3]{:=}x}$
and $2\pointsto \QUOTE{\SYN{free(-1)}}$, then the above program crashes. 
Although the code in $c$ does not call the crashing code 
$\QUOTE{\SYN{free(-1)}}$ in $2$, it copies $\QUOTE{\SYN{free(-1)}}$ 
into $3$, which is possible due to the ``laundered'' specification of $2$ in the triple for $c$. 

Again, this shows how essential it is that $P_1[\_]\otimes R $ is equivalent to $R$ which forces $R$ to be recursively defined to actually allow the copying to be performed. This version of the counterexample uses the (\textsc{DeepFrameRule}) rather than (\textsc{ModusPonens}) and (\textsc{$\otimes$-Mono}), and its pattern  is more likely to appear in ``naturally occurring'' examples.
 \end{remark}
 
As Proposition~\ref{prop:DeepFrameUnsound} shows, we cannot include (\textsc{DeepFrameAxiom}) in the proof system. 
Fortunately, 
the second best choice of frame axioms leads to a consistent proof system:
\begin{prop}
Both the inference rule version of the frame rule for $\circ$ and the axiom version for $*$
are sound.
In fact, the  following more general version (\textsc{$\otimes$-Frame}) of the rule for $\circ$ holds:
\begin{align*}
\inferrule
{
  \X;\Gamma \;\adash\; P
}{
  \X;\Gamma \;\adash\; P\otimes R
}
\end{align*}
\end{prop}
We will prove this proposition in Section~\ref{sec:Semantics} by a model construction. 

\begin{example}[Application of (\textsc{$\otimes$-Frame})]
\label{example:DFR} 
Recall our specification
\begin{align}\label{exDFR_eq}
\triple{c\, {\pointsto}\_ * f\,{\pointsto}\triple{\EMP}{\_}{\EMP} * R_{\mathit{it}}}
{\QUOTE{C_{\mathit{it},f,c}}}
{c\,{\pointsto}0 * f\,{\pointsto}\triple{\EMP}{\_}{\EMP} * R_{\mathit{it}}}
\end{align}
of the iteration command in Example~\ref{example:spec}, 
where $R_{\mathit{it}}$ is a recursive specification for the iterator itself:
$$R_{\mathit{it}} \equiv \mu X.\, \mathit{it} \, {\pointsto}\,
 \triple{c\,{\pointsto}\_ * f\,{\pointsto}\triple{\EMP}{\_}{\EMP} * X}
 {\_}
{c\,{\pointsto}0 * f\,{\pointsto}\triple{\EMP}{\_}{\EMP} * X}
$$
Assume this triple has been already proven (cf.\ Example~\ref{example:eval} below).
If the code $C_{\mathit{it},f,c}$ is to be used  on a procedure $f$ that 
 needs some state $I$, e.g.\ $I \equiv a{\pointsto}\_$, then we need to show
\[\begin{array}{l}
\{c\, {\pointsto}\_ * f\,{\pointsto}\triple{ I}{\_}{I}   * (R_{\mathit{it}} \otimes I) *I \}
{\QUOTE{C_{\mathit{it},f,c}}}
\{c\,{\pointsto}0 * f\,{\pointsto}\triple{I}{\_}{ I} * (R_{\mathit{it}} \otimes I) *  I  \}  
\end{array}
\]
This triple could be established by a proof similar to the one for the triple \ref{exDFR_eq} above, just carrying around the extra assumption $I$. If we want to \emph{reuse} this proof though, or even more importantly, if we do not have the proof of the above triple because it is part of a module for which we do not have the actual code, then we can use rule (\textsc{$\otimes$-Frame}) on triple \eqref{exDFR_eq} to derive:
$$
\Bigl(
\triple{c\, {\pointsto}\_ * f\,{\pointsto}\triple{\EMP}{\_}{\EMP} * R_{\mathit{it}}}
{\QUOTE{C_{\mathit{it},f,x}}}
{c\,{\pointsto}0 * f\,{\pointsto}\triple{\EMP}{\_}{\EMP} * R_{\mathit{it}}}\Bigr) \otimes I
$$
A \textsc{Conseq}$^r$ step  using the equivalence of the first axiom in  Figure~\ref{fig:DistRules}  in both directions for the pre- and postcondition, respectively,  thus gives us the triple:
\[\begin{array}{l}
\{(c\, {\pointsto}\_\otimes I) * (f\,{\pointsto}\triple{\EMP}{\_}{\EMP}  \otimes I) * (R_{\mathit{it}} \otimes I) * I \}\\
{\QUOTE{C_{\mathit{it},f,x}}}\\
\{(c\,{\pointsto}0 \otimes I) * (f\,{\pointsto}\triple{\EMP}{\_}{\EMP} \otimes  I)* (R_{\mathit{it}} \otimes  I) *  I\}  
\end{array}
\]
which by another four applications of distribution axioms yields the required triple.
Note that the rule~\textsc{(RUnique)}   would be needed to show   that $R_{\mathit{it}} \otimes  I$ is equivalent to the recursive assertion
$$
\mu Y.\,  \mathit{it} \, {\pointsto}\,
 \triple{c\,{\pointsto}\_ *  f\,{\pointsto}\triple{I}{\_}{I} * I *Y}
 {\_}
{c\,{\pointsto}0 * f\,{\pointsto}\triple{I}{\_}{I} *I* Y}\ .
$$
\end{example}

\subsection{Rule for executing stored code}
\label{subsec:eval-rule}
An important and challenging 
part of the design of a program logic for higher-order store is the
design of a proof rule for $\UNQUOTE{[e]}$, the command
that executes code stored at $e$.
Indeed, the rule should overcome two challenges directly
related to the recursive nature of higher-order store: (1)
implicit recursion through the store (i.e., Landin's knot), and (2)
extensional specifications of stored code.

These two challenges are addressed, using the expressiveness of our
assertion language, by the following rule for $\UNQUOTE{[e]}$:
$$
\inferrule[Eval]{
  \X;\Gamma,k \adash R[k] \Rightarrow \triple{P\,{*}\,e\,{\pointsto}\,R[\_]}{k}{Q}
}{
  \X;\Gamma \adash \triple{P\,{*}\,e\,{\pointsto}\,R[\_]}{\QUOTE{\UNQUOTE{[e]}}}{Q}
}
$$
This rule states that in order to prove $\triple{P*e\,{\pointsto}\,R[\_]}{\QUOTE{\UNQUOTE{[e]}}}{Q}$ for executing stored code in $[e]$ under the assumption that $e$ points to arbitrary code $k$ (expressed by the $\_$ which is an abbreviation for $\exists k. e\mapsto R[k]$), it suffices to show that the specification $R[k]$ implies  that $k$ itself fulfils triple 
$\triple{P*e\,{\pointsto}\,R[\_]}{k}{Q}$. 

In the above rule we do not make any assumptions about what code $e$ actually
points to, as long as it fulfils the specification $R$. It may even be updated between recursive calls.  However, for recursion through the store, $R$ must be recursively defined as it needs to maintain itself as an invariant of the code in $e$.

\begin{example}[Recursion through the store with the iterator]\label{example:eval}
As seen in the iterator Example~\ref{example:spec} one would like to prove
$$
\triple{\, c\,{\pointsto}\_ * f\,{\pointsto}\triple{\EMP}{\_}{\EMP} *R_{\mathit{it}}\, }
{ \QUOTE{\UNQUOTE{[\mathit{it}]}}}{\,c\, {\pointsto}0 * f\,{\pointsto} \triple{\EMP}{\_}{\EMP} * R_{\mathit{it}} \, }
$$
with the help of \textsc{(Eval)}. First we set 
$$R  \equiv 
 \triple{c\,{\pointsto}\_ * f\,{\pointsto}\triple{\EMP}{\_}{\EMP} * R_{\mathit{it}}}
 {\_}
{c\,{\pointsto}0 * f\,{\pointsto}\triple{\EMP}{\_}{\EMP} * R_{\mathit{it}}}
$$
such that $R_{\mathit{it}}$ is the same as $\mathit{it}\,{\pointsto}R[\_]$. We are now in a position to  apply \textsc{(Eval)} obtaining the following proof obligation 
$$ R[k] \Rightarrow \triple{ c\,{\pointsto}\_ * f\,{\pointsto}\triple{\EMP}{\_}{\EMP} * \mathit{it}\,{\pointsto}\,R[\_]}{k}
{c\, {\pointsto}0 * f\,{\pointsto} \triple{\EMP}{\_}{\EMP} * R_{\mathit{it}} }
$$
which can be seen to be identical to $R[k] \Rightarrow R[k]$ which holds trivially.
\end{example}

The ({\sc Eval}) rule crucially relies on the expressiveness of
our assertion language, especially the presence of nested triples
and recursive assertions. In our previous work, we did not consider nested 
triples. As a result, we had to reason explicitly with stored code,
rather than properties of the code, as illustrated by one of our previous rules
for $\SYN{eval}$ \cite{Birkedal:Reus:Schwinghammer:Yang:08}:
$$
\inferrule[OldEval]{
  \X;\Gamma \adash
  \triple{P}{\QUOTE{\UNQUOTE [e]}}{Q} \,\Rightarrow\, \triple{P}{\QUOTE{C}}{Q}
}{
  \X;\Gamma\vdash
  \triple{P*e\,{\mapsto}\, {\QUOTE C}}{\QUOTE{\UNQUOTE [e]}}{Q * e\,{\mapsto}\,{\QUOTE C}}
}
$$
Here the actual code $C$ is specified explicitly in the pre- and post-conditions of the triple.
In both rules the intuition is that the premise states
that the body of the recursive procedure fulfils the triple, under the assumption that the recursive call already does so. 
In the ({\sc Eval}) rule this is done without direct reference to the code itself, using the variable $k$ to stand for arbitrary code satisfying $R$.
The soundness proof of ({\sc OldEval}) proceeded along the lines of Pitts' method for establishing relational properties of domains~\cite{Pitts:96}.
On the other hand, as we will show in Section~\ref{sec:Semantics}, ({\sc Eval}) relies on the availability of recursive assertions,
the existence of which is guaranteed by Banach's fixpoint theorem.

From the ({\sc Eval}) rule one can easily derive the axioms of Figure~\ref{fig:DerivedRulesHO}. The first two axioms are for non-recursive calls. This can be seen from the fact that in the pre-condition of the nested triples $e$   does not appear at all or does not have a specification, respectively. Only the third axiom  ({\sc EvalRec}) allows for recursive calls. 
The idea of this axiom is that one  assumes that  the code in $[e]$ fulfils the required triple   provided the code that $e$ points to  at call-time fulfils the triple as well.
\begin{figure*}[t]
\hrule
\begin{gather*}
\inferrule[EvalNonRec1]{ 
}
{\X;\Gamma\adash \triple{P*e\pointsto \forall \vec{y}.\, \triple{P}{\_}{Q}}{\QUOTE{\UNQUOTE{[e]}}}{Q*e\pointsto \forall \vec{y}.\, \triple{P}{\_}{Q}}
}\\[2ex]
\inferrule[EvalNonRecUpd]{
}{
 \X;\Gamma\adash \triple{P*e\pointsto \forall \vec{y}.\, \triple{P* e\pointsto \_}{\_}{Q}}{\QUOTE{\UNQUOTE{[e]}}}{Q} 
} 
\\[2ex]
\inferrule[EvalRec]{
 }{
 \X;\Gamma\adash \triple{P\circ R}{\QUOTE{\UNQUOTE{[e]}}} {Q\circ R}
 }   (\mbox{where}\ R = \mu X.(e\pointsto \forall \vec{y}.\,  \triple{P}{\_}{Q} * P_0)\otimes X)
 \end{gather*}
\hrule
\caption{Derived rules from {\sc Eval}}
\label{fig:DerivedRulesHO}
\end{figure*}
 Let us look at the actual derivation of ({\sc EvalRec}) to make this evident.  We write 
$$S[k] \equiv \forall \vec{y}.\, \triple{P\,{\circ}\, R}{k}{Q\,{\circ}\, R}$$
 such that for the original 
 $$R = \mu X.(e\pointsto \forall \vec{y}.\,  \triple{P}{\_}{Q} * P_0)\otimes X$$
  of the rule  ({\sc EvalRec}) we obtain with the help of Axiom~\eqref{eqn:unfolding-axiom}:
\begin{equation} \label{eq:unfold_rec} R\Leftrightarrow (e\,{\pointsto}\,S[\_]) * (P_0\otimes R)
\end{equation}
 Note that  in the derivation below $\Gamma$ contains the variables $\vec{y}$ which may appear freely in $P$ and $Q$.
\[ \inferrule*[Right=FOL]{
}{
\inferrule*[Right=Def.\ of $S$] 
{  \X;\Gamma,k \adash (\forall \vec{y}.\, \triple{P\circ R}{k}{Q\circ R}) \Rightarrow  \triple{P\circ R} {k} {Q\circ R} }
{\inferrule*[Right=Csub]{ 
 \X;\Gamma,k \adash  S[k] \Rightarrow  \triple{P\circ R} {k} {Q\circ R}
}
{
\inferrule*[Right=Conseq$^r$]{ 
\inferrule*[Right=Eval]{ 
 \X;\Gamma,k\adash S[k]  \Rightarrow  \triple{(P\otimes R) * e\pointsto S[\_] * (P_0\otimes R)} {k} {Q\circ R}
 } 
 {
  \X;\Gamma\adash \triple{(P\otimes R) *  e\pointsto S[\_]* (P_0\otimes R)} {\QUOTE{\UNQUOTE{[e]}}} {Q\circ R}
 } 
 } 
 {  \X;\Gamma\adash \triple{P\circ R}{\QUOTE{\UNQUOTE{[e]}}} {Q\circ R}
}
}
}}\vspace{0.2cm}
\]
In the derivation tree above, the axiom used at the top is simply a first-order axiom for $\forall$ elimination. The quantified variables $\vec{y}$ are substituted by the variables  with the same name from the context.
After an application of rule ({\sc EvalRec}), those variables $\vec{y}$ can then be substituted further.
Step \textsc{Csub} abbreviates the following derivation where contexts have been omitted for clarity:
 
\[ 
\inferrule*[Right=T$\Rightarrow$]
{
 \inferrule{\vdots} {S[k]  \Rightarrow     \triple{P\circ R} {k} {Q\circ R}    } \\
 \inferrule*[Right=Conseq] {  
 \inferrule*[Right=unfold]{ 
  \inferrule*[Right=Def.\ $\circ$] 
   { \inferrule*[Right=R$\Rightarrow$] {   {\large \phantom{x}}
   }{P\circ R \Rightarrow P\circ R }
   }
   {(P\otimes R) * R \Rightarrow P\circ R }
 }
 {(P\otimes R) *  e\pointsto S[\_]* (P_0\otimes R) \Rightarrow P\circ R } \qquad\qquad\quad\
  \inferrule*[Right=R$\Rightarrow$]{ }{ Q\circ R \Rightarrow  Q\circ R }
  }
   {{\Huge \phantom{g_{f_f}} } \triple{P\circ R} {k} {Q\circ R}  \Rightarrow  \triple{(P\otimes R) *  e\pointsto S[\_] * (P_0\otimes R)} {k} {Q\circ R}  
  } 
}
{  
S[k]  \Rightarrow  \triple{(P\otimes R) *  e\pointsto S[\_] * (P_0\otimes R)} {k} {Q\circ R} 
}
\vspace{0.2cm}
\]
In the above derivation, (\textsc{R}$\Rightarrow$) and (\textsc{T}$\Rightarrow$) denote reflexivity and transitivity of implication, respectively, and step \textsc{unfold} denotes the following sub-derivation:

\[ 
\inferrule*[Right=T$\Rightarrow$]{
\inferrule*[Right=$\star$-Mono]{
\inferrule*[Right=R$\Rightarrow$]{  }{P\otimes R \Rightarrow P\otimes R}\qquad\ 
\inferrule*[Right=\eqref{eq:unfold_rec}]{  }{ e\pointsto S[\_]* (P_0\otimes R) \Rightarrow  R}
}
{(P\otimes R) *  e\pointsto S[\_]* (P_0\otimes R)\Rightarrow(P\otimes R)*R}
\qquad\qquad 
\inferrule*[Right= ]{\vdots}{(P\otimes R)*R \Rightarrow P\circ Q}  
}
{
(P\otimes R) *  e\pointsto S[\_]* (P_0\otimes R) \Rightarrow P\circ R
}\vspace{0.1cm}
\]
 The use of recursive specification 
 $$R = \mu X.(e\pointsto \forall \vec{y}.\,  \triple{P}{\_}{Q} * P_0)\otimes X$$
  is essential here as it allows us to unroll the definition (see equivalence \eqref{eq:unfold_rec}) so that the ({\sc Eval}) rule can be applied. Note that in the logic of \cite{Honda:Yoshida:Berger:05}, which also uses nested triples  
 but  features neither a specification logic nor  any  frame rules or axioms,  recursive specifications do not exist. 
Avoiding them, one loses an elegant specification mechanism to allow for code updates during   recursion. Such updates are indeed possible as   \SYN{eval} uses a pointer to call code from the (obviously changeable) heap. In the logic of  \cite{Honda:Yoshida:Berger:05}   specifications would have to refer to other means to deal with such code updates, like e.g.\ families of code with uniform specifications. But  it is unclear to what extent such a formulation would allow for modular extensions. For modular reasoning one must not rely on concrete families of code in proofs,  otherwise these proofs are not reusable when the family has to be changed to allow for additional code.   Assuming the code in $e$ does \emph{not} change, the  recursively defined $R$ above can be expressed without recursion (we can omit the $P_0$ now, as this is only needed for mutually recursively defined triples) as follows: 
 $$e\pointsto \triple{e\pointsto k * P}{k}{e\pointsto k * Q}.$$
 The question however remains how the assertion  can be proved for some concrete   $\QUOTE{C}$ that is stored in $[e]$. In  \cite{Honda:Yoshida:Berger:05}  this is done by an induction on some appropriate argument, which is possible since only  total correctness  is considered there.    
 In our logic,    ({\sc OldEval}) is strikingly  similar to a fixpoint induction  rule in ``de~Bakker and Scott'' style and ({\sc Eval}) even allows one to    abstract away from concrete code. These rules are elegant and simple  to use. Not only do they allow for recursion through the store, (\textsc{Eval}) also disentangles the reasoning from the concrete code stored in the heap, supporting modularity and extensibility.

 Figure~\ref{fig:RulesHigherOrder}  summarizes a particular choice of
  proof-rule set from the current and previous subsections.  
Soundness is proved  in Section~\ref{sec:Semantics}.

\begin{figure*}[!t]
\hrule
\begin{gather*}
\inferrule[$\otimes$-Frame]{
  \X;\Gamma \adash P
}{
  \X;\Gamma  \adash P\,{\otimes}\, R
}
\ \  \,  
\inferrule[$*$-Frame]{
  \,
}{
  \X;\Gamma \adash \triple{P}{e}{Q} \Rightarrow \triple{P\,{*}\,R}{e}{Q\,{*}\,R}
}
\  \ \,  
\inferrule[Eval]{
  \X;\Gamma,k \adash R[k] \Rightarrow \triple{P\,{*}\,e\,{\pointsto}\,R[\_]}{k}{Q}
}{
  \X;\Gamma \adash \triple{P\,{*}\,e\,{\pointsto}\,R[\_]}{\QUOTE{\UNQUOTE{[e]}}}{Q}
}
\end{gather*}
\hrule
\caption{Proof rules specific to higher-order store}
\label{fig:RulesHigherOrder}
\end{figure*}

\subsection{Nested triples and classical assertion logic}
\label{subsec:ClassicalAssertionLogic}
One may wonder why we insist on an intuitionistic program logic. 
Unfortunately, as the following proposition shows, it is not possible to use a classical version of our logic; more precisely, the combination of a classical specification logic and   rule (\textsc{$\otimes$-Frame}) is not sound. Thus, by our identification of assertion and specification language, we cannot have a classical assertion logic either. 

\begin{prop}
\label{prop:classicalUnsound}
Adding rule  (\textsc{$\otimes$-Frame})  to a classical specification logic is not sound.
\end{prop}
\proof
Assuming the rule for the elimination of double negation, 
we can derive the   problematic triple $$\triple{\True}{\QUOTE{\SYN{skip}}}{\False}\ .$$
Assume $\neg\triple{\True}{\QUOTE{\SYN{skip}}}{\False}$,
using the abbreviation $\neg\varphi$ for $\varphi\Rightarrow\False$. With   rule (\textsc{$\otimes$-Frame}) to frame in $\False$ we can derive the triple 
$(\neg\triple{\True}{\QUOTE{\SYN{skip}}}{\False})\otimes \False$ from $\neg\triple{\True}{\QUOTE{\SYN{skip}}}{\False}$. 
Since $\True * \False \Leftrightarrow \False$ and $\False * \False\Leftrightarrow\False$, rule (\textsc{Conseq}) and the distribution axioms then let us derive  $\neg\triple{\False}{\QUOTE{\SYN{skip}}}{\False}$. On the other hand, rule (\textsc{Skip}) derives  the triple $\triple{\False}{\QUOTE{\SYN{skip}}}{\False}$.
 Thus, we have shown that from the assumption  $\neg\triple{\True}{\QUOTE{\SYN{skip}}}{\False}$ we can derive $\False$, i.e.\ we have shown  
$\neg \neg\triple{\True}{\QUOTE{\SYN{skip}}}{\False}$. 
By eliminating the double negation we can now derive the triple $\triple{\True}{\QUOTE{\SYN{skip}}}{\False}$. \qed

Note that this derivation does not use nested triples, and also applies to the specification logics used in \cite{BirkedalL:semslt-lmcs,Birkedal:Reus:Schwinghammer:Yang:08}.

\section{Semantics of Nested Triples}
\label{sec:Semantics}

This section develops a model for the programming language and logic we have presented. 
The semantics of programs, given in Subsection~\ref{subsec:semantics_exp_com} using an untyped domain-theoretic model, is standard. 
The following semantics of the logic is, however, unusual; it
is a possible world semantics where the worlds live in a  recursively defined {\em metric\/} space. 
Before we begin with the technical devlopment proper we give a brief overview of the main ideas employed.

\subsection{Overview of the technical development}
\label{sec:overview-of-technical-development}

In earlier work, Birkedal, Torp-Smith, and
Yang~\cite{BirkedalL:semslt-lmcs,Birkedal:Yang:parsepl-journal} showed
how to model a specification logic with higher-order frame rules but
for a language with first-order store. There, the assertion and specification 
logic were kept distinct. Assertions were modelled as semantic predicates
$\Pred=P(H)$, with $H$ the set of heaps,
and specifications as world-indexed truth values $W\to 2$. (These latter maps were
restricted to be monotone in a certain sense, but that does not matter for the present
explanation.)
The informal idea was that the set of worlds would consist of invariants that 
had been framed in and thus worlds consisted of semantic predicates, $W=\Pred$. Here, with 
higher-order store and nested triples and the collapse of assertion and
specification logic, assertions will be modelled as world-indexed predicates.
So we get $\Pred= W\to P(H)$. Worlds will still consist of semantic predicates,
so $W=\Pred$. Thus we see that the set of worlds $W$ should be recursively
defined.  This captures the idea that any assertion can serve as an invariant to 
be framed in via a frame rule.

The idea of using such a Kripke model over a recursively defined set
of worlds comes from~\cite{Birkedal:Stovring:Thamsborg:09}, where this
idea was used to define a model of a type system with general ML-like
references (hence higher-order store). Following~\cite{Birkedal:Stovring:Thamsborg:09} we show how to
find a solution to the recursive world equation in a category of
complete bounded ultra-metric spaces (the definition of which we
recall below). This is possible by restricting the subsets of $H$ that
we use to so-called uniform admissible subsets of $H$. The set $\UAdm$
of all such forms a complete bounded ultra-metric space and thence we
can solve the recursive world equation.  Having solved that, we show how
to define a world extension operator $\otimes$ (which will be used to model
the syntactic $\otimes$ operator used earlier), as a fixed point of a suitable contractive
operator. 
Moreover, we show that the subset $\UAdm$ of $P(H)$ is a complete Heyting algebra with a commutative and monotone monoid structure, 
as needed for the interpretation of separation logic. 

Having defined semantic predicates in certain metric spaces allows us
to interpret recursively defined assertions via application of
Banach's fixed point theorem.

The final core idea in the development is the interpretation of
triples. Here we bake in the frame rules to the model by including
suitable quantifications over future worlds, following ideas from
earlier work~\cite{Birkedal:Reus:Schwinghammer:Yang:08}. To ensure
that nested triples are modelled as semantic predicates, we also force
the interpretation of triples to be metrically non-expansive in the worlds
argument. In particular,    predicates involving nested triples can be used
in recursive definitions of assertions.

\subsection{Semantics of expressions and commands}
\label{subsec:semantics_exp_com}

\begin{figure*}[t]
\hrule
\vspace{-2mm}
\begin{align*}
\den{\SYN{skip}}_\eta h & 
\defeq h
\\[-1.5mm]
\den{C_1\SYN{;}C_2}_\eta h &
\defeq \key{if}\;\den{C_1}_\eta h\,{\in}\,\{\bot,\ERR\}\;\key{then}\,\den{C_1}_\eta h\;\key{else}\,\den{C_2}_\eta\! (\den{C_1}_\eta h)
\\[-1.5mm]
\den{\SYN{if}\,e_1{=}e_2\,\SYN{then}\,C_1\,\SYN{else}\,C_2}_\eta h &
\defeq
\key{if}~\{\EXPden{e_1}_\eta, \EXPden{e_2}_\eta\}\subseteq\lift{\Com}~\key{then}~\bot~\\[-1.5mm]
&\quad\;\;\key{else}~\key{if}~(\EXPden{e_1}_\eta{=}\EXPden{e_2}_\eta)~\key{then}~\den{C_1}_\eta h~\key{else}~\den{C_2}_\eta h
\\[-1.5mm]
\den{\SYN{let}\,x{=}\SYN{new}\,e_1,...,e_{n}\,\SYN{in}\,C}_\eta h &
\defeq \key{let}~\l = 
\min \{\l \mid 
  \forall \l'\ignore{\in\Nats}.\, (\l {\leq} \l'{<} \l{+}n) \IMPLIES \l' \notin \dom{h}\}\\[-1.5mm]
&\quad\;\; \key{in}~\den{C}_{\eta[x\mapsto\l]}
(h\COMB\record{\l{=}\EXPden{e_1}_\eta, \ldots, \l{+}n{-}1{=}\EXPden{e_n}_\eta})
\\[-1.5mm]
\den{\SYN{free}~e}_\eta h &
\defeq 
\key{if}~\EXPden{e}_\eta\notin\dom h~\key{then}~\ERR\\[-1.5mm]
&\quad\;\; \key{else}~(\key{let}~h'~\text{s.t.}~h=h'\COMB\record{\EXPden{e}_\eta{=}h(\EXPden{e}_\eta)}~\key{in}~h')
\\[-1.5mm]
\den{[e_1]\SYN{:=}e_2}_\eta h &
\defeq   
\key{if}~\EXPden{e_1}_\eta\notin\dom h~\key{then}~\ERR~\key{else}~(h[\EXPden{e_1}_\eta{\mapsto}\EXPden{e_2}_\eta])
\\[-1.5mm]
\den{\SYN{let}~x\SYN{=}[e]~\SYN{in}~C}_\eta h &
\defeq
\key{if}~\EXPden{e}_\eta\notin\dom h~\key{then}~\ERR~\key{else}~\den{C}_{\eta[x\mapsto \recselect{h}{\EXPden{e}_\eta}]}h
\\[-1.5mm]
\den{\UNQUOTE [e]}_\eta h &
\defeq
\key{if}~(\EXPden{e}_\eta\notin \dom h \vee \recselect{h}{\EXPden{e}_\eta} \notin \Com)~\key{then}~\ERR\\[-1.5mm]
&\quad\;\;\key{else}~
(h(\EXPden{e}_\eta))(h)
\end{align*}
\vspace{-4mm}
\hrule
\caption{\label{fig:Interpretation}Interpretation of commands $\den{C}_\eta\in\Heap \sto \TERR(\Heap)$}
\end{figure*}

The interpretation of the programming language is given in the category $\Cppo_\bot$ of pointed cpos and strict continuous functions\footnote{As usual, $\sqsubseteq$    denote the partial order of a cpo and   $\bot$  denotes the least element of a pointed cpo, ie.\ $\bot \sqsubseteq d$ for any $d$.}  and is the same as in our previous work \cite{Birkedal:Reus:Schwinghammer:Yang:08}. That is, commands denote strict continuous functions 
$\den{C}_\eta\in \Heap \sto \TERR(\Heap)$
where 
\begin{align}
\label{eqn:domain-eqn}
\Heap &= \Rec\Val 
&
\Val &= \lift{\Ints} \,{\oplus}\, \lift\Com 
&
\Com &= \Heap\,{\sto}\,\TERR(\Heap) 
\end{align}
In these equations,  $\TERR(D) = D\oplus\lift{\{\ERR\}}$ denotes the 
error monad, and $\Rec D$ denotes records with entries from $D$
and labelled by positive natural numbers. 
{Formally, 
$\Rec{D} = \lift{\bigl(\textstyle{\sum_{N\subseteq_\mathit{fin}\Nats}(N{\to}\unlift D)}\bigr)}$ where $(N{\to}\unlift D)$ is the cpo of maps from the finite
address set $N$ to the cpo $\unlift D = D{-}\{\bot\}$  of non-bottom 
elements of~$D$.}
We use some evident record notation, such as 
$\record{\l_1{=}d_1,\ldots, \l_n{=}d_n}$
for the record mapping label $\l_i$ to $d_i$, and $\dom r$ for the set of labels of a record $r$. The \emph{disjointness predicate} $r\DISJ r'$ on records holds if $r$ and $r'$ are not $\bot$ and have disjoint domains, and a 
partial \emph{combining operation} $r\COMB r'$ is defined~by 
\begin{align*}
r\COMB r' &\;\;\defeq\;\; 
\text{if $r\DISJ r'$ then $r\cup r'$ else $\bot$ }.
\end{align*} 
The interpretation of commands is repeated in Figure~\ref{fig:Interpretation} (assuming $h\neq\bot$) and below we
point out where this interpretation deviates from the norm.
Firstly, the $\SYN{new}$ statement uses a deterministic allocator which, however, can not be controlled by the programmer\footnote{This
means that there is no way to stipulate what the new location is as this must  depend solely on the already allocated locations.} 
which is important to ensure that allocation respects the frame rule. Any deterministic allocator would work here, but note that in our denotational semantics
we can only work with deterministic allocation.
The semantics of the \SYN{if} statement is divergence if one of the expressions in the test is a command.
If we wanted to raise an $\ERR$ in this case (which is more appropriate), we would have to include type checking into the logic
 due to our fault avoiding semantics of triples. We decided not do this here as it would
clutter the  rules with  type checking assertions like $\mathsf{int}( e)$ or $\mathsf{com}( e)$ which  are true
in case expression $e$ is an integer valued expression or a command, respectively.

The interpretation of  expressions is entirely standard with the exception  of the quote operation, $\QUOTE C$, that
uses the injection of $\Com$ into $\Val$. Thus, the semantic equations for expressions are omitted.

A solution to equation \eqref{eqn:domain-eqn} for $\Heap$ can be obtained by the usual inverse limit construction \cite{Plotkin:Smyth:82} in the category $\Cppo_\bot$. This solution is an SFP domain (e.g., \cite{Streicher:06}), and thus comes equipped with an increasing chain $\pi_n:\Heap\to\Heap$ of continuous projection maps, satisfying 
$\pi_0 = \bot$, 
 $\bigsqcup_{n\in\omega} \pi_n = \id_\Heap$, 
and $\pi_n\circ\pi_m = \pi_{\min\{n,m\}}$ .
The image of each $\pi_n$ is finite, hence each $\pi_n(h)$ is a compact element of $\Heap$. 
Moreover, the projections are compatible with composition of heaps: we have  $\pi_n(h\COMB h') = \pi_n(h)\cdot\pi_n(h')$ for all $h,h'$. 

\subsection{Semantic domain for assertions}
\label{subsec:Assertion-Domain}

A subset $p\subseteq\Heap$ is \emph{admissible} if $\bot\in p$ and if $p$ is closed under taking least upper bounds of $\omega$-chains. It is \emph{uniform} \cite{Birkedal:Stovring:Thamsborg:09} if it is closed under the projections, i.e., if    $h\in p$ implies  $\pi_n(h)\in p$ for all $n$. 
We write $\UAdm$ for the set of all uniform admissible subsets of $\Heap$. For $p\in\UAdm$, $\RANK{p}n$ denotes the image of $p$ under $\pi_n$. 
Note that uniformity means $\RANK{p}{n}\subseteq p$, and that $\RANK{p}n\in\UAdm$. 
We may regard any subset $p\subseteq\Heap$ (not necessarily uniform or admissible) as a subset of $\TERR(\Heap)$ in the evident way.

The uniform admissible subsets will form the basic building block when interpreting the assertions of our logic. As we have already described informally above, assertions in general depend on invariants for stored code. Thus, 
the space of semantic predicates $\Pred$ will consist of
functions $W\to\UAdm$ from a set of ``worlds,'' describing the invariants,
to the collection of uniform admissible subsets of heaps. 
But, the invariants for stored code are themselves 
semantic predicates, and the interaction between $\Pred$ and $\W$ is governed by (the semantics of) $\otimes$. Hence we seek a space of worlds $W$  that is
``the same'' as $W\to\UAdm$.  
We 
obtain such 
a $W$ using metric spaces.


Recall that a 1-bounded ultrametric space $(X,d)$ is  a 
metric space where the distance function  $d:X\times X\to\RR$ takes values in the closed interval $[0,1]$ and satisfies the strong triangle inequality $d(x,y)\leq \max\{d(x,z),d(z,y)\}$, for all $x,y,z\in X$. 
An (ultra-) metric space is complete if every Cauchy sequence has a limit. 
A function $f:X_1\to X_2$ between metric spaces $(X_1,d_1)$ and $(X_2,d_2)$ is \emph{non-expansive} if for all $x,y\in X_1$, $d_{2}(f(x),f(y))\leq d_1(x,y)$. It is \emph{contractive} if for some $\delta<1$, $d_{2}(f(x),f(y))\leq \delta\cdot d_1(x,y)$ for all $x,y\in X_1$. 
By the Banach fixed point theorem, every contractive function $f:X\to X$ on a non-empty and complete metric space $(X,d)$ has a unique fixed point.

The complete, 1-bounded, non-empty ultrametric spaces and non-expansive functions between them form a Cartesian closed category $\CBUlt$. Products in $\CBUlt$ are given by the set-theoretic product where the distance is the maximum of the componentwise distances. The exponentials are given by the non-expansive functions equipped with the $\sup$-metric, i.e.,  
the exponential $(X_1, d_1) \to (X_2, d_2)$ has the set of
non-expansive functions from $(X_1, d_1)$ to $(X_2, d_2)$ as
underlying set, and 
distance function:
$d_{X_1 \to X_2}(f,g) = \sup\{d_2(f(x),g(x)) \mid x \in X_1 \}$. 
A functor $F:\CBUlt^\textit{op}\times\CBUlt\longrightarrow\CBUlt$ is \emph{locally non-expansive} 
if $d(F(f,g),F(f',g'))\leq \max\{d(f,f'),d(g,g')\}$ for all non-expansive $f,f',g,g'$, 
and it is \emph{locally contractive} 
if $d(F(f,g),F(f',g'))\leq \delta\cdot \max\{d(f,f'),d(g,g')\}$ for some $\delta<1$.
 The functor that results from composing a locally non-expansive functor with a locally contractive one is locally contractive. 
By multiplication of the distance function of an ultrametric space $(X,d)$ with a shrinking factor $\delta <1$ one obtains a new ultrametric space, $\delta\cdot (X,d) = (X,d')$ where $d'(x,y) = \delta\cdot d(x,y)$. Using this operation, a locally contractive functor $(\delta\cdot F)(X_1,X_2) = \delta\cdot (F(X_1,X_2))$ can be obtained from any locally non-expansive functor $F$.

The set $\UAdm$ of uniform admissible subsets of $\Heap$ becomes a complete, 1-bounded ultrametric space when equipped with the following distance function:
\begin{align*}
d(p,q) &= 
\begin{cases}
2^{-\max\{i\in\omega\;|\;\RANK{p}i = \RANK{q}i\}} &\text{if $p\neq q$}\\
0 &\text{otherwise} 
\end{cases}
\end{align*}
Note that $d$ is well-defined: first, because $\pi_0 = \bot$ and $\bot\in p$ for all $p\in\UAdm$ the set $\{i\in\omega\;|\;\RANK{p}i = \RANK{q}i\}$ is non-empty; second,  this set is finite, because $p\neq q$ implies $\RANK{p}i \neq \RANK{q}i$ for all sufficiently large $i$ by the uniformity of $p,q$ and the fact that the limit of the projections $\pi_i$ is the identity on $\Heap$.

\begin{theorem}[Existence of recursive worlds]
\label{thm:ultrametric-existence}
There exists an ultrametric space $W$ and an isomorphism 
$\FOLD$ from $\frac 1 2\cdot(W\to\UAdm)$ to $W$
in $\CBUlt$. 
\end{theorem}
\proof
  By an application of America \& Rutten's existence theorem for fixed points of locally contractive 
  functors~\cite{America-Rutten:JCSS89}, applied to the functor $F(X,Y)= \frac 1 2\cdot(X\to\UAdm)$ on $\CBUlt$. 
  See~\cite{Birkedal:Stovring:Thamsborg:09} for details of a similar recent 
  application.  
\qed

We write $\Pred$ for $\frac 1 2\cdot(\W\to\UAdm)$ and $\UNFOLD: W\cong \Pred$ for the
inverse to $\FOLD$.

\begin{definition}[Approximate equality,  \cite{Birkedal:Stovring:Thamsborg:09}]
For an ultrametric space $(X,d)$ and $n\in\omega$ we use the notation $x \nequiv n y$ to mean that $d(x,y)\leq 2^{-n}$. 
\end{definition}

We conclude this subsection with a number of simple  but useful observations, which will be used repeatedly in the following proofs. 
By the ultrametric inequality, each $\nequiv n$ is an equivalence relation on $X$. 
Moreover, if $n\leq m$ then  ${\nequiv n} \supseteq {\nequiv m}$, and $x=y$ if and only if $x\nequiv n y$ for all $n\in\omega$. 
%
Since all non-zero distances in $\UAdm$ are of the form $2^{-n}$ for some $n\in\omega$, this is also the case for the distance function on $\W$. 
Therefore, to show that a map is non-expansive it suffices to show that $f(x)\nequiv n f(y)$ whenever $x\nequiv n y$. 
Finally, the definition of $\Pred$ has  the following consequence:   
for $p,q\in\Pred$, $p\nequiv n q$ holds if and only if $p(w)\nequiv{n-1} q(w)$ for all $w\in\W$.

\subsection{Separating conjunction and invariant extension}

For $p,q\in\UAdm$, the separating conjunction $p*q$ is defined as usual, by 
\begin{align*}
h \in p*q\ \defiff\ \ 
\exists h_1,h_2.\  \ h = h_1\COMB h_2\ \wedge\ h_1 \in p\ \wedge\ h_2\in q.
\end{align*}
This operation is lifted  to non-expansive functions $p_1,p_2\in\Pred$ pointwise, by letting 
$(p_1 \,{*}\, p_2)(w) \,=\, p_1(w) \,{*}\, p_2(w)$. 
This lifting is well-defined, and moreover determines a non-expansive operation on the space $\Pred$:
\begin{lemma}[Separating conjunction]
\label{lem:separation:non-expansiveness}
If $p,q\in\Pred$ then  $p*q\in\Pred$. Moreover, the assignment of $p, q$ to $p*q$ is a non-expansive operation on $\Pred$.
\end{lemma}
\proof
As a preliminary step one shows that separating conjunction on $\UAdm$ is well-defined, i.e., if $p,q\in\UAdm$ then so is $p*q$: 
The admissibility of $p*q$ follows from $\bot = \bot\COMB\bot$, and from the fact that (non-$\bot$) heaps are only comparable with   respect to the order on $\Heap$ if they have equal (finite) domains. More precisely, any chain $h_0\sqsubseteq h_1\sqsubseteq\ldots$ in $p*q$ 
must have a subsequence $(h_{i_k})_k = (h_{i_k}'\COMB h_{i_k}'')_k$ that splits into chains $h_{i_1}'\sqsubseteq h_{i_2}'\sqsubseteq\ldots$  in $p$ and  $h_{i_1}''\sqsubseteq h_{i_2}''\sqsubseteq\ldots$  in $q$. The combination of their respective lubs in $p$ and $q$ is the lub of the $h_k$'s, and therefore in $p*q$ by the admissibility of $p$ and $q$. 
The uniformity of $p*q$ is a consequence of the equation $\pi_n(h_1\COMB h_2) = \pi_n(h_1)\COMB \pi_n(h_2)\in p*q$.
 
We now show that for $p,q\in\Pred$, $p*q$ is a non-expansive function. Suppose $w,w'\in\W$ such that $w\nequiv n w'$, and suppose $\pi_n(h)\in(p * q)(w)=p(w)*q(w)$. We must show that $\pi_n(h)\in (p * q)(w')$. 
By definition of $*$ on $\UAdm$ there exist $h_1\in p(w)$ and $h_2\in q(w)$ such that $\pi_n(h)=h_1\COMB h_2$. 
By uniformity, we also have $\pi_n(h_1)\in p(w)$ and $\pi_n(h_2)\in q(w)$. 
Since we assumed $w\nequiv n w'$, this yields 
\[\pi_n(h_1\COMB h_2) = \pi_n(h_1)\COMB \pi_n(h_2)\in p(w')*q(w') = (p*q)(w').\] 
Finally, since $\pi_n(h) = \pi_n(\pi_n(h)) = \pi_n(h_1\COMB h_2)$, the statement $\pi_n(h)\in (p * q)(w')$ follows. 

To see that separating conjunction is non-expansive, assume that $p\nequiv n p'$ and $q\nequiv n q'$ for arbitrary $p,p',q,q'\in\Pred$. We must show that 
$p*q\nequiv n p'*q'$. Since $\Pred = \frac 1 2 \cdot(\W\to\UAdm)$ we can equivalently show that 
$p(w)*q(w)\nequiv {n-1} p'(w)*q'(w)$ for all $w\in\W$. 
This follows from the assumption that $p\nequiv{n}p'$
and $q\nequiv{n}q'$ and the fact that 
$\pi_{n{-}1}(h) \,{=}\, \pi_{n{-}1}(h_1)\COMB\pi_{n{-}1}(h_2)$ whenever $h\,{=}\,h_1\COMB h_2$. \qed

The corresponding unit for the lifted separating conjunction   is the non-expansive function $\EMP \,{=}\,\lambda w.\{\record{},\bot\}$, i.e., $p * \EMP =\EMP * p = p$ holds for all $p\in\Pred$.
We let the world $\wemp \defeq \FOLD(\EMP)$ be its image under the isomorphism. 

The following lemma introduces semantic analogues of the syntactic invariant extension operation $P\otimes R$ and the invariant combination $R\circ R'$. 

\begin{lemma}
[Invariant combination and invariant extension] 
\label{lem:existence-tensor-and-circ}
There exists a non-expansive map $\circ : W\times W\to W$  and a map 
$\otimes : \Pred\times W \to\Pred$ 
that is non-expansive in its first and contractive in its second argument, satisfying the equations
\[
r\circ r' = \FOLD(\UNFOLD(r)\otimes r' *\UNFOLD(r'))
\quad \text{and}\quad
(p\otimes r)(w) =  p(r\circ w)\] 
for all $p\,{\in}\,\Pred$ and $r,r'\,{\in}\,\W$.
\end{lemma}
\proof
The defining equations of both operations give rise to contractive maps, which have (unique) fixed points by Banach's fixed point theorem. 
More precisely, consider the endofunction $\PHI\cdot$ on the  function  space $W\times W\to W$, defined for all $\circ\in(W\times W\to W)$ and all $r,r'\in\W$ by 
\begin{align*}
r \PHI{\circ} r' &= \FOLD((\lambda w.\UNFOLD(r)(r'\circ w)) * \UNFOLD(r'))\ .
\end{align*}
Note that $\PHI{\circ}$ is indeed a non-expansive function, i.e., an element of the function space $(W\times W\to W)$: 
if $r\nequiv n s$ and $r'\nequiv n s'$ then $r'\circ w\nequiv n s'\circ w$ holds in $W$, for all $w\in\W$, and $\UNFOLD (r)\nequiv n \UNFOLD(s)$ and $\UNFOLD (r')\nequiv n \UNFOLD(s')$   holds in $\Pred$. Since separating conjunction is non-expansive by Lemma \ref{lem:separation:non-expansiveness}, the approximate equality
\begin{align*}
(\lambda w.\UNFOLD(r)(r'\circ w)) * \UNFOLD(r')
\nequiv{n-1}
(\lambda w.\UNFOLD(s)(s'\circ w)) * \UNFOLD(s')
\end{align*}
holds in $\W\to\UAdm$, so that $r\PHI{\circ}r' \nequiv n s\PHI\circ s'$ in $\W$. 

We show that the function $\PHI\cdot$ is contractive. Assume that $\circ_1\nequiv n\circ_2$ holds in $W\times W\to W$; we must show that $\PHI{\circ_1}\nequiv {n+1}\PHI{\circ_2}$. Let $r,r'\in\W$ be arbitrary. Then by the $\sup$-metric on $\W\times \W\to\W$ it suffices to prove that $r\PHI{\circ_1}r' \nequiv {n+1} r\PHI{\circ_2} r'$ holds in $\W$, or equivalently, that 
\begin{align*}
(\lambda w.\UNFOLD(r)(r'\circ_1 w)) * \UNFOLD(r')
\nequiv{n}
(\lambda w.\UNFOLD(r)(r'\circ_2 w)) * \UNFOLD(r')
\end{align*}
holds in $\W\to\UAdm$. 
By the non-expansiveness of separating conjunction (Lemma~\ref{lem:separation:non-expansiveness}) and the $\sup$-metric on $\W\to\UAdm$, this follows since $r'\circ_1 w\nequiv n r'\circ_2 w$ holds for all $w\in\W$ by the assumption that $\circ_1\nequiv n\circ_2$, and hence $\UNFOLD(r)(r'\circ_1 w)\nequiv n\UNFOLD(r)(r'\circ_2 w)$  holds. 

By contractiveness of $\PHI\cdot$ and the Banach fixed point theorem, there exists a unique non-expansive map $\circ$ satisfying $r\circ r' = r\PHI\circ r'$. 
We can now define the operation $\otimes : \Pred\times W\to\Pred$ by 
$p\otimes r \defeq \lambda w. p(r\circ w)$ for all $p\in\Pred$ and $r\in\W$, from which the required equivalences follow: 
\begin{align*}
r\circ r'\  =\ r\PHI\circ r'
\ =\ \FOLD(\lambda w.\UNFOLD(r)(r'\circ w) * \UNFOLD(r'))
\ =\ \FOLD(\UNFOLD(r)\otimes r'\, * \UNFOLD(r'))
\end{align*}
Finally, we note that if $p\nequiv n p'$ and $r\nequiv m r'$ then $p\otimes r\nequiv k p'\otimes r'$ for $k = \min\{n,m+1\}$, i.e., the operation is non-expansive in its first argument and contractive in its second argument. To see this, suppose $p\nequiv n p'$ holds in $\Pred$ and $r\nequiv m r'$ holds in $\W$. Without loss of generality we may assume $n>0$, so that $p\nequiv {n-1} p'$ holds in $\W\to\UAdm$. By non-expansiveness of $\circ$ it follows that $r\circ w \nequiv m r'\circ w$ for all $w$, and therefore 
$\lambda w.p(r\circ w)\nequiv{\min\{n{-}1,m\}}\lambda w.p'(r'\circ w)$ in $\W\to\UAdm$. 
Hence $p\otimes r\nequiv{\min\{n,m{+}1\}} p'\otimes r'$ holds in $\Pred$ as required. \qed

The following lemma establishes key properties of the two operations $\circ$ and $\otimes$ that we defined in Lemma~\ref{lem:existence-tensor-and-circ}. These properties provide a semantic explanation of the distribution axioms given in Figure~\ref{fig:DistRules}. 
  
\begin{lemma}
[Monoid structure and monoid action]
\label{lem:monoid-structure}
  $(W,\circ,\wemp)$ is a monoid in $\CBUlt$. Moreover, $\otimes$ is an action of 
  this monoid on $\Pred$.
\end{lemma}
\proof
First, $\wemp$ is a left-unit for $\circ$, since 
\begin{align*}
\wemp\circ r 
\ =\ \FOLD((\lambda w.\UNFOLD(\wemp)(r\circ w)) * \UNFOLD (r))
\ =\ \FOLD(\UNFOLD(r)) 
\ =\ r\ .
\end{align*} 
Using this fact, it is easy to prove that it is also a right-unit for the $\circ$ operation: 
\begin{align*}
r\circ\wemp 
\ =\ \FOLD(\lambda w.\UNFOLD(r)(\wemp\circ w) * \UNFOLD(\wemp))
\ =\ \FOLD(\lambda w.\UNFOLD(r)(w) * \EMP)
\ =\ r\ .
\end{align*}
Next, we prove by induction that for all $n\in\omega$, $\circ$ is associative up to distance $2^{-n}$, from which associativity follows. By the 1-boundedness of $W$ the base case is clear. For the inductive step $n>0$,  by definition of the distance function on $\Pred$ it suffices to show that for all $w\in\W$, $\UNFOLD((r\circ s)\circ t)(w) \nequiv {n-1} \UNFOLD(r\circ (s\circ t))(w)$. This equation follows from the definition of $\circ$ as follows:
\begin{align*}
\UNFOLD((r\circ s)\circ t)(w)
&= \UNFOLD(r\circ s)(t\circ w) * \UNFOLD(t)(w)\\
&= \UNFOLD(r)(s\circ(t\circ w)) * \UNFOLD(s)(t\circ w) * \UNFOLD(t)(w)\\
&=  \UNFOLD(r)(s\circ(t\circ w)) * \UNFOLD(s\circ t)(w)\\
&\nequiv{n-1}   \UNFOLD(r)((s\circ t)\circ w) * \UNFOLD(s\circ t)(w)\\
&= \UNFOLD(r\circ(s\circ t))(w)\ .
\end{align*}
The second last step in this derivation is by the inductive hypothesis, using the non-expansiveness of $\UNFOLD(r)$. 

That $\otimes$ forms an action of $\W$ on $\Pred$ follows from these properties of $\circ$. First, $p\otimes\wemp = \lambda w.p(\wemp\circ w) = p$ since $\wemp$ is a unit for $\circ$. Second,
\begin{align*}
(p\otimes r)\otimes s 
= \lambda w.p(r\circ(s\circ w))
=\lambda w.p((r\circ s)\circ w) 
= p\otimes(r\circ s)
\end{align*}
by the associativity of $\circ$. \qed

\subsection{Semantics of triples and assertions}
\label{subsec:Assertion-Interpretation}

Since assertions appear in the pre- and post-conditions of Hoare triples, and triples can be nested inside assertions, the interpretation of assertions and the validity of triples must be defined simultaneously. To achieve this, we first define a notion of fault-avoiding semantic triple. 

\begin{definition}[Semantic triple] A \emph{semantic Hoare triple} consists of predicates  $p,q\in \Pred$ and a strict continuous function $c\in\Heap\sto\TERR(\Heap)$, written $\triple pcq$. For $w\in\W$, a semantic triple  $\triple pcq$ is \emph{forced by $w$}, written $w\models\triple{p}{c}{q}$, if for all $r\in \UAdm$ and all $h\in\Heap$:
\begin{align*}
h\in p(w) * \UNFOLD(w)(\wemp) *r\ \IMPLIES\ 
c(h)\in \DCl(q(w) * \UNFOLD(w)(\wemp) * r) ,  
\end{align*}
where $\DCl(r)$ denotes the least downward closed and admissible set of heaps containing $r$. 
A semantic triple is \emph{valid}, written $\models\triple pcq$, if $w\models\triple{p}{c}{q}$ for all $w\in\W$. We extend semantic triples from $\Com = \Heap\sto\TERR(\Heap)$ to all $d\in\Val$, by $w\models\triple{p}{d}{q}$ iff $d = c$ for some command $c\in\Com$ and $w\models\triple{p}{c}{q}$. \\
A triple holds \emph{approximately up to level $k$}, $w\models_k \triple{p}{d}{q}$, if $w\models \triple{p}{\pi_k;d;\pi_k}{q}$.
\end{definition}

Thus, semantic triples bake in the first-order frame property (by conjoining $r$), and ``close'' the ``open'' recursion (by applying the world $w$, on which the triple implicitly depends, to $\wemp$). The semantics also ensures that if a triple holds the command in question must not have produced $\ERR$ as result. One calls such a   semantics   \emph{fault-avoiding} and this is one of the intrinsic features of Separation Logic.  In our case fault-avoidance follows directly from the fact that semantics of assertions indexed by worlds lives in $\UAdm$ that ranges over heaps and does not include value $\ERR$.
The admissible downward closure that is applied to the entire post-condition is in line with a partial correctness interpretation of triples. In particular, it entails that the sets 
$\{c\in\Com\;|\;w\models_k \triple{p}{c}{q}\}$ and $\{c\in\Com\;|\;w\models \triple{p}{c}{q}\}$ 
are admissible and downward closed subsets of $\Com$. 

Since there is a closure operation applied to the post-condition of semantic triples, but no similar closure used in the pre-condition, it may not be immediate that   proved commands compose. The following characterisation is helpful, for instance when proving soundness of the rule of sequential composition. 
\begin{lemma}[Closure]
\label{lem:precondition-closure}
If $f{:}D\,{\sto}\,D'$ is a strict continuous function, $q\subseteq D'$ is an admissible and downwards closed subset of $D'$, and $p\subseteq D$ is an arbitrary subset of $D$, 
then $f(p)\subseteq q$ implies $f(\DCl(p))\subseteq q$. 
\end{lemma}
\proof
Since $f$ is continuous, the pre-image $f^{-1}(q)$ of $q$ is admissible and downward closed. 
From the assumption that $f(p)\,{\subseteq}\, q$ it follows that $p\,{\subseteq}\,f^{-1}(q)$, and thus $\DCl(p)\,{\subseteq}\,f^{-1}(q)$ as the former is by definition the least admissible and downward closed subset of $D$ containing $p$. 
Thus, if $h\,{\in}\,\DCl(p)$ then $f(h)\,{\in}\,q$. \qed

Observe that $w\models_k \triple{p}{d}{q}$ provides indeed an approximation of the judgement $w\models \triple{p}{c}{q}$, in the sense that $w\models \triple{p}{c}{q}$ is equivalent to  $\forall  k\in\omega.\, w\models_k \triple{p}{c}{q}$. 
Finally, semantic triples are non-expansive, in the sense that if 
$w {\nequiv{n+1}} w'$ and $w\,{\models}_n\,\triple pcq$, then $w'\,{\models}_{n}\,\triple p{
c}q$; they are similarly non-expansive in the pre- and post-conditions $p$ and $q$. 
This observation plays a key role in the following definition of the semantics of nested triples. 
\begin{lemma}[Non-expansiveness of semantic triples] 
\label{lem:semantic-triples-key}
Let $w,w'\in\W$ such that $w\,{\nequiv{n{+}1}}\,w'$. 
Let $p,p',q,q'\in\Pred$ be such that $p\nequiv n p'$ and $q\nequiv n q'$. 
If $w\,{\models}_n\,\triple pcq$, then $w'\,{\models}_{n}\,\triple {p'}{
c}{q'}$. 
\end{lemma}
\proof
Let $w,w',p,p',q$ and $q'$ be as in the statement of the lemma, and 
let $c:\Heap\sto\TERR(\Heap)$ be such that $w\models_n\triple pcq$. 
To prove that $w'\models_n\triple p{c'}q$, suppose $r\in \UAdm$ and $h\in\Heap$ are such that 
$h\in p'(w') * \UNFOLD(w')(\wemp) *r$. We have to show that 
$\pi_n(c(\pi_n\,h))\in \DCl(q'(w') * \UNFOLD(w')(\wemp) *r)$. 

Since $w\,{\nequiv{n{+}1}}\,w'$ holds by assumption, we have $\UNFOLD(w')(\wemp)\nequiv{n }\UNFOLD(w)(\wemp)$.  
Hence, by the non-expansiveness of $p$, by the assumption $p\nequiv n p'$, and by the compatibility of the heap combination operation with projections, we have 
$\pi_{n }(h)\in p(w) * \UNFOLD(w)(\wemp) *r$. 
By the assumption that $w\models_n\triple pcq$ and since $\pi_n\circ\pi_n=\pi_n$, this yields 
$\pi_n(c(\pi_{n}\,h))\in \DCl(q(w) * \UNFOLD(w)(\wemp) *r)$. 
Using the non-expansiveness of $q$, the assumption $q\nequiv n q'$,  uniformity of $r$, and the fact that $\UNFOLD(w)(\wemp)\nequiv{n }\UNFOLD(w')(\wemp)$, we know that  
$\pi_{n }(h')\in q'(w') * \UNFOLD(w')(\wemp) *r$ holds whenever $h'\in q(w) * \UNFOLD(w)(\wemp) *r$. 
Thus, using $\pi_n\circ\pi_n=\pi_n$ again, $\pi_{n }(c(\pi_{n }(h)))\in \DCl(q'(w') * \UNFOLD(w')(\wemp) *r)$ holds by Lemma~\ref{lem:precondition-closure} and the  continuity of the projection $\pi_{n }$. \qed 

Assertions (without free relation variables) are interpreted as elements  
$\den{P}_\eta\in\Pred$. 
More generally,  assume that the free relation variables of $P$ are contained in $\X = X_1,\ldots,X_n$, where the arity of $X_i$ is $n_i$. 
Then $P$ denotes  a non-expansive function from $\prod_{X_i\in\X}\Pred^{(\Val^{n_i})}$ to $\Pred$. 
Note that $(\UAdm,\subseteq)$ is a complete Heyting algebra (as shown in Appendix \ref{app:subsec:assertions}, 
Lemma~\ref{lem:UAdm-Heyting}). Using the pointwise extension of the operations of this algebra to the set of non-expansive functions 
$\W\to\UAdm$, we also obtain a complete Heyting  algebra on $\Pred = \frac 1 2\cdot(\W\to\UAdm)$ which soundly models the intuitionistic 
predicate  part of the assertion logic. (See Appendix \ref{app:subsec:assertions}, Lemma~\ref{lem:UAdm-Heyting-II} for details.)  
The monoid action of $\W$ on $\Pred$ serves to model the invariant extension of the assertion logic. 

\begin{remark}\label{remark:wand}
While $\UAdm$ (and hence $\Pred$) is a complete Heyting algebra, it is not a complete Heyting \emph{BI} algebra, as usually assumed for the interpretation of the assertion language in separation logic \cite{Pym:OHearn:Yang:04}. 
More precisely, 
what is missing is the right adjoint (``magic wand'') for the monoid operation $*$: the candidate operation, 
\[
p\mimp q = \{h\;|\;\forall n\in\omega.\forall h'\in \Heap.\ \text{if $\pi_n(h')\in p\ \wedge\ \pi_n(h)\DISJ \pi_n(h')$ then $\pi_n(h\COMB h')\in q$}\}\ ,
\] 
 alas,  fails to be \emph{non-expansive}. This is a particularly annoying shortcoming of our model since this spatial implication is important when dealing with shared memory. For instance, $(P\mimp Q) * (P\mimp R) * P$ expresses that $R$ and $Q$ overlap in shared part  $P$. 
 Recently, we have constructed an alternative model of our logic, based on an operational semantics of the programming language and using the ideas of step-indexing, where the right adjoint does exist. 
 \end{remark}

In order to define an  interpretation of nested triples we   use the following definition:
\begin{definition}[Rank of  a heap]
If $h$ is a compact element of $\Heap$, then the least $n$ for which $\pi_n(h) = h$ is the \emph{rank} of $h$, abbreviated $\textit{rnk}(h)$, otherwise the rank is undefined.
 \end{definition}

\begin{figure}[t]
\hrule
\begin{align*}
\den{X(\vec e)}_{\eta,\rho} w & =   \rho(X)(\den{\vec e}_\eta)w\\
\den{\False}_{\eta,\rho} w & =   \{\bot\}\\
\den{\True}_{\eta,\rho} w & =  \Heap\\
\den{P\vee Q}_{\eta,\rho} w & =   \den{P}_{\eta,\rho} w\cup\den{Q}_{\eta,\rho} w\\
\den{P\wedge Q}_{\eta,\rho} w & =  \den{P}_{\eta,\rho} w\cap\den{Q}_{\eta,\rho} w\\
\den{P \Rightarrow Q}_{\eta,\rho} w & = \{h\;|\;\forall n\in\omega.\ \pi_n(h)\in\den{P}_{\eta,\rho} w\ \text{implies}\ \pi_n(h)\in\den{Q}_{\eta,\rho} w\}\\
\den{\forall x.P}_{\eta,\rho} w & = \textstyle{\bigcap_{d\in\Val}} \den{P}_{\eta[x:=d],\rho}w\\
\den{\exists x.P}_{\eta,\rho} w & =  \{h\;|\; \forall n\in\omega.\ \pi_n(h)\in \textstyle{\bigcup_{d\in\Val}} \den{P}_{\eta[x:=d],\rho}w\}\\
\den{e_1=e_2}_{\eta,\rho} w & =\{h\;|\;h\neq\bot\Rightarrow\,\den{e_1}_\eta = \den{e_2}_\eta\}\\
\den{e_1\,{\mapsto}\, e_2}_{\eta,\rho} w & = \{h\;|\;h\sqsubseteq  \record{\den{e_1}_\eta  = \den{e_2}_{\eta}}\}\\
\den{\EMP}_{\eta,\rho} w & = \{\record{},\bot\}\\
\den{P*Q}_{\eta,\rho} w & =  \den{P}_{\eta,\rho} w * \den{Q}_{\eta,\rho} w\\
\den{\triple{P}{e}{Q}}_{\eta,\rho} w & = \DCl \{ h \in \Heap \ |\ rnk(h)>0 \Rightarrow  w\models_{rnk(h)-1} \triple{\den{P}_{\eta,\rho}}{\den{e}_\eta}{\den{Q}_{\eta,\rho}} \}\\
\den{P\otimes Q}_{\eta,\rho} w & =  ( \den{P}_{\eta,\rho}\otimes\FOLD(\den{Q}_{\eta,\rho}) ) w\\
\den{(\mu X(\vec x).P)(\vec e)}_{\eta,\rho} w & =  \textit{fix}(\lambda q,\vec d. \den{P}_{\eta[\vec x:=\vec d],\rho[X:=q]}) (\den{\vec e}_\eta)w
\end{align*}
\hrule
\caption{\label{fig:assertion-semantics}Semantics of assertions}
\end{figure}

The interpretation of assertions is spelled out in detail in Figure~\ref{fig:assertion-semantics}. 
The interpretation of  a nested triple $\triple{P}{e}{Q}$  is \emph{not} independent of the heap,    unlike the  (more traditional) semantics     of ``top-level'' triples, i.e.\  $\models \triple{p}{c}{q}$. More precisely, the definition in Figure~\ref{fig:assertion-semantics} means that triples as assertions depend on the  {rank} of the current heap. This is necessary to provide a 
 \emph{non-expansive} function from $\W$ to $\UAdm$. 
 Simpler definitions of the   interpretation of triples, like $\{ h\in \Heap\ |\, w \models \triple{\den{P}_{\eta,\rho
 }}{\den{e}_\eta}{\den{Q}_{\eta,\rho}}  \}$, are heap independent but   not non-expansive. 
A similar approach has been taken in \cite{Birkedal:Stovring:Thamsborg:09} to force non-expansiveness for a reference type constructor for ML-style references. 
We discuss the ramifications of this choice in Section~\ref{sec:discussion-of-proof-rules}. 
Note also that the only atomic assertions that depend on the world $w$ are triples, as they are the only ones that are affected by invariants.

\begin{lemma}[Well-definedness]
The interpretation in Figure~\ref{fig:assertion-semantics} is well-defined: 
\begin{enumerate}[\em(1)]
\item If the free relation variables of $P$ are contained in $\X= X_1,\ldots,X_n$ then $\den{P}_{\eta}$ denotes a non-expansive function from $\prod_{X_i\in\X}\Pred^{(\Val^{n_i})}$ to $\Pred$. 
\item If $P$ is formally contractive in $X$ then the functional $\lambda q.\den{P}_{\eta,\rho[X:=q]}$  is a contractive map from $\Pred^{(\Val^n)}$ to $\Pred$. 
\end{enumerate}
\end{lemma}

\proof[Proof sketch]
Both parts are proved simultaneously by induction on the structure of $P$. 
The second part is used to show the well-definedness of recursive specifications, using the fact that the fixed point operator itself is non-expansive. 
Details are given in Appendix \ref{subsec:app:interpretation-assertions}. \qed

As a  consequence of the interpretation of triples, the axiom  
$\triple{\triple{A}{e}{B} \wedge A}{e}{B}$ does not hold; the inner triple  is only approximately valid up to the level of the rank of the argument heap. Similarly, the following rule
\[
\frac{\triple{A}{e}{B} \Rightarrow \triple{P}{e'}{Q}} {\triple{\triple{A}{e}{B} \wedge P}{e'}{Q}}
\]
 is not validated by our semantics (the opposite direction actually holds; see Section~\ref{sec:discussion-of-proof-rules}). 
Axioms and rules like these are  used, 
 e.g., by Honda et al.\ \cite{Honda:Yoshida:Berger:05}, in   proofs for recursion through the store; instead we use ({\textsc{Eval}}).

\subsection{Soundness of the axioms and proof rules}

We prove soundness of  the axioms and proof rules listed in 
Sections~\ref{sec:syntax} and \ref{sec:ProgramLogic}.
We start by defining  a notion of validity for judgements and rules with respect to which the soundness will be shown. 

\begin{definition}[Validity of judgements]
A judgement $\X;\Gamma \,\adash\,
\triple{P}{\QUOTE{C}}{Q}$ is \emph{valid} if, and only if, for all $\eta \in \Env $ such that 
$\dom\eta\supseteq \Gamma$ and for all $\rho \in \prod_{X_i\in\X}\Pred^{(\Val^{n_i})}$
such that  $n_i$ is the arity of $X_i$ we have
$\models \triple{\den{P}_{\eta,\rho}}{\den{C}_{\eta}}{\den{Q}_{\eta,\rho}}$.
A   rule $\frac{J_1}{J_2}$ is then called \emph{sound} if validity of judgement $J_1$
implies the validity of judgement $J_2$. Similarly, an axiom $J$ is called \emph{sound}
if  judgement $J$ is valid.
\end{definition}
Below we prove the most interesting rules of our logic sound.  Where proofs are parametric in  the assertions
  we will directly work with semantic Hoare triples. 

Let us first consider the distribution axioms for $-\otimes R$ given 
in Figure~\ref{fig:DistRules}.

\begin{lemma}[Distribution axioms]
\label{lem:soundness:distribution}
The  distribution axioms for $-\otimes R$ 
are valid.
\end{lemma}
\proof
We consider the case of invariant extension and triples:
\begin{iteMize}{$\bullet$}
\item The validity of $(P\,{\otimes}\,Q)\,{\otimes}\,R\IFF P\,{\otimes}\,(Q\,{\circ}\,R)$ is an instance of the fact that $\otimes$ is a monoid action (Lemma~\ref{lem:monoid-structure}). 
\item The validity of $\triple{P}{e}{Q}\,{\otimes}\, R\IFF \triple{P\,{\circ}\,R}{e}{Q\,{\circ}\,R}$  follows from the following claim: for all $p,q,r\in\Pred$, strict continuous $c:\Heap\sto\TERR(\Heap)$ and all $w\in\W$,
$\FOLD(r)\circ w \models\triple{p}{c}{q}$ if and only if  
$w\models\triple{p\otimes\FOLD(r)\,*r}{c}{q\otimes\FOLD(r)\,*r}$.
The proof of this claim uses the property 
\begin{align*}
\forall p.\ (p\otimes\FOLD(r) * r)(w) * \UNFOLD(w)(\wemp) 
&= p(\FOLD(r)\circ w) * \UNFOLD(\FOLD(r)\circ w)(\wemp) \ .
\end{align*}
This property  is a  consequence of the definitions of $\otimes$ and $\circ$: 
\begin{align*}
(p\otimes\FOLD(r) * r)(w) * \UNFOLD(w)(\wemp) 
&= (p\otimes\FOLD(r))(w) * r(w) * \UNFOLD(w)(\wemp) \\
&= p(\FOLD(r)\circ w) * r(w\circ\wemp) * \UNFOLD(w)(\wemp)\\
&= p(\FOLD(r)\circ w) * (r\otimes w)(\wemp) * \UNFOLD(w)(\wemp)\\
&= p(\FOLD(r)\circ w) * (r\otimes w * \UNFOLD(w))(\wemp)\\
&= p(\FOLD(r)\circ w) * \UNFOLD(\FOLD(r)\circ w)(\wemp)\ .
\end{align*}
\end{iteMize} 
The proofs of the remaining distribution axioms are easy since the  logical connectives are interpreted   pointwise, and since $\EMP$ 
and $(e_1\pointsto e_2)$ are constant. \qed

Next, we consider the proof rules for higher-order store given in  
Figure~\ref{fig:RulesHigherOrder}. 

\begin{lemma}[$\otimes$-Frame]
\label{lem:higher-order-frame-rule}
The {\sc $\otimes$-Frame} rule is sound: if 
 $h\in {p} (w)$ for all $h\in\Heap$ and $w\in\W$,
then  $h\in (p\otimes \FOLD(r)) (w)$ 
for all $h\in\Heap$, $w\in\W$ and $r\in\Pred$. 
\end{lemma}
\proof
Assume that $h\,{\in}\,{p}(w)$ holds for all $h\in\Heap$ and $w\in\W$. Let $r\,{\in}\,\Pred$,
$w\,{\in}\,\W$ and $h\,{\in}\,\Heap$. 
We show $h\,{\in}\,(p\,{\otimes}\,\FOLD(r)) (w)$. 
Note that we have $(p\,{\otimes}\, \FOLD(r)) (w)  \,{=}\, p(\FOLD(r)\,{\circ}\, w)$
by the definition of $\otimes$.
So,  for $w'\,{\defeq}\,\FOLD(r)\,{\circ}\,w$, the assumption yields $h\,{\in}\, p(w') 
\,{=}\, (p\,{\otimes}\,\FOLD(r)) (w)$.
\qed

The rule (\textsc{$\otimes$-Mono}), which expresses the monotonicity of $\otimes$ in its left-hand argument, is in fact derivable from ($\otimes$-Frame) and the distribution axioms. Thus, its soundness is a consequence of Lemmas~\ref{lem:soundness:distribution} and \ref{lem:higher-order-frame-rule}.

\begin{lemma}[$*$-Frame]
\label{lem:star-frame}
The axiom $\triple{P}{e}{Q} \,{\Rightarrow}\, \triple{P\,{*}\,R}{e}{Q\,{*}\,R}$ is valid 
for all $P,Q,R,e$.
\end{lemma}
\proof
We show that for all  worlds $w \,{\in}\, \W$,  predicates $p,q,r \,{\in}\, \Pred$ and commands  $c\,{\in}\,\Com$,
if $w\,{\models}\,\triple pcq$, then $w\,{\models}\,\triple{p\,{*}\,r}{c}{q\,{*}\,r}$. This implies the lemma as follows. If 
$k>0$ is the rank of $\pi_n(h)$ and
$\pi_n(h) \,{\in}\, \den{\triple{P}{e}{Q}}_{\eta,\rho}w$, 
 then $w\models_{k-1} \triple {\den{P}_{\eta,\rho}}{\den{e}_\eta}{\den{Q}_{\eta,\rho}}$.
This lets us conclude $w\models_{k-1}\triple {\den{P*R}_{\eta,\rho}}{\den{e}_\eta}{\den{Q*R}_{\eta,\rho}}$,
which in turn implies that $\pi_n(h)$ is in 
$\den{\triple{P*R}{e}{Q*R}}_{\eta,\rho}w$.

To prove the claim, assume $w\,{\models}\,\triple pcq$. We must show that $w\models\triple{p*r}{c}{q*r}$.
Let $r'\in\UAdm$ and 
assume 
\[
h\in (p \,{*}\, r)(w) \,{*}\,\UNFOLD(w)(\wemp) \,{*}\, r' 
= p(w)\,{*}\,\UNFOLD(w)(\wemp) \,{*}\, (r(w)\,{*}\,r')\ .
\] 
Since  $w\,{\models}\,\triple pcq$, it follows that 
\[
c(h)\,{\in}\, \DCl(q(w)\,{*}\,\UNFOLD(w)(\wemp) \,{*}\, (r(w)\,{*}\,r')) 
= \DCl((q\,{*}\,r)(w)\,{*}\,\UNFOLD(w)(\wemp) \,{*}\, r')\ ,
\] 
which establishes $w\,{\models}\,\triple{p\,{*}\,r}{c}{q\,{*}\,r}$.  \qed

\begin{lemma}[Eval] 
\label{lem:Eval}
Suppose that  $R[k] \Rightarrow \triple{P*e\,{\pointsto}\,R[\_]}{k}{Q}$ is a valid implication. 
Then, if there are no free occurrences of $k$, also  $\triple{P*e\,{\pointsto}\,R[\_]}{\QUOTE{\UNQUOTE{[e]}}}{Q}$ is valid. 
\end{lemma}

\proof
Let $w\in\W$, $\eta\in\Env$ and $r\in\UAdm$. Let $\rho$ be a suitable assertion environment. 
Let $h\in\den{P*e\pointsto R[\_]}_{\eta,\rho} w *\UNFOLD(w)(\wemp) *r$, so that 
$h = h'\COMB h''$ for some $h'$ and $h''$ such that 
\begin{align}
\label{eqn:eval-1}
h'\in\den{e\pointsto R[\_]}_{\eta,\rho} w\ \text{ and }\ h''\in\den{P}_{\eta,\rho} w*\UNFOLD(w)(\wemp) *r . 
\end{align}
We must show that $\den{\UNQUOTE{[e]}}_\eta h\in \DCl(\den{Q}_{\eta,\rho} w*\UNFOLD(w)(\wemp) *r)$. 
Recall that $e\pointsto R[\_]$ abbreviates $\exists k. e\pointsto k \wedge R[k]$ for fresh $k$. 
By \eqref{eqn:eval-1} we have for all $n\geq 0$ such that $\pi_n(h')\neq\bot$:
\begin{gather}
\label{eqn:eval-2}
\den{e}_\eta\in\dom{\pi_n(h')} = \dom{h'} \subseteq\dom{h}\\
\label{eqn:eval-3}
\exists d_n.\  \pi_n(h)(\den{e}_\eta) = \pi_n(h')(\den{e}_\eta) \sqsubseteq d_n\ \text{ and }\ \pi_n(h')\in\den{R[k]}_{\eta[k:=d_n],\rho} w
\end{gather}
Let us denote $\eta[k:=d_n]$ by $\eta_n$. The assumption that $R[k] \Rightarrow \triple{P*e\,{\pointsto}\,R[\_]}{k}{Q}$ is  valid yields:
\begin{align*}
\pi_n(h')\in \den{R[k]}_{\eta_n,\rho}w\ \text{ implies }\ \pi_n(h')\in\den{\triple{P*e\,{\pointsto}\,R[\_]}{k}{Q}}_{\eta_n,\rho}w
\end{align*}
Therefore, by \eqref{eqn:eval-3}, $\pi_n(h')\in\den{\triple{P*e\,{\pointsto}\,R[\_]}{k}{Q}}_{\eta_n,\rho}w$ holds for all $n$ sufficiently large. 
Let $r_n$ be the rank of $\pi_n(h')$. Since $\pi_n(h')\neq\bot$ we have $r_n>0$. 
It follows that 
\begin{align*}
\forall n.\ w\models\triple{\den{P*e\,{\pointsto}\,R[\_]}_{\eta_n,\rho}}{\pi_{r_n-1};d_n;\pi_{r_n-1}}{\den{Q}_{\eta_n,\rho}}\ .
\end{align*}
Since  
\begin{align}
\label{eqn:eval-key-equation}
\pi_n(h')(\den{e}_\eta)= \pi_{r_n}(\pi_n(h'))(\den{e}_\eta) \sqsubseteq \pi_{r_n-1};d_n;\pi_{r_n-1}\ , 
\end{align}
the downward closure  of semantic triples  in the command argument gives
\begin{align*}
\forall n.\ w\models\triple{\den{P*e\,{\pointsto}\,R[\_]}_{\eta_n,\rho}}{\pi_n(h')(\den{e}_\eta)}{\den{Q}_{\eta_n,\rho}}\ .
\end{align*}
Since $k$ was chosen fresh,   by the admissibility of semantic triples  we thus obtain 
\begin{align}
\label{eqn:eval-4}
\forall n.\ w\models\triple{\den{P*e\,{\pointsto}\,R[\_]}_{\eta,\rho}}{h'(\den{e}_\eta)}{\den{Q}_{\eta,\rho}}\ .
\end{align}
In particular, \eqref{eqn:eval-4} entails that $h(\den{e}_\eta) = h'(\den{e}_\eta)\in\Com$, and thus  $\den{\UNQUOTE{[e]}}_\eta h = h'(\den{e}_\eta)(h)$. 
Since we assumed that 
$h\in\den{P*e\pointsto R[\_]}_{\eta,\rho} w *\UNFOLD(w)(\wemp) *r$, 
we can conclude  
$\den{\UNQUOTE{[e]}}_\eta h\in\DCl(\den{Q}_{\eta,\rho} w *\UNFOLD(w)(\wemp) *r)$ 
by \eqref{eqn:eval-4}. \qed

The soundness of the standard Hoare logic rules is   straightforward. We illustrate this for the sequencing rule next. 

\begin{lemma}[Sequencing] 
Provided $\triple{P}{\QUOTE {C}}{R}$ and $\triple{R}{\QUOTE{D}}{Q}$ are valid, then so is $\triple{P}{\QUOTE{C;D}}{Q}$.  
\end{lemma}

\proof
Let $\eta\in\Env$, $w\in\W$, let $\rho$ be an assertion environment, and let  $r\in\UAdm$. 
Let  $h\in\den{P}_{\eta,\rho}(w)*\UNFOLD(w)(\wemp)*r$. 
We must show that $\den{C;D}_\eta h\in\DCl(\den{Q}_{\eta,\rho}(w)*\UNFOLD(w)(\wemp)*r)$. 
First note that $\den{C}_\eta h\in\DCl(\den{R}_{\eta,\rho}(w)*\UNFOLD(w)(\wemp)*r)$, by the assumption that $\triple{P}{\QUOTE {C}}{R}$ is valid. In particular, $\den{C}_\eta h\neq\ERR$. Moreover, in the case where $\den{C}_\eta h = \bot$ we also have 
$\den{C;D}_\eta h = \bot$ by the semantics of sequential composition, so that the admissibility of $\DCl(\den{Q}_{\eta,\rho}(w)*\UNFOLD(w)(\wemp)*r)$ gives the result. 

Thus, we can assume that $\den{C;D}_\eta h = \den{D}_\eta(\den{C}_\eta h)$. 
From the assumption that $\triple{R}{\QUOTE{D}}{Q}$ is valid it follows that $\den{D}_\eta$ maps the set $(\den{R}_{\eta,\rho}(w)*\UNFOLD(w)(\wemp)*r)$ into $\DCl(\den{Q}_{\eta,\rho}(w)\mathop*\UNFOLD(w)(\wemp)\mathop*r)$. 
Since $\den{C}_\eta h\in \DCl(\den{R}_{\eta,\rho}(w)\mathop*\UNFOLD(w)(\wemp)\mathop*r)$ we   obtain 
$\den{D}_\eta(\den{C}_\eta h)\,{\in}\, \DCl(\den{Q}_{\eta,\rho}(w)\mathop{*}\UNFOLD(w)(\wemp)\mathop{*}r)$ by Lemma~\ref{lem:precondition-closure} and   continuity of $\den{D}_\eta$. \qed

The proofs for the remaining rules from Figure~\ref{fig:SLRules} are similar, and given in  Appendix  \ref{app:subsec:HoareLogicRules}. An exception is the rule of consequence: 
The soundness
proof of rule~(\textsc{Conseq}) is slightly
different from those of the others because (\textsc{Conseq})  involves
 an implication between triples, whereas the other rules are inference rules
for transforming valid Hoare triples. 
Due to the pointwise interpretation of implication and the inclusion of the approximations 
in the interpretation of triples, this form of the consequence rule 
could be potentially problematic. Our proof of (\textsc{Conseq}) overcomes this potential problem, 
by exploiting the fact that
the rule is ``parametric'' in the command, i.e.,  it is the same command that appears in 
all the triples of the rule. {Two further cases that are similar in this respect are the axioms (\textsc{ExistAux}) for the elimination of auxiliary variables and (\textsc{Disj}); see Appendix  \ref{app:subsec:HoareLogicRules}.}

\begin{lemma}[Consequence]
If $P'\Rightarrow P$ and $Q\Rightarrow Q'$ are valid implications, 
then so is $\triple{P}{e}{Q}\Rightarrow\triple{P'}{e}{Q'}$. 
\end{lemma}

\proof
Let $\eta\in\Env$, $\rho$ an assertion environment,  and fix $w\in\W$ and $n\geq 0$. 
Let $p = \den{P}_{\eta,\rho}$, $p' = \den{P'}_{\eta,\rho}$, $q = \den{Q}_{\eta,\rho}$ and $q' = \den{Q'}_{\eta,\rho}$, and assume that $\pi_n(h)\in\den{\triple{P}{e}{Q}}_\eta w$.
We must prove that $\pi_n(h)\in\den{\triple{P'}{e}{Q'}}_\eta w$. 
 
Let $k$ denote the rank of $\pi_n(h)$. Without loss of generality, we can assume $k>0$. Let
$c$ denote the command $\pi_{k-1};\den{e}_\eta;\pi_{k-1}$. 
Then the assumption yields 
$w\models \triple{p}{c}{q}$, 
and it suffices to establish  $w\models \triple{p'}{c}{q'}$.
For this, suppose that $r\in\UAdm$ and let $h'\in p'(w) * \UNFOLD(w)(\wemp) * r$. 
We must show that $c(h')\in \DCl(q'(w) * \UNFOLD(w)(\wemp) * r)$. 
By the assumption that $P'\Rightarrow P$ is valid, we also have $h'\in p(w) * \UNFOLD(w)(\wemp) * r$ by the monotonicity of $*$. 
By assumption, $c(h')\in \DCl(q(w) * \UNFOLD(w)(\wemp) * r)$. 
By the assumption that $Q\Rightarrow Q'$ is valid, and using monotonicity of $*$ and $\DCl(\cdot)$, we obtain 
$c(h')\in \DCl(q'(w) * \UNFOLD(w)(\wemp) * r)$
as required.  \qed

\section{Proof Rules involving different Nesting Levels}
\label{sec:discussion-of-proof-rules}

The soundness of rule~(\textsc{Eval}) as shown in Lemma~\ref{lem:Eval}
involves an assertion $R$ that is used at different nesting levels in its hypothesis and conclusion. 
In this section we discuss two further proof rules that relate nested triples to top-level implications in a similar way:
\begin{align*}
\inferrule[Out-T]{ 
\X;\Gamma \adash \triple{ \triple{A}{d}{B} \wedge P}{e}{Q}}
{ \X;\Gamma \adash \triple{A}{d}{B} \Rightarrow \triple{ P} {e} {Q}
}
\qquad 
 \inferrule[In-T]{ 
 \X;\Gamma \adash \triple{A}{d}{B} \Rightarrow \triple{P} {e} {Q}
}
{\X;\Gamma \adash \triple{ \triple{A}{d}{B} \wedge P}{e}{Q}}
 \end{align*}
While, at first glance, both rules may seem reasonable, we will show below that in our model rule~(\textsc{Out-T}) is valid
but   rule~(\textsc{In-T}) is not. 
We begin by making some observations regarding the semantics of nested triples.

\begin{lemma} \label{lem:triple-sem-observation}
 For any $w\in W, p,q\in \Pred$ and $c\in \Com$ we have that $w\models_k \triple{p}{c}{q}$ if and only if for all $r\in \UAdm$, $n\leq k$ and all $h\in\Heap$:
\begin{align*}
\pi_n(h)\in p(w) * \UNFOLD(w)(\wemp) *r\ \IMPLIES\ 
\pi_n(c(\pi_n(h)))\in \DCl(q(w) * \UNFOLD(w)(\wemp) * r). 
\end{align*}
\end{lemma}
\proof
For the direction from left to right, 
let $n\leq k$. 
Using the assumption and $ \pi_n(h)\in p(w) * \UNFOLD(w)(\wemp) *r$ 
we obtain 
$\pi_k(c(\pi_k(\pi_n(h))))\in \DCl(q(w) * \UNFOLD(w)(\wemp) * r)$,
 thus $\pi_k(c(\pi_n(h)))\in \DCl(q(w) * \UNFOLD(w)(\wemp) * r)$ 
 and by downward-closure also $\pi_n(c(\pi_n(h)))\in \DCl(q(w) * \UNFOLD(w)(\wemp) * r)$.

For the direction from right to left, let $h\in p(w) * \UNFOLD(w)(\wemp) *r$. 
By uniformity we know that for all $n\in \omega$ 
also  $\pi_n(h)\in p(w) * \UNFOLD(w)(\wemp) *r$. 
We thus know by assumption that $\pi_n(c(\pi_n(h)))\in \DCl(q(w) * \UNFOLD(w)(\wemp) * r)$
 for $n\leq k$ and thus in particular for $n=k$. \qed


\begin{definition}
A predicate $p\in \Pred$ is \emph{\ALMOSTPURE} if
for all  $h,h' \in \Heap$ such that $\rnk{h} = \rnk{h'}$ and all $w\in W$ we have that $h \in p(w)$ if, and only if, $h'\in p(w)$. 
An assertion is \ALMOSTPURE if its denotation is a  \ALMOSTPURE predicate.
\end{definition}

In the following, $\phi$ will always stand for an assertion that is \ALMOSTPURE. Note that the typical examples
for \ALMOSTPURE assertions are  triples. Obviously, every pure (i.e., entirely heap-independent) assertion
is trivially also \ALMOSTPURE. 
Assertions that depend on the shape and content of the heap itself, e.g.\ $x\mapsto \_$, are not \ALMOSTPURE.
We also observe that the interpretation of a \ALMOSTPURE assertion is 
  downward closed in the rank itself:
 
 \begin{lemma} \label{lem:trank-downward-closure}
 For any \ALMOSTPURE assertion $p$, and any heaps $h$ and $h'$, if $\rnk{h'} \leq  \rnk{h}$ then $h\in p(w)$ implies $h'  \in p(w)$.
 \end{lemma}
\proof  
Suppose $h\in p(w)$, and let $n = \rnk{h'}\leq \rnk h$. 
Thus we have $\rnk{\pi_n(h)} = n$. 
Since $\pi_n(h)\in p(w)$ by uniformity, we can conclude $h'\in p(w)$ from the assumption that $p$ is \ALMOSTPURE. \qed

With the definition of \ALMOSTPURE in place, we can now generalise the rules (\textsc{Out-T}) and (\textsc{In-T}) in the following way:
\begin{align*}
\inferrule[Out]{ 
\X;\Gamma \adash \triple{ \phi \wedge P}{e}{Q}}
{ \X;\Gamma \adash \phi \Rightarrow \triple{ P} {e} {Q}
} \mbox{($\phi$ \ALMOSTPURE)}
\qquad 
 \inferrule[In]{ 
 \X;\Gamma \adash \phi \Rightarrow \triple{P} {e} {Q}
}
{\X;\Gamma \adash \triple{ \phi \wedge P}{e}{Q}}
\mbox{($\phi$ \ALMOSTPURE)}
 \end{align*}
\begin{prop} 
The above rule (\textsc{Out})  is sound.
\end{prop}
\proof
Assume   environments $\eta$ and $\rho$, let $n\in \omega, w\in W$ and $h\in\Heap$ be 
such that 
\begin{equation}\label{eq:assumption_out_rule}
\pi_n(h)\in \den{\phi}_{\eta,\rho}\, w
\end{equation}
We have to show that $\pi_n(h)\in \den{\triple{P}{e}{Q}}_{\eta,\rho} \, w$. 
Let  $k$ denote  the rank of $\pi_n(h)$. If $k= 0$ we are done. Otherwise we have to show that    $w\models_{k-1} \triple{\den{P}_{\eta,\rho}}{\den{e}_\eta}{\den{Q}_{\eta,\rho}}$. 
But  by the observation in Lemma~\ref{lem:triple-sem-observation}, it suffices to show for any heap $h'$ and any $l\leq k-1$ that if $\pi_l(h') \in \den{P}_{\eta,\rho}(w) * \UNFOLD(w)(\wemp) *r$ then $\pi_{l}(c(\pi_{l}(h')))\in \DCl(\den{Q}_{\eta,\rho}(w) * \UNFOLD(w)(\wemp) * r)$.  From the interpretation of the premise of the rule using $\eta, w$ and $l$ we get the desired result if we can show that
$\pi_l(h') \in \den{P}_{\eta,\rho}(w) * \UNFOLD(w)(\wemp) *r$ implies $\pi_l(h') \in \den{\phi \wedge P}_{\eta,\rho}(w) * \UNFOLD(w)(\wemp) *r$.
Yet, $\pi_l(h') \in \den{\phi}_{\eta,\rho}(w)$ follows from Lemma~\ref{lem:trank-downward-closure}
 due to assumption \eqref{eq:assumption_out_rule}, the fact that $\phi$ is \ALMOSTPURE, and
the fact that $\rnk{\pi_l(h')} \leq l< k =\rnk{\pi_n(h)}$. \qed

\begin{prop}
\label{prop:InUnsoundness}
The rule (\textsc{In}) does not hold in our model. 
\end{prop}
\proof
Assuming that (\textsc{In}) holds in our semantics we can derive an invalid triple as follows.  
Let $R$ abbreviate the recursive assertion $\mu X.\triple{X}{\QUOTE{\SYN{skip}}}{\False}$. Then, from the tautology $R\Rightarrow R$ we obtain 
$R\Rightarrow\triple{R}{\QUOTE{\SYN{skip}}}{\False}$ by unfolding the recursive definition of $R$. 
Applying (\textsc{In}) and the consequence rule thus gives
\begin{align}
\label{eqn:InUnsound}
\adash \triple{ R}{\QUOTE{\SYN{skip}}}{\False}
\end{align}
Our model validates the implication $\EMP \Rightarrow R$: 
By definition of implication, it suffices to prove that $\rnk{\record{}}=n$ implies $w\models_{n{-}1}\triple{\den{R}}{\den{\SYN{skip}}}{\den{\False}}$. 
Since the empty heap has rank 1, this implication holds trivially for \emph{any} triple on the right hand side, 
in particular $R$.
From \eqref{eqn:InUnsound} and this implication we conclude that the triple
$\triple{ \EMP}{\QUOTE{\SYN{skip}}}{\False}$ holds, which is clearly not the case by definition of the semantics of triples. 
We conclude that rule (\textsc{In}) cannot hold with respect to our semantics. \qed

It is worth looking more closely at the reason why rule (\textsc{In}) does not hold in our semantics. 
Essentially, to show the triple in the conclusion at level $k$, one needs to show that in the hypothesis the formula $\phi$ holds for a heap with rank $k+1$. But this property cannot be established in general from the assumptions of the triple in the conclusion at level $k$.\footnote{Rule \textsc{(In)} does hold in the special case when $\phi$ is pure. } 
Note that, in the case of (\textsc{Eval}), the corresponding property can be established since the heap access of the \SYN{eval} command offsets the increase in the rank   (cf.\ equation~\eqref{eqn:eval-key-equation} in the proof of Lemma~\ref{lem:Eval}). 

In the case where the command is arbitrary (i.e., not \SYN{eval}), one can express the   upwards shift of levels   explicitly with the help of a modal operator $\Diamond P$ (``previous $P$,'' or ``$P$ one level up''). This operator is defined by  
$h\in \den{\Diamond P}_{\eta,\rho}\, w$
if and only if 
\begin{iteMize}{$\bullet$}
\item $\rnk{h} = \infty$ and $h\in \den{P} _{\eta,\rho}\, w$ or
\item $\rnk{h} = k < \infty$ and there exists $h'  \in  \den{P} _{\eta,\rho}\, w$ such that
  $\rnk{h'} = k+1$ and
    $\pi_k(h') = h$, 
\end{iteMize} 
     and thus $\Diamond P$
    denotes a downward closed, admissible predicate.
With the help of the modality, we can give variants of the above rules that keep track of the rank information:
\[ 
\inferrule[$\Diamond$Out]{ 
\X;\Gamma \adash \triple{ \phi \wedge P}{e}{Q}}
{ \X;\Gamma \adash \phi \Rightarrow \Diamond \triple{ P} {e} {Q}
}
\qquad 
\inferrule[$\Diamond$In]{ 
 \X;\Gamma \adash \phi \Rightarrow \Diamond \triple{P} {e} {Q}
}
{\X;\Gamma \adash \triple{ \phi \wedge P}{e}{Q}}
\qquad
\inferrule[$\Diamond$E]{\ignore{x} }
{ \X;\Gamma \adash \Diamond P \Rightarrow P}
\]
In our semantics,  which still satisfies (\textsc{$\Diamond$Out}) and  (\textsc{$\Diamond$E}), even this strengthened variant (\textsc{$\Diamond$In}) does not hold. This is due to the following simple observation, which means that ranks are not preserved by the separating conjunction that is used in the interpretation of triples.

\begin{lemma} Given a heap $h = h_1 \cdot h_2$ with rank $n$,  such that $\rnk{h_1} = n_1$ and $\rnk{h_2} = n_2$ it may well be the case that $n_1 < n$ or $n_2 < n$.
\end{lemma}

However, the modal rules can be proved sound with the help of a step-indexed model. In such a model, the ranks are replaced by an explicit natural number index that gives a lower bound on the number of steps that can be safely taken  in an operational semantics without invalidating a given assertion. The slightly unintuitive  implications $\EMP \Rightarrow \triple{P}{e}{Q}$ will  not hold in the step-indexed model either. However, also the step-indexed model does not validate (\textsc{In}), and we conjecture that this rule renders the logic inconsistent.
More details about step-indexed models can be found in \cite{BirkedalRSSTY11}.

 It is   worth pointing out that not only   unintuitive  implications $\EMP \Rightarrow \triple{P}{e}{Q}$ do hold in our model, but also that 
 the so-called invariance rule\footnote{This should not be confused with the stronger conjunction rule which is known to be inconsistent with higher-order frame rules \cite{OHearn:Yang:Reynolds:04}).}
 
 \[ 
\inferrule[Invariance]{ 
\X;\Gamma \adash \triple{P}{e}{Q}}
{ \X;\Gamma \adash \triple{P\wedge \triple{A}{k}{B}}{e}{Q\wedge \triple{A}{k}{B}}
}
\]

\noindent
does not hold. It is only valid  for invariants that are  pure, so it does  not hold for $\triple{A}{k}{B}$ nor any other \ALMOSTPURE invariant.
This can be easily seen by considering the triple
 $$\triple{\EMP}{\QUOTE{\SYN{let}\, x = \SYN{new}\, 0 \ \SYN{in}\ [x]\SYN{:=} \QUOTE{\SYN{skip}}\ }}{\exists x.\, x{\mapsto}\triple{\EMP}{\_}{\EMP}}$$
  with invariant  $\triple{\EMP}{\SYN{skip}}{\False}$, since the latter only holds for  heaps with rank $1$, for instance  the empty heap.
  Unfortunately, not even the following restricted form of invariance  holds:
 
  \[ 
\inferrule[InvarianceR]{ 
\X;\Gamma \adash \triple{P*e_1\mapsto e_2}{e}{Q * e_1\mapsto e_2}}
{ \X;\Gamma \adash \triple{P* (e_1\mapsto e_2 \wedge \phi)}{e}{Q * (e_1\mapsto e_2 \wedge \phi)} }(\mbox{$\phi$ \ALMOSTPURE})
\]

\noindent
since the semantics of triples and of $\mapsto$ does not guarantee that the data stored at $\den{e_1}$, and thus the rank of any heap fulfilling $\den{e_1\mapsto e_2 \wedge \phi}$, is invariant. It could still be the case that the result heap meeting the postcondition 
$\den{e_1\mapsto e_2 \wedge \phi}$ has a higher rank than the pre-execution heap  meeting the  same condition. 
The only way to guarantee that   invariance involving triples or other \ALMOSTPURE assertions holds is to ensure that the rank (or even the content) of the heap cells in question does not change during execution. 
Because of this issue we needed another update rule for programs that copy code:
\[ \inferrule[UpdateInv]{ 
}{
  \X;\Gamma\,{\adash}\,\triple{e\,{\mapsto}\,\_ \,{*}\, (e_1{\mapsto}e_0 \wedge \phi)}{\QUOTE{[e] \,{:=}\, e_0}}{(e\,{\mapsto}\,e_0\wedge  \phi) \,{*}\, (e_1{\mapsto}e_0 \wedge \phi) }
}(\mbox{$\phi$ \ALMOSTPURE})
\] 
%
%
\noindent
Note that $\phi$ may contain the expression $e_0$ (which can also be a variable) and will typically be a triple $\triple{A}{k}{B}$. The soundness of this rule follows from the soundness of  the assignment rule (\textsc{Update}) and the fact that the rank of the heap with domain $\den{e_1}$ is not changed by the command. Consequently,  the heap with domain $\den{e}$, which is identical to the one with domain $\den{e_1}$ after execution, satisfies $\phi$. But this axiom is not derivable from (\textsc{Update}) as, for a \ALMOSTPURE $\phi$, the axiom
\[
{e\,{\mapsto}\,e_0 \,{*}\, (e_1{\mapsto}e_0 \wedge \phi)} \Rightarrow
{(e\,{\mapsto}\,e_0 \wedge \phi) \,{*}\, (e_1{\mapsto}e_0 \wedge \phi)}  \]
does not hold for the same reasons as (\textsc{InvarianceR}) does not hold.
%


 \section{Conclusion}

 In this article we have investigated a separation logic for a simple programming language with higher-order store. As our counterexamples illustrate, the design of such a logic is not straightforward: 
 \begin{iteMize}{$\bullet$}
 \item  In the presence of recursive assertions, unrestricted use of a deep frame axiom permits the ``laundering'' of code, which allows for the derivation of insufficient memory footprints (Proposition~\ref{prop:DeepFrameUnsound}).
 \item Higher-order frame rules are inconsistent with a classical
 specification logic (and hence in our case, due to the identification of assertion and specification language, with a classical assertion logic; Proposition~\ref{prop:classicalUnsound}).
 \item In the presence of recursive assertions, one cannot move global  assumptions of triples, expressed as implications, into pre-conditions
 (Proposition~\ref{prop:InUnsoundness}).
 \end{iteMize}
Note that the first two points are  independent  of any choice of model whereas this is not clear for the third point.

In our model, we use recursively defined Kripke worlds to interpret the invariant extension $P\otimes R$. 
In a logic without recursive assertions (and assertion variables), like the one considered by Birkedal et al.\ for Idealized Algol \cite{BirkedalL:semslt-lmcs}, the invariant extension operation can be considered essentially as a syntactic abbreviation. In particular, it need not be treated as a primitive operation and recursive worlds are not needed. 
In a logic with second-order quantification, frame conditions can be made explicit in a specification, which gives rise to a modular proof pattern without explicit deep frame rule; this idea is discussed and used in, e.g., \cite{Benton:06,CharltonReusLola10}. 

Recursive worlds similar to the ones employed here can be used to construct a model for Pottier's anti-frame rule, a proof rule for hiding local state from the context \cite{Pottier:08}. 
In that case, predicates must depend on the worlds in a {monotonic} way (with respect to an order on worlds defined from the composition operation $\circ$), which complicates the model construction considerably \cite{schwinghammerYBPR10,Schwinghammer:Birkedal:Stovring:11}.

   During the process of writing this article,  it has been discovered that one can also build a model for the presented logic, including deep frame rules and recursive assertions,  with the help of step-indexing \cite{Appel:McAllester:01} based on an operational semantics for the programming language. We have already  pointed out differences regarding both models  throughout the paper but here is a short summary. The domain model in our work uses 
   ranks of heaps in order to equip semantic assertions with an ultrametric.  Whereas  steps are counted separately in the step-indexed approach,  heaps, and thus their ranks, are manipulated by programs. This leads to  some contamination of the assertion semantics
   that the step-index model does not share. First of all, we do not get a BI algebra, more 
   precisely we do not get spatial implication. Secondly, triples are not pure but \ALMOSTPURE. This, in turn, means that the invariance rule for triples is not valid and holds only for programs that do not change the rank of the heap in question (as expressed in (\textsc{UpdateInv})). Moreover,
    some unwanted implications between triples are validated. The (\textsc{In-T}) rule  does not hold in either of the two models but it holds in \cite{Honda:Yoshida:Berger:05}. The (\textsc{$\Diamond$In}) rule, on the other hand, does hold in the step-indexed model but not the presented one. Despite the complications caused by the ranks of heaps, the  denotational model has some upsides as well. 
From earlier work one knows that it represents a way to combine some equational reasoning  with Hoare style logics. Equational reasoning has been used to some extent to prove properties of the model, in particular the soundness of the presented rules.
It remains to be seen whether the denotational models have more advantages over the operational step-indexed ones regarding binary relations, e.g.\ in order to prove parametricity results.

A detailed description of the step-indexed model and its applications will appear elsewhere in due course.

The work of Honda et al.~\cite{Honda:Yoshida:Berger:05} also presents a logic for higher-order functions and general references, including even observational completeness, i.e.\ two programs are equal if they fulfill the same triples. The main differences with respect to  the logic presented here are as follows. In  \cite{Honda:Yoshida:Berger:07} a logic for total correctness is given. Therefore, there is no  need for a specific rule handling recursion through the store, since procedures are always proved sound using induction  on a termination measure that the verifier needs to guess. Moreover,
 local reasoning is ignored so there are no frame rules. The follow-up work \cite{Honda:Yoshida:Berger:07} addressed this issue, but using content quantification and not separation logic. There does not appear to be an implementation of the logic of  \cite{Honda:Yoshida:Berger:07} either.

A variant of our logic, for a language with recursive procedures and the possibility of partial application, has been implemented in the Crowfoot tool \cite{Crowfoot11}.  This verification tool is mainly targeted to prove memory safety for programs with stored procedures automatically. In its current state it does not yet cover a full-fledged first-order logic. Some example specifications for nested triples and recursive assertions can be found e.g.\ in \cite{Charlton:Reus:11}.

 \section*{Acknowledgment}
We would like to thank 
Nathaniel ``Billiejoe'' Charlton, Fran\c{c}ois Pottier, Kristian St{\o}vring and Jacob Thamsborg for helpful discussions. Kristian suggested that $\otimes$ is a monoid action. We are grateful for the suggestions of the anonymous referees to improve the paper.
Partial support has been provided by FNU project 272-07-0305
 ``Modular reasoning about software'' (Birkedal), EPSRC  projects  EP/G003173/1 
``From reasoning principles for function pointers to logics for self-configuring programs'' (Reus),
EP/E053041/1 ``Scalable program analysis for software verification'' 
and EP/H008373/1 ``Resource reasoning'' (Yang).

\bibliographystyle{abbrv}
\bibliography{nested-long}

\newpage
\appendix
\input appendix-long.tex

\end{document}

%% file: appendix-long.tex
\section{Summary of Proof Rules}
Figure~\ref{fig:proofrulesummary} summarizes the proof rules that we have proved sound with respect to our model. 
Not shown are the standard proof rules for (intuitionistic) first-order logic (for instance, see \cite{Sorensen:Urzyczyn:06}) 
and the distribution axioms for $\otimes$ that appear in Figure~\ref{fig:DistRules}.

\begin{figure}
\hrule
\begin{align*}
&\textsc{$*$-Assoc}&&
\inferrule{}{\X;\Gamma\vdash P*(Q *R) \IFF (P*Q)*R}
\\[.5ex]
&\textsc{$*$-Comm}&&
\inferrule{}{\X;\Gamma\vdash P*Q\IFF Q*P}
\\[.5ex]
&\textsc{$*$-Unit}&&
\inferrule{}{\X;\Gamma\vdash P * \EMP\IFF P}
\\[.5ex]
&\textsc{$*$-Zero}&&
\inferrule{}{\X;\Gamma\vdash P*\False\IFF \False}
\\[.5ex]
&\textsc{$*$-Overlap}&&
\inferrule{}{\X;\Gamma\vdash (e\mathop{\mapsto} e_1 \,\mathop*\, e\mathop{\mapsto} e_2)\IFF \False}
\\[.5ex]
&\textsc{$*$-Mono}&&
\inferrule{\X;\Gamma\vdash P\IMPLIES P'\\ \X;\Gamma\vdash Q\IMPLIES Q' }{\X;\Gamma\vdash P * Q\IMPLIES P' * Q'}
\\[.5ex]
&\textsc{$\otimes$-Mono}&&
\inferrule{\X;\Gamma\vdash P\IMPLIES P' }{\X;\Gamma\vdash P \otimes R\IMPLIES P' \otimes R}
\\[2ex]
&\textsc{Deref}&&
\inferrule{
  \X;\Gamma,x\,{\adash}\,\triple{P\,{*}\,e\,{\mapsto}\,x}{\QUOTE{C}}{Q}
}{
  \X;\Gamma\,{\adash}\, \triple{\exists x.P\,{*}\,e\,{\mapsto}\,x}{\QUOTE{\SYN{let}\,{x{=}[e]}\,\SYN{in}\,{C}}}{Q}
}\ (x \not\in \fvar(e,Q))
\\[.5ex]
& \textsc{Update} &&
\inferrule{}{
  \X;\Gamma\,{\adash}\,\triple{e\,{\mapsto}\,\_ \,{*}\, P}{\QUOTE{[e] \, \SYN{:=}\, e_0}}{e\,{\mapsto}\,e_0 \,{*}\, P}
}
\\[.5ex]
& \textsc{UpdateInv} &&
{\mbox{($\phi$ \ALMOSTPURE)}}\\ &&&
\inferrule{}{\X;\Gamma\,{\adash}\,\triple{e\,{\mapsto}\,\_ \,{*}\, (e_1{\mapsto}e_0 \wedge \phi)}{\QUOTE{[e] \,{:=}\, e_0}}{(e\,{\mapsto}\,e_0\wedge  \phi) \,{*}\, (e_1{\mapsto}e_0 \wedge \phi)} 
} 
\\[.5ex]
&\textsc{New} &&
\inferrule{
  \X;\Gamma,x \,{\adash}\, \triple{P*x\,{\mapsto}\,e}{\QUOTE{C}}{Q}
}{
  \X;\Gamma \,{\adash}\, \triple{P}{\QUOTE{\SYN{let}\,{x {=} \SYN{new}\,e}\,\SYN{in}\,{C}}}{Q}
}\ (x \not\in \fvar(P,e,Q))
\\[.5ex]
&\textsc{Free} &&
\inferrule{}{
  \X;\Gamma\adash\triple{e\,{\mapsto}\,\_ * P}{\QUOTE{\SYN{free}(e)}}{P}
}
\\[.5ex]
&\textsc{If} &&
\inferrule{
  \X;\Gamma\adash \triple{P \,{\wedge}\, e_0{=}e_1}{\QUOTE{C}}{Q} \quad \X;\Gamma\adash \triple{P \,{\wedge}\, e_0{\not=}e_1}{\QUOTE{D}}{Q}
}{
  \X;\Gamma\adash \triple{P}{\QUOTE{\SYN{if}\;(e_0{=}e_1)\;\SYN{then}\;C\;\SYN{else}\;D}}{Q}
}
\\[.5ex]
&\textsc{Skip} &&
\inferrule{}{
  \X;\Gamma\adash\triple{P}{\QUOTE{\SYN{skip}}}{P}
}
\\[.5ex]
&\textsc{Seq} &&
\inferrule{
  \X;\Gamma\adash \triple{P}{\QUOTE{C}}{R} \quad \Gamma\adash \triple{R}{\QUOTE{D}}{Q}
}{
  \X;\Gamma\adash \triple{P}{\QUOTE{C;D}}{Q}
}
\\[.5ex]
&\textsc{Eval} &&
\inferrule{
  \X;\Gamma,k \adash R[k] \Rightarrow \triple{P\,{*}\,e\,{\pointsto}\,R[\_]}{k}{Q}
}{
  \X;\Gamma \adash \triple{P\,{*}\,e\,{\pointsto}\,R[\_]}{\QUOTE{\UNQUOTE{[e]}}}{Q}
}
\\[2ex]
&\textsc{Conseq} &&
\inferrule{
  \X;\Gamma \,\adash\, P'{\Rightarrow}\,P 
  \quad \X;\Gamma \,\adash\, Q\,{\Rightarrow}\,Q'
}{
  \X;\Gamma \,\adash\, {\triple{P}{e}{Q}} \Rightarrow {\triple{P'}{e}{Q'}}
}
\\[.5ex]
&\textsc{Disj}&& 
\inferrule{}{
  \X;\Gamma \;\adash\;
  (\triple{P}{e}{Q}\wedge{\triple{P'}{e}{Q'}} )\Rightarrow \triple{P\vee P'}{e}{Q\vee Q'}
}
\end{align*}
\hrule
\caption{\label{fig:proofrulesummary} Axioms and proof rules. Rule \textsc{$\otimes$-Mono} is in fact a derived rule.}
\end{figure}

\begin{figure}
\ContinuedFloat
\hrule
\begin{align*}
&\textsc{ExistAux}&&
\inferrule{}{
  \X;\Gamma \;\adash\; (\forall x.\triple{P}{e}{Q})\Rightarrow\triple{\exists x.P}{e}{\exists x.Q}
}\qquad(x \not\in \fvar(e))
\\[.5ex]
&\textsc{Invariance}&&
\inferrule{}{
 \X;\Gamma \;\adash\; \triple{P}{e}{Q}
\Rightarrow
\triple{P\wedge \psi}{e}{Q \wedge \psi}
  } \mbox{\quad ($\psi$ is pure)}
  \\[.5ex]
&\textsc{$\otimes$-Frame} &&
\inferrule{
  \X;\Gamma \adash P
}{
  \X;\Gamma  \adash P\,{\otimes}\, R
}
\\[.5ex]
&\textsc{$*$-Frame} &&
\inferrule{}{
  \X;\Gamma \adash \triple{P}{e}{Q} \Rightarrow \triple{P\,{*}\,R}{e}{Q\,{*}\,R}
}
\\[2ex]
&\textsc{RUnique}&&
\inferrule{
\X;\Gamma\adash R\Leftrightarrow P[X:=R]\\ 
\X;\Gamma\adash  S\Leftrightarrow P[X:=S]\\ 
  }
{ \X;\Gamma\adash R \Leftrightarrow S }\ \text{($P$ formally contr.\ in $X$)}
\end{align*}
\hrule
\caption{ Axioms and proof rules (cont.). }
\end{figure}

\section{Proofs}

This section contains the proofs omitted from the main part of the paper. 

\subsection{Heyting algebra structure of uniform admissible subsets}
\label{app:subsec:assertions}

\begin{lemma}[Heyting algebra]
\label{lem:UAdm-Heyting}
Let $I=\{\record{},\bot\}$. 
Then $(\UAdm,\subseteq)$ is a complete Heyting algebra
with a (monotone) commutative monoid structure $(\UAdm,*,I)$.
All the algebra operations are non-expansive with respect to the metric defined in Section~\ref{subsec:Assertion-Domain}. 
\end{lemma}

\begin{proof}
Since admissibility and uniformity are preserved by taking arbitrary intersections, $\UAdm$ is a complete lattice, with meets given by set-theoretic intersection, least element $\{\bot\}$ and greatest element $\Heap$. Binary joins are given by set-theoretic union, and arbitrary joins by $\bigsqcup_i p_i = \bigcap\{p \in \UAdm\;|\;p\supseteq\bigcup_i p_i\}$.

The join is described more explicitly  as $\bigsqcup_i p_i = \{h\;|\;\forall n\in\omega.\ \pi_n(h)\in\bigcup_i p_i\}$. 
First, note that the  right hand side $r\defeq   \{h\;|\;\forall n\in\omega.\ \pi_n(h)\in\bigcup_i p_i\}$ is an element of $\UAdm$: 
$r$ is uniform, i.e., $h\in r$ implies $\pi_m(h)\in r$ for all $m\in\omega$, since $\pi_n\cdot \pi_m = \pi_{\min\{n,m\}}$. To  show that $r$ is also admissible suppose $h_0\sqsubseteq h_1\sqsubseteq\ldots$ is a chain in $r$, and let $h$ be the lub of this chain. We must show that $\pi_n(h)\in \bigcup_i p_i$ for all $n\in\omega$. By compactness, $\pi_n(h)\sqsubseteq h_k\sqsubseteq h$ for some $k$, and hence $\pi_n(h) = \pi_n(h_k)\in\bigcup p_i$ using the idempotency of $\pi_n$ and the fact that $h_k\in r$. 
To see the inclusion $r\subseteq\bigsqcup_i p_i$, 
note that for all $h$, if $\pi_n(h)\in\bigcup_i p_i\subseteq p$ for all $n\in\omega$ and some arbitrary $p\in\UAdm$, 
then also $h = \sqcup_n \pi_n(h) \in p$ by admissibility, and hence $h\in\bigsqcup_i p_i$ follows. 
For the other inclusion, we claim that the right hand side $r\defeq   \{h\;|\;\forall n\in\omega.\ \pi_n(h)\in\bigcup_i p_i\}$ is one of the elements appearing in the intersection; from this claim it is immediate that $r\supseteq\bigsqcup_i p_i$.  
The claim follows since $r\supseteq\bigcup_i p_i$ by the uniformity of the $p_i$'s.  

The implication  of this complete lattice $\UAdm$ is described by  $p\Rightarrow q\defeq  \{h\;|\;\forall n\in\omega.\ \text{if $\pi_n(h)\in p$ then $\pi_n(h)\in q$} \}$: 
Using $\pi_n\cdot \pi_m = \pi_{\min\{n,m\}}$ it is easy to see that $p\Rightarrow q$ is uniform. Admissibility follows analogously to the case of joins: if $h_0\sqsubseteq h_1\sqsubseteq\ldots$ is a chain in $p\Rightarrow q$ with lub $h$, and if $n\in\omega$ is such that $\pi_n(h)\in p$ then we must show that $\pi_n(h)\in q$. Since $\pi_n(h)\sqsubseteq h$ is compact, there is some $k$ such that $\pi_n(h)\sqsubseteq h_k\sqsubseteq h$, and thus the required $\pi_n(h) = \pi_n(h_k)\in q$ follows from $h_k\in p\Rightarrow q$. 
Next, to see that $p\Rightarrow q$ is indeed the implication in $\UAdm$, first note that we have $p\cap(p\Rightarrow q)\subseteq q$, using the uniformity of $p$ and the admissibility of $q$. If $p\cap r\subseteq q$ for some $r\in\UAdm$, and $h\in r$ and $\pi_n(h)\in p$ for some $n\in\omega$, then the uniformity of $r$ yields $\pi_n(h)\in q$. Thus we obtain $p\cap r\subseteq q \IFF r\subseteq p\Rightarrow q$. 

That $*$ is an operation on $\UAdm$ is established in the proof of Lemma~\ref{lem:separation:non-expansiveness}. It is easy to check that $*$ is  commutative and associative and  that it is monotone, i.e., if $p\subseteq p'$ and $q\subseteq q'$ then $p*q\subseteq p'*q'$. Moreover, we have $I\in\UAdm$, and the fact that $p * I=p = I * p$ follows from the definition of the heap combination $h\COMB h'$. 
 
For the non-expansiveness of the algebra operations, we  only consider the case of meets  as an example. Assume $p\nequiv{n}p'$ and $q\nequiv{n}q'$, then whenever $h\in p\cap q$ we have $\pi_n(h)\in p'$ and $\pi_n(h)\in q'$ by assumption. Thus also $p\cap q\nequiv{n} p'\cap q'$.
\end{proof}

\begin{lemma}[Heyting   algebra, II]
\label{lem:UAdm-Heyting-II}
The set of non-expansive functions $W\to\UAdm$,  ordered pointwise, forms a complete Heyting algebra with a (monotone) commutative monoid structure. 
The operations are given by the pointwise extension of the corresponding ones on $\UAdm$, and  they are non-expansive with respect to the $\sup$-metric on $W\to\UAdm$.
\end{lemma}
\begin{proof}
We begin by showing that all the claimed algebra operations on $W\to\UAdm$ are well-defined, i.e., that the pointwise definitions give rise to non-expansive functions from $W$ to $\UAdm$. 
The cases of the various units are given by constant functions and thus non-expansive:
\begin{align*}
\top(w)&=\Heap 
&
\bot(w)&=\{\bot\}
&
I(w) &=\{\record{},\bot\}
\end{align*}
Next, consider the case of meets. Let $(p_i)_{i\in I}$ be a family of functions $p_i$ in $W\to\UAdm$ and $w,w'\in W$ such that $w\nequiv n w'$, we have
\[
(\bigsqcap_{i\in I}p_i)(w) = \bigcap_{i\in I} p_i(w)
\nequiv n \bigcap_{i\in I} p_i(w') = (\bigsqcap_{i\in I} p_i)(w')
\]
by the non-expansiveness of each $p_i$. Well-definedness for the other operations is shown analogously. 

We now show that the operations are non-expansive. Again, we consider the case of meets only, as the remaining cases are similar. 
Let $(p_i)_{i\in I}$ and $(q_i)_{i\in I}$ be two families of non-expansive functions such that $p_i\nequiv n q_i$ holds for all $i\in I$. 
To see that $\bigsqcap_i p_i\nequiv n \bigsqcap_i q_i$ holds,  by definition of the $\sup$-metric it suffices to prove 
$(\bigsqcap_i p_i)(w)\nequiv n (\bigsqcap_i q_i)(w)$ for all $w\in W$. 
This follows from the pointwise definition since $p_i(w)\nequiv n q_i(w)$ holds for every $i\in I$ by assumption.
\end{proof}

\subsection{Interpretation of assertions}
\label{subsec:app:interpretation-assertions}

\begin{lemma}[Non-expansiveness of fix, \cite{Birkedal:Stovring:Thamsborg:09}]
\label{lem:fix-non-expansiveness}
Let $(X,d)$ be an object in $\CBUlt$, and let $f,g : X\to X$ be contractive functions on $X$. 
Then $d(\textit{fix}\,f,\textit{fix}\,g)\leq \sup_{x\in X} d(f(x),g(x))$. 
\end{lemma}

\begin{lemma}[Well-definedness]
The interpretation in Fig.~\ref{fig:assertion-semantics} is well-defined. More precisely, let $P$ be an assertion with free relation variables in $\X = X_1,\ldots,X_k$, where the arity of $X_i$ is $n_i$. Then:
\begin{enumerate}
\item for every $\eta\in\Val^\Var$ and $\rho\in\prod_{X_i\in\X}\Pred^{(\Val^{n_i})}$, $\den{P}_{\eta,\rho}$ is an element of $\Pred$, i.e., a non-expansive function $W\to\UAdm$; 
\item $\den{P}_{\eta}$ denotes a non-expansive function from $\prod_{X_i\in\X}\Pred^{(\Val^{n_i})}$ to $\Pred$; 
\item   If $P$ is formally contractive in $X$ then the functional $\lambda q.\den{P}_{\eta,\rho[X:=q]}$  is a contractive map from $\Pred^{(\Val^n)}$ to $\Pred$, where $X$ is an $n$-ary relation variable. 
\end{enumerate}
\end{lemma}

\begin{proof}
The claims are proved simultaneously by induction on the structure of $P$. 
Note that the composition of non-expansive functions is again a non-expansive function, and that the composition of a contractive function with a non-expansive function is again a contractive function. 
\begin{iteMize}{$\bullet$}
\item For the logical connectives, the claims follow from the inductive hypothesis and Lemmas~\ref{lem:UAdm-Heyting} and \ref{lem:UAdm-Heyting-II} respectively.
 
\item The case of invariant extension, $P\otimes R$, follows from Lemma~\ref{lem:existence-tensor-and-circ}. In particular, $q\mapsto \den{P\otimes R}_{\eta,\rho[X:=q]}$ is  a contractive function whenever $P$ is formally contractive in $X$. 

\item The case of a relation variable, $X_i(\vec e)$, follows from the assumption that $\rho(X_i)$ is a non-expansive function from $\Val^{n_i}$ to $\Pred$. 

\item In the case of recursive assertions, $(\mu X(\vec x).P)(\vec e)$, the well-formedness requirement that $P$ be formally contractive in $X$ means that 
$\lambda q. \den{P}_{\eta ,\rho[X:=q]}$ is contractive, by part (3) of the induction hypothesis. Hence, $\lambda q,\vec d. \den{P}_{\eta[\vec x:=\vec d],\rho[X:=q]}$ is a contractive endofunction on $\Pred^{\Val^n}$. 
In particular,  the fixed point in the definition of 
$\den{(\mu X(\vec x).P)(\vec e)}$ 
 is well-defined, and by Lemma~\ref{lem:fix-non-expansiveness}, 
\begin{align*}
\den{(\mu X(\vec x).P)(\vec e)}_{\eta} = \lambda\rho.(\textit{fix} (\lambda q,\vec d.\den{P}_{\eta[\vec x:=\vec d],\rho[X:=q]}))(\den{\vec e}_\eta)\end{align*}
 is a non-expansive function. 

Similarly, if $P$ is formally contractive in $Y\neq X$, then 
$\lambda q . \den{(\mu X(\vec x).P)(\vec e)}_{\eta,\rho[Y:=q]}$ is contractive by Lemma~\ref{lem:fix-non-expansiveness} and the inductive hypothesis that 
$q\mapsto\den{P}_{\eta,\rho'[Y:=q]}$ is contractive for any $\rho'$. 

\item It remains to consider the case of (nested) triples. Note that the interpretation of triples is defined in terms of the admissible downward closure, so it is clear that $\den{\triple{P}{e}{Q}}_{\eta,\rho}w$ is uniform and admissible. 
We first prove claim (1), i.e., the non-expansiveness of $\den{\triple{P_1}{e}{Q_1}}_{\eta,\rho}$.  
To this end, assume that $w\nequiv n w'$, and let $h\in \den{\triple{P}{e}{Q}}_{\eta,\rho} w$. We must show that $\pi_n(h)\in \den{\triple{P}{e}{Q}}_\eta w'$. 
By the downward closure, we also know that $\pi_n(h)\in \den{\triple{P}{e}{Q}}_{\eta,\rho} w$. 
Since $k\defeq \rnk {\pi_n(h)}\leq n$, we also have $w\nequiv k w'$. Without loss of generality we can assume that $k>0$, and thus must have 
$w\models_{k-1}\triple{\den{P}_{\eta,\rho}}{\den{e}_\eta}{\den{Q}_{\eta,\rho}}$. 
By Lemma~\ref{lem:semantic-triples-key} this implies  
$w'\models_{k-1}\triple{\den{P}_{\eta,\rho}}{\den{e}_\eta}{\den{Q}_{\eta,\rho}}$, and thus also 
$\pi_n(h)\in\den{\triple PeQ}_{\eta,\rho}$.

We now prove the following claim which implies the non-expansiveness and contractiveness properties stated in conditions (2) and (3):
\begin{align*}
\rho\nequiv n\rho'\ \Rightarrow\ 
\den{\triple PeQ}_{\eta,\rho}\nequiv{n{+}1} \den{\triple PeQ}_{\eta,\rho'}
\end{align*}
For the proof of this claim, assume $\rho\nequiv n\rho'$ and $h\in\den{\triple PeQ}_{\eta,\rho}w$ for some $w$. 
We must show that $\pi_{n{+}1}(h)\in \den{\triple PeQ}_{\eta,\rho'}w$. 
Let $k\defeq\rnk{\pi_{n{+}1}(h)}\leq n+1$. Without loss of generality we can assume $k>0$ (and hence $k-1\leq n$), and thus obtain 
$w\models_{k-1}\triple{\den{P}_{\eta,\rho}}{\den{e}_\eta}{\den{Q}_{\eta,\rho}}$. 
By induction hypothesis, $\den{P}_\eta$ and $\den{Q}_\eta$ are non-expansive, and thus
$\den{P}_{\eta,\rho}\nequiv {k{-}1}\den{P}_{\eta,\rho'}$ and 
$\den{Q}_{\eta,\rho}\nequiv {k{-}1}\den{Q}_{\eta,\rho'}$. 
By Lemma~\ref{lem:semantic-triples-key} we obtain 
$w\models_{k-1}\triple{\den{P}_{\eta,\rho'}}{\den{e}_\eta}{\den{Q}_{\eta,\rho'}}$. 
This yields $\pi_{n{+}1}(h)\in \den{\triple PeQ}_{\eta,\rho'}w$.
\end{iteMize}
\end{proof}

\subsection{Soundness of standard rules from separation logic}
\label{app:subsec:HoareLogicRules}

The following lemmas show that the usual rules of separation logic, 
expressed using triples containing quoted commands as shown in Figure~\ref{fig:SLRules}, are sound. 

\begin{lemma}[Skip] 
The axiom $\triple{P}{\QUOTE{\SYN{skip}}}{P}$ is valid. 
\end{lemma}

\begin{proof}
This follows from the fact that $\den{\SYN{skip}}_\eta h = h$ for all $h\in\Heap$, and that $\DCl(\cdot)$ is a closure operation. 
\end{proof}

\begin{lemma}[Conditional]
If $\triple{P \,{\wedge}\, e_0{=}e_1}{\QUOTE{C}}{Q}$ 
and $\triple{P \,{\wedge}\, e_0{\not=}e_1}{\QUOTE{D}}{Q}$ are both valid, 
then so is $\triple{P}{\QUOTE{\SYN{if}\;(e_0{=}e_1)\;\SYN{then}\;C\;\SYN{else}\;D}}{Q}$. 
\end{lemma}

\begin{proof}
Let $w\in\W$ and $r\in\UAdm$ and suppose $h\in\den{P}_{\eta,\rho} w * \UNFOLD(w)(\wemp) *r$. From the semantics of 
the conditional, we can assume without loss of generality that $\den{e_0}_\eta$ and $\den{e_1}_\eta$ 
are not both in $\Com_\bot$. We must show that 
\begin{align*}
c(h)\in\DCl(\den{Q}_{\eta,\rho} w * \UNFOLD(w)(\wemp) *r) ,
\end{align*}
where $c(h) = \key{if}~(\EXPden{e_0}_\eta{=}\EXPden{e_1}_\eta)~\key{then}~\den{C}_\eta h~\key{else}~\den{D}_\eta h$.
Depending on whether the statement $\EXPden{e_0}_\eta{=}\EXPden{e_1}_\eta$ hold, we have $\den{e_0{=}e_1}_\eta w = \Heap$ or $\den{e_0 {\not=} e_1}_\eta w = \Heap$. Therefore, the claim follows from either the first or the second assumed triple. 
\end{proof}

\begin{lemma}[Update]
The axiom $\triple{e\,{\mapsto}\,\_ \,{*}\,P}{\QUOTE{[e] \, \SYN{:=}\, e_0}}{e\,{\mapsto}\,e_0\,{*}\,P}$ is valid. 
\end{lemma}

\begin{proof}
By Lemma~\ref{lem:star-frame}, it suffices to prove the validity of
\[
\triple{e\,{\mapsto}\,\_}{\QUOTE{[e] \,{:=}\, e_0}}{e\,{\mapsto}\,e_0}\ .
\]
Let $\eta\in\Env$, $\rho\in\Pred^\X$, $p = \den{e\,{\mapsto}\,\_}_{\eta,\rho}$, $q = \den{e\,{\mapsto}\,e_0}_{\eta,\rho}$ and $c = \den{[e] \,\SYN{:=}\, e_0}_\eta$. 
We will show that $w\models\triple{p}{c}{q}$ holds for all $w\in\W$. 

Let $w\in\W$ and $r\in\UAdm$, and suppose $h\in p(w) * \UNFOLD(w)(\wemp) * r$. 
We may assume that $h\neq\bot$, for otherwise $c(h) = \bot\in q(w) * \UNFOLD(w)(\wemp) * r$ is immediate. 
Thus, $h = h'\COMB h''$ such that $h'\in p(w)$ and $h''\in \UNFOLD(w)(\wemp) * r$. 
In particular, since $h'\in p(w) = \den{e\,{\mapsto}\,\_}_{\eta,\rho} w$, we obtain that $\den{e}_\eta\in\dom{h'}\subseteq\dom{h}$. 
Therefore, from the semantics of the assignment command,  
$c(h) = h[\den{e}_\eta\mapsto\den{e_0}_\eta]$. 
But this heap is the same as $\record{\den{e}_\eta = \den{e_1}_\eta}\COMB h''$, 
and therefore $c(h) \in q(w) * \UNFOLD(w)(\wemp) * r $. The latter set is contained in $\DCl(q(w) * \UNFOLD(w)(\wemp) * r)$ since $\DCl(\cdot)$ is a closure operation.
\end{proof}

\begin{lemma}[UpdateInv]
The axiom
\[ \inferrule[UpdateInv]{ 
}{
  \X;\Gamma\,{\adash}\,\triple{e\,{\mapsto}\,\_ \,{*}\, (e_1{\mapsto}e_0 \wedge \phi)}{\QUOTE{[e] \,\SYN{:=}\, e_0}}{(e\,{\mapsto}\,e_0\wedge  \phi) \,{*}\, (e_1{\mapsto}e_0 \wedge \phi) }
} \mbox{\quad ($\phi$ \ALMOSTPURE)}
\] 
 is valid. 
\end{lemma}

\begin{proof} 
Consider  $\eta\in\Env$, $\rho\in\Pred^\X$, $c = \den{[e] \,\SYN{:=}\,
  e_0}_\eta$, $p = \den{e\,{\mapsto}\,\_
  \,{*}\,e_1{\mapsto}e_0 \wedge \phi}_{\eta,\rho}$ and $q =
\den{(e\,{\mapsto}\,e_0\wedge \phi) \,{*}\, (e_1{\mapsto}e_0 \wedge
  \phi)}_{\eta,\rho}$.  We
will show that $w\models\triple{p}{c}{q}$ holds for all $w\in\W$.

Let $w\in\W$ and $r\in\UAdm$, and suppose $h\in p(w) * \UNFOLD(w)(\wemp) * r$. 
We may assume that $h\neq\bot$, for otherwise $c(h) = \bot\in q(w) * \UNFOLD(w)(\wemp) * r$ is immediate. 
Thus, $h = h'\COMB h''$ such that $h'\in p(w)$ and $h''\in \UNFOLD(w)(\wemp) * r$. 
In particular, since $h'\in p(w) = \den {e\,{\mapsto}\,\_ \,{*}\, (e_1{\mapsto}e_0 \wedge \phi)}_{\eta,\rho} w$, we obtain that
$h' = h_1\COMB h_2$ such that 
 $\{\den{e}_\eta \}= \dom{h_1}\subseteq\dom{h'}\subseteq\dom{h}$ and 
 $\{\den{e_1}_\eta\} =\dom{h_2}\subseteq\dom{h'}\subseteq\dom{h}$ and
 $h_2\in \den{\phi}_{\eta,\rho} w$.
Therefore, from the semantics of the assignment command,  
$c(h) = h[\den{e}_\eta\mapsto\den{e_0}_\eta]$. 
But this heap is the same as $(\record{\den{e}_\eta = \den{e_0}_\eta}\COMB \record{\den{e_1}_\eta = \den{e_0}_\eta})  \COMB h_2$.
Now the rank of heap $\record{\den{e}_\eta = \den{e_0}_\eta}$ is obviously identical to the rank of
$ \record{\den{e_1}_\eta = \den{e_0}_\eta}$ and thus $\record{\den{e}_\eta = \den{e_0}_\eta} \in \den{\phi}_{\eta,\rho}$ as $\phi$ is \ALMOSTPURE and $ \record{\den{e_1}_\eta = \den{e_0}_\eta} = h_2\in  \den{\phi}_{\eta,\rho} w$.
Therefore $c(h) \in q(w) * \UNFOLD(w)(\wemp) * r $. The latter set is contained in $\DCl(q(w) * \UNFOLD(w)(\wemp) * r)$ since $\DCl(\cdot)$ is a closure operation.
\end{proof}

\begin{lemma}[Free] 
The axiom $\triple{e\,{\mapsto}\,\_ * P}{\QUOTE{\SYN{free}(e)}}{P}$ is valid. 
\end{lemma}

\begin{proof}
By Lemma~\ref{lem:star-frame}, it suffices to prove the validity of
\[
\triple{e\,{\mapsto}\,\_}{\QUOTE{\SYN{free}(e)}}{\EMP}\ .
\]
Let $\eta\in\Env$, $\rho\in\Pred^\X$, $p = \den{e\,{\mapsto}\,\_}_{\eta,\rho}$, $q = \den{\EMP}_{\eta,\rho}$ and $c = \den{\SYN{free}(e)}_\eta$. 
We will prove that $w\models\triple{p}{c}{q}$ holds for all $w\in\W$. 

Let $w\in\W$, let $r\in\UAdm$ and suppose $h\in p(w) * \UNFOLD(w)(\wemp) * r$. 
Since $q(w)$ is the unit for $*$ and $\DCl(\cdot)$ is a closure operation, we must only show $c(h)\in\UNFOLD(w)(\wemp) * r$. 
We may assume that $h\neq\bot$, for otherwise $c(h) = \bot\in \UNFOLD(w)(\wemp) * r$ is immediate. 
Thus, $h = h'\COMB h''$ such that $h'\in p(w)$ and $h''\in \UNFOLD(w)(\wemp) * r$. 
In particular, since $h'\in p(w) = \den{e\,{\mapsto}\,\_}_{\eta,\rho} w$, we obtain that $\{\den{e}_\eta\} = \dom{h'}\subseteq\dom{h}$. 
Therefore, from the semantics of the deallocation command, 
$c(h) = h''$. It follows that $c(h)\in\UNFOLD(w)(\wemp) * r$. 
\end{proof}

\begin{lemma}[Deref] 
If $\triple{P\,{*}\,e\,{\mapsto}\,x}{\QUOTE{C}}{Q}$ is valid and $x$ is not free in $e$ and $Q$, 
then $\triple{\exists x.P\,{*}\,e\,{\mapsto}\,x}{\QUOTE{\SYN{let}\,{x{=}[e]}\,\SYN{in}\,{C}}}{Q}$ is also valid. 
\end{lemma}

\begin{proof}
Assume that $\triple{P\,{*}\,e\,{\mapsto}\,x}{\QUOTE{C}}{Q}$ is valid, and pick $\eta\in\Env$ and $\rho\in\Pred^\X$. 
Let $c=\den{\SYN{let}\,{x{=}[e]}\,\SYN{in}\,{C}}_\eta$. 
We will show that $w \models\triple{\den{\exists x.P\,{*}\,e\,{\mapsto}\,x}_{\eta,\rho}}{c}{\den{Q}_{\eta,\rho}}$ for all $w \in \W$.

Let $w\in\W$, $r\in\UAdm$ and $h\in\den{\exists x.P\,{*}\,e\,{\mapsto}\,x}_{\eta,\rho}(w) * \UNFOLD(w)(\wemp) * r$.  
We must show that $c(h)\in \DCl(\den{Q}_{\eta,\rho}(w) *\UNFOLD(w)(\wemp) * r)$. 
By definition there are heaps $h',h''$ such that $h = h'\COMB h''$ and 
$h'\in   \den{\exists x.P\,{*}\,e\,{\mapsto}\,x}_{\eta,\rho}(w)$ 
and 
$h''\in\UNFOLD(w)(\wemp) * r$.  
By definition this means that
\begin{align*}
\forall n.\ \exists d_n\in\Val.\ \pi_n(h')\in \den{P\,{*}\,e\,{\mapsto}\,x}_{\eta[x:=d_n],\rho}(w).
\end{align*}
Let us write $\eta_n$ for $\eta[x:=d_n]$. In the remainder of the proof,
we will prove that
\[
\forall n.\ c(\pi_n(h)) \in \DCl(\den{Q}_{\eta,\rho} * \UNFOLD(w)(\wemp) * r),
\]
because then, by admissibility and the continuity of $c$, we obtain 
the required
$c(h)\in \DCl(\den{Q}_{\eta,\rho} * \UNFOLD(w)(\wemp) * r)$.

Without loss of generality we can assume that $\pi_n(h) \not= \bot$, so that $\pi_n(h')\neq\bot$
as well.
Then, since $x\notin\fv{e}$, we have in particular $\den{e}_\eta\in\dom{\pi_n(h')}\subseteq\dom h$ and $\pi_n(h')(\den{e}_\eta)\sqsubseteq d_n$. 
Using the monotonicity of commands with respect to the environment, this gives 
\[
c(\pi_n(h))\ =\ \den{C}_{\eta[x := \pi_n(h')(\den{e}_\eta)]}(\pi_n(h))\ 
\sqsubseteq\ \den{C}_{\eta_n}(\pi_n(h))
\]
By uniformity of $\UNFOLD(w)(\wemp) * r$, we have 
$\pi_n(h)\in \den{P * e\pointsto x}_{\eta_n,\rho} * \UNFOLD(w)(\wemp) * r$, so that the assumption gives us 
\begin{align*}
c(\pi_n(h))\sqsubseteq \den{C}_{\eta_n}(\pi_n(h)) \in\DCl(\den{Q}_{\eta_n,\rho} * \UNFOLD(w)(\wemp) * r) .
\end{align*}
Since $\DCl(p')$ is a downward-closed set for every predicate $p'$, the above
formula implies that $c(\pi_n(h))$ belongs to the set on the right hand side.
Furthermore, since $x\notin\fv{Q}$, we have $\den{Q}_{\eta_n} = \den{Q}_\eta$. The combination
of these two facts
gives the desired $c(\pi_n(h)) \in \DCl(\den{Q}_{\eta,\rho} * \UNFOLD(w)(\wemp) * r)$.
\end{proof}

\begin{lemma}[New] 
If\ $\,\triple{P\,{*}\,x\,{\mapsto}\,e}{\QUOTE{C}}{Q}$ is valid and $x$ is not free in $P$, $Q$ and $e$, then 
$\triple{P}{\QUOTE{\SYN{let}\,{x {=} \SYN{new}\,e}\,\SYN{in}\,{C}}}{Q}$ is valid. 
\end{lemma}

\begin{proof}
Let $w\in\W$, $\eta\in\Env$, $\rho\in\Pred^\X$ and $r\in\UAdm$. Suppose $h\in\den{P}_{\eta,\rho}(w) * \UNFOLD(w)(\wemp) * r$.  
We must show that $c(h)\in\DCl(\den{Q}_{\eta,\rho}(w) * \UNFOLD(w)(\wemp) * r)$. 
Consider the following environment $\eta'$ and heap $h'$:
\begin{align*}
\eta' &\defeq \eta[x:=\l]
&
h' &\defeq h\COMB\record{\l = \den{e}_\eta}
\end{align*}
where $\l$ is the least natural number not contained in $\dom h$. 
Since $x$ is not free in $e$ and $P$, we have $\den{e}_\eta = \den{e}_{\eta'}$ and $\den{P}_\eta = \den{P}_{\eta'}$. Thus by the assumption on $h$ we obtain:
\begin{align*}
h'\in\den{P *x\,{\mapsto}\,e}_{\eta',\rho}w * \UNFOLD(w)(\wemp) * r .
\end{align*}
Then the 
assumption that $\triple{P*x\,{\mapsto}\,e}{\QUOTE{C}}{Q}$ is valid implies:
\begin{align*}
\den{C}_{\eta'}h'\in\DCl(\den{Q}_{\eta',\rho}(w) * \UNFOLD(w)(\wemp) * r) .
\end{align*}
Using the fact that 
$\den{\SYN{let}\,{x {=} \SYN{new}\,e}\,\SYN{in}\,{C}}_\eta(h) = \den{C}_{\eta'}h'$ and 
since $\den{Q}_{\eta'} = \den{Q}_\eta$, this proves the statement. 
\end{proof}

\begin{lemma}[Auxiliary variable]
Assume that $x$ is not free in $e$. Then the  axiom 
\[
\inferrule[ExistAux]{
}{
  \Gamma \;\adash\; (\forall x.\triple{P}{e}{Q})\Rightarrow\triple{\exists x.P}{e}{\exists x.Q}
}
\]
is valid. 
\end{lemma}

\begin{proof}
Let $\eta\in\Env$, $\rho\in\Pred^\X$, and fix $w\in\W$. 
For each $d\in\Val$, let $\eta_d = \eta[x{:=}d]$,  $p_d = \den{P}_{\eta_d,\rho}$ 
and $q_d = \den{Q}_{\eta_d,\rho}$.
Since $x$ is not free in $e$, we have 
$\den{e}_{\eta_d} = \den{e}_\eta$. Thus, a similar reasoning
with rank as that in the proof of Consequence implies that it is
sufficient to prove the following claim:
\begin{align*}
\textstyle{
\text{
for all $c$,
if $w\models\triple{p_d}{c}{q_d}$ for every $d$, then $w\models\triple{\bigsqcup_d p_d}{c}{\bigsqcup_d q_d}$} .
}
\end{align*}
Assume $w\models\triple{p_d}{c}{q_d}$, let $r\in\UAdm$ and $h\in(\bigsqcup_d p_d)(w) * \UNFOLD(w)(\wemp) * r$. 
We must show that $c(h)\in\DCl( (\bigsqcup_d q_d)(w) * \UNFOLD(w)(\wemp) * r)$. 
By definition, $h = h'\COMB h''$ where $h'\in (\bigsqcup_d q_d)(w)$ and $h''\in \UNFOLD(w)(\wemp) * r$. 
Thus, for each $n$ there exists $d\in\Val$ such that $\pi_n(h')\in p_d(w)$, and therefore 
$\pi_n(h)\in p_d(w) *   \UNFOLD(w)(\wemp) * r$ by the uniformity of $\UNFOLD(w)(\wemp) * r$. 
From the assumption $w\models\triple{p_d}{c}{q_d}$ we then obtain 
that for each $n$, 
\[
c(\pi_n(h)) \,\;\in\;\, \DCl( q_d(w) \,{*}\,   \UNFOLD(w)(\wemp) \,{*}\, r)
            \,\;\subseteq\;\, \DCl( (\textstyle{\bigsqcup_d q_d})(w) \,{*}\, \UNFOLD(w)(\wemp) \,{*}\, r). 
\] 
Using the admissibility of $\DCl((\bigsqcup_d q_d)(w) *   \UNFOLD(w)(\wemp) * r)$ and the continuity of $c$, it follows that $c(h)\in\DCl( (\bigsqcup_d q_d)(w) * \UNFOLD(w)(\wemp) * r)$. 
\end{proof}

\begin{lemma}[Invariance]
Then the  axiom 
\[\inferrule[Invariance]{}{
 \X;\Gamma \;\adash\; \triple{P}{e}{Q}
 \Rightarrow \triple{P\wedge \psi}{e}{Q \wedge \psi}
  } \mbox{\quad ($\psi$ is pure)}
\]is valid. 
\end{lemma}

\begin{proof} 
Let $\eta\in\Env$, $\rho\in\Pred^\X$, and fix $w\in\W$. 
For each $d\in\Val$,let $p= \den{P}_{\eta,\rho}$ 
and $q = \den{Q}_{\eta,\rho}$ and $f= \den{\psi}_{\eta,\rho}$.
A similar reasoning
with rank as that in the proof of (\textsc{Conseq}) implies that it is
sufficient to prove the following claim:
\begin{align*}
\textstyle{
\text{
for all $c$,
if $w\models\triple{p}{c}{q}$  then $w\models\triple{p\cap f}{c}{q\cap f}$} .
}
\end{align*}
But since  $\psi$ is pure, either $f\, w = \Heap$ for all $w\in W$ or $f\, w = \emptyset$ for all $w\in W$.  In the former case, the above implication reduces to the identity axiom, in the latter case  
 $w\models\triple{p\cap f}{c}{q\cap f}$ always holds.
 \end{proof}

\begin{lemma}[Disjunction]
For all $P,P',Q,Q'$ and $e$, the axiom 
$$
\inferrule[Disj]{
}{
  \triple{P}{e}{Q}\wedge\triple{P'}{e}{Q'} \Rightarrow \triple{P\vee P'}{e}{Q\vee Q'}
}
$$ 
is valid.
\end{lemma}

\begin{proof}
Let $\eta\in\Env$, $\rho\in\Pred^\X$, and fix $w\in\W$. 
Let $p = \den{P}_{\eta,\rho}$, $p' = \den{P'}_{\eta,\rho}$, $q = \den{Q}_{\eta,\rho}$ and $\den{Q'}_{\eta,\rho}$. 
As in the  preceding proofs, it suffices to show that 
\begin{align*}
\text{
for all $c$,
if $w\models\triple{p}{c}{q}$ and $w\models\triple{p'}{c}{q'}$, then $w\models \triple{p\cup p'}{c}{q\cup q'}$} .
\end{align*}
For this, suppose that $r\in\UAdm$ and let $h\in(p\cup p')(w) * \UNFOLD(w)(\wemp) * r$. 
We must show that $c(h) \in (q\cup q')(w) * \UNFOLD(w)(\wemp) * r$. 
Note that $h\in(p\cup p')(w) * \UNFOLD(w)(\wemp) * r$ entails that 
$h\in p(w) * \UNFOLD(w)(\wemp) * r$ or 
$h\in p'(w) * \UNFOLD(w)(\wemp) * r$. 
Therefore, by the assumption we know that 
$c(h)\in \DCl(q(w) * \UNFOLD(w)(\wemp) * r)$ or 
$c(h)\in \DCl(q'(w) * \UNFOLD(w)(\wemp) * r)$, from which 
it follows that $c(h) \in \DCl((q\cup q')(w) * \UNFOLD(w)(\wemp) * r)$ by the monotonicity of $*$ and of the closure operation. 
\end{proof}